\documentclass[preprint,12pt]{elsarticle}

\usepackage{wrapft}
\usepackage{subfigure}
\usepackage{ulem}

\usepackage[]{graphicx,psfrag,color}
\usepackage{comment}
\usepackage{amsmath,amssymb}
\usepackage{type1cm}
\usepackage{bm}

\journal{Nuclear Physics A}

\makeatother

\newcommand{\bra}[1]{\left\langle #1 \right|}
\newcommand{\brared}[1]{\langle #1 ||}

\newcommand{\ket}[1]{\left| #1 \right\rangle}
\newcommand{\ketred}[1]{|| #1 \rangle}

\newcommand{\Dhatp}{\hat{D}^{(+)}}
\newcommand{\Dp}{D^{(+)}}

\newcommand{\Hhat}{\hat{H}}

\newcommand{\Ihat}{\hat{I}}

\newcommand{\Hc}{{\cal H}}

\newcommand{\Jc}{{\cal J}}

\renewcommand{\b}{\beta}
\newcommand{\g}{\gamma}
\newcommand{\be}{$\beta$}
\newcommand{\ga}{$\gamma$}
\newcommand{\bg}{\beta,\gamma}
\newcommand{\kr}[1]{$^{#1}$Kr}

\newcommand{\Nhat}{\hat{N}}
\newcommand{\Nt}{\widetilde{N}}

\newcommand{\Dc}{{\cal D}}

\newcommand{\Dhat}{\hat{D}}

\newcommand{\Qhat}{\hat{Q}}

\newcommand{\Phat}{\hat{P}}

\newcommand{\del}{\partial}

\newcommand{\beq}{\begin{equation}}
\newcommand{\beqa}{\begin{eqnarray}}
\newcommand{\eeq}{\end{equation}}
\newcommand{\eeqa}{\end{eqnarray}}

\renewcommand{\thanks}{\footnote}

\begin{document}

\begin{frontmatter}

\title{Shape mixing dynamics in the low-lying states of proton-rich Kr isotopes}


\author{Koichi Sato$^{\rm a,b}$ and Nobuo Hinohara$^{\rm b}$}

\address{$^{\rm a}$Department of Physics, Kyoto University, Kyoto 606-8502, Japan}
\address{$^{\rm b}$RIKEN Nishina Center, Wako 351-0198, Japan}

\begin{abstract}
We study the oblate-prolate shape mixing in the low-lying states of proton-rich Kr isotopes
using the five-dimensional quadrupole collective Hamiltonian.
The collective Hamiltonian is derived microscopically
by means of the CHFB (constrained Hartree-Fock-Bogoliubov) + 
Local QRPA (quasiparticle random phase approximation) method, 
which we have developed recently 
on the basis of the adiabatic self-consistent collective coordinate method. 
The results of the numerical calculation show the importance of 
large-amplitude collective vibrations in the triaxial shape degree of freedom  
and rotational effects on the oblate-prolate shape mixing dynamics 
in the low-lying states of these isotopes.
\end{abstract}


\end{frontmatter}

\section{Introduction}
Atomic nuclei exhibit different shapes depending on the numbers of proton and neutron, 
the excitation energies, or angular momentum.
Shape coexistence phenomena, in which an excited band with a shape different 
from the shape in the ground band exists close to the ground band, 
are widely observed all over the nuclear chart. 
From the mean-field viewpoint, shape coexistence
indicates that there are two equilibrium points in the mean field, 
and appearance of a low-lying excited $0^+$ state is 
one of typical indications of the shape coexistence.

In proton-rich Kr isotopes, it has been conjectured that the oblate and prolate shapes
coexist in their low-lying states,  
since the ground bands quite different from regular rotational spectra  
\cite{Piercey1981,Varley1987} and
the low-lying excited $0^+$ states \cite{Chandler1997,Bouchez2003} were measured.
Since then, much experimental data have been accumulated 
for proton-rich Kr isotopes \cite{Poirier2004,Clement2007,Gade2005,Fischer2003}
and they support the interpretation as the oblate-prolate shape coexistence. 
The spectroscopic quadrupole moments \cite{Clement2007} suggest 
the prolate ground state in \kr{74} and \kr{76},
while the ground state of \kr{72} is assumed to be oblate from the properties of the 
$E2$ transition probabilities \cite{Gade2005}.
The systematics of the electric monopole transition strengths \cite{Chandler1997,Bouchez2003,Giannatiempo1995} 
and the excitation energies of the first excited $0^+$ states  
are consistent with the interpretation 
that a shape transition from the oblate ground state in \kr{72} 
to the prolate ground states in \kr{74} and \kr{76} takes place.

The description of these isotopes
can be a touchstone for nuclear structure models, 
and many attempts were done from several theoretical approaches 
\cite{Nazarewicz1985,Langanke2003,Petrovici2000,Bonche1985,Yamagami2001,Bender2006,Prochniak2009}. 
Recently Girod et al. \cite{Girod2009} have reproduced the shape transition
from the oblate ground state in \kr{72}  to the prolate in \kr{76}
with the HFB (Hartree-Fock-Bogoliubov)-based GCM (generator coordinate method) + 
GOA (Gaussian overlap approximation) calculation. 
Making a comparison with the result of the axial GCM calculation done 
by Bender et al. \cite{Bender2006},
they have pointed out that 
the triaxial shape plays an essential role in the shape
coexistence and the shape transition in the light Kr isotopes.
The importance of large-amplitude collective vibrations in the 
triaxial shape degree of freedom has also been shown in the previous works 
for \kr{72} \cite{Hinohara2008} and neighboring $^{68-72}$Se \cite{Hinohara2009}  
on the basis of the (1+3) dimensional calculation using 
the adiabatic self-consistent collective coordinate (ASCC) method \cite{Matsuo2000,Hinohara2007}.

In this paper, we study the low-lying states of proton-rich Kr isotopes
using a new method we have developed recently \cite{Hinohara2010}.
In this method, we determine the five-dimensional (5D) quadrupole collective Hamiltonian 
\cite{Bohr1975,Belyaev1965,Kumar1967,ProchniakRohozinski2009} 
microscopically on the basis of the ASCC method.
Local normal modes on top of each constrained HFB (CHFB) state at each point 
on the ($\bg$) plane is the main concept.
After solving the CHFB equations imposing the constraints on the deformation 
and the particle numbers, 
we solve the local QRPA (LQRPA) equations, 
which is an extension of the usual QRPA (quasiparticle random phase approximation)  
to non-HFB-equilibrium points, on top of the CHFB states.
Therefore, we call this method the CHFB+LQRPA method.
Using this method, we derive the seven quantities in the collective Hamiltonian:
the collective potential, three vibrational inertial masses, and
three rotational moments of inertia. 

One of the advantages of the CHFB+LQRPA method is that
the vibrational and rotational masses (inertial functions) determined with this method 
contain the contributions from the time-odd components of the mean-field
unlike the widely-used Inglis-Belyaev (IB) cranking masses \cite{Inglis1954,Beliaev1961}.
It is well known that the ignorance of the time-odd components in the IB masses 
breaks the self-consistency of the theory \cite{Baranger1978,Dobaczewski1981}.
Nevertheless, even in recent microscopic studies by means of 
the 5D quadrupole collective Hamiltonian 
\cite{Girod2009,Niksic2009,Li2009,Li2010,Delaroche2010}, 
the IB cranking masses are still used at least for vibrational inertial functions.
In this study, we use the pairing-plus-quadrupole (P+Q) force model 
\cite{Bes1969,Baranger1968} 
including the quadrupole-pairing force and take into account 
the contributions from the time-odd components of the mean field 
to the vibrational and rotational masses.
Inclusion of the quadrupole pairing force is essential because it is the only
term which gives the time-odd components of the mean field \cite{Hinohara2006}.

This paper is organized as follows.
In Section \ref{sec:method}, 
we summarize the procedure of deriving the 5D quadrupole collective Hamiltonian 
by means of the CHFB+LQRPA method. 
In Section \ref{sec:result}, we calculate the vibrational and rotational masses.  
By solving the collective Schr\"odinger equation, 
we calculate excitation spectra, $B(E2)$, spectroscopic quadrupole moments 
and monopole transition matrix elements  
for the low-lying states in $^{72,74,76}$Kr and discuss 
the properties of the oblate-prolate shape coexistence/mixing 
in these nuclei. 
Conclusion is given in Section \ref{sec:conclusion}.

\section{Microscopic derivation of the 5D quadrupole collective Hamiltonian}
\label{sec:method}

In this section, we summarize the procedure of microscopically deriving 
the 5D quadrupole collective Hamiltonian and the collective Schr\"odinger equation 
by means of the CHFB+LQRPA method.

\subsection{5D quadrupole Hamiltonian and collective Schr\"odinger equation}

The 5D quadrupole collective Hamiltonian is written 
in terms of the magnitude $\beta$ and the degree of triaxiality $\gamma$  
of quadrupole deformation, 
and their time derivatives, $\dot{\beta}$ and $\dot{\gamma}$, as
\begin{align}
 \Hc_{\rm coll}&= T_{\rm vib} + T_{\rm rot} + V(\bg), \label{eq:Hc} \\
 T_{\rm vib}   &= \frac{1}{2} D_{\beta\beta}(\bg)\dot{\beta}^2 +
 D_{\beta\gamma}(\bg)\dot{\beta}\dot{\gamma}
+ \frac{1}{2}D_{\gamma\gamma}(\bg)\dot{\gamma}^2, \label{eq:Tvib}\\
 T_{\rm rot} &= \frac{1}{2}\sum_{k=1}^3 \Jc_k(\bg) \omega^2_k, \label{eq:collH_BM}, 
\end{align}
where $T_{\rm vib}$, $T_{\rm rot}$ and $V$ represent  
the vibrational, rotational and collective potential energies, respectively.
The 5D quadrupole collective Hamiltonian has seven quantities to be determined: 
three vibrational inertial masses, three rotational
moments of inertia, and the collective potential.
The moments of inertia can be parametrized as
\begin{align}
 \Jc_k(\bg)  = 4 \beta^2 D_k(\bg) \sin^2 \gamma_k  \quad (k = 1, 2, 3)
\label{eq:Jc} 
\end{align}
with $\gamma_k = \gamma - (2\pi k/3)$.  
The vibrational and rotational masses are, in general, functions of the deformation parameters.
How to determine these inertial masses and the collective potential is discussed  
in the following subsection.
 
As the collective Hamiltonian (\ref{eq:Hc}) is classical, 
we quantize it according to the Pauli prescription
to obtain the collective Schr\"odinger equation
\begin{align}
 \{\hat{T}_{\rm vib} + \hat{T}_{\rm rot} + V \}
 \Psi_{\alpha IM}(\bg,\Omega) = E_{\alpha I} \Psi_{\alpha IM}(\bg,\Omega), 
\label{eq:Schroedinger}
\end{align}
where  
\begin{align}
 \hat{T}_{\rm vib} = \frac{-1}{2\sqrt{WR}} 
 \left\{ 
   \frac{1}{\beta^4} 
   \left[
     \left(
       \del_\beta 
       \beta^2 \sqrt{\frac{R}{W}} D_{\gamma\gamma} \del_\beta 
     \right)  
     - \del_\beta 
     \left(
       \beta^2 \sqrt{\frac{R}{W}} D_{\beta\gamma} \del_\gamma
     \right)
   \right] 
 \right.  \\
 \left.
 +  \frac{1}{\beta^2 \sin 3\gamma} 
 \left[
   -\del_\gamma 
   \left( 
     \sqrt{\frac{R}{W}} \sin 3\gamma D_{\beta\gamma} \del_\beta 
   \right) 
   + \del_\gamma 
   \left(
     \sqrt{\frac{R}{W}} \sin 3\gamma
     D_{\beta\beta} \del_\gamma 
   \right)
 \right] 
\right\}
\end{align}
and
\begin{align}
\hat{T}_{\rm rot} = \sum_k \frac{ \Ihat_k^2}{2\Jc_k}. 
\end{align}
Here, $R(\b,\g)$ and $W(\b,\g)$ are defined as
\begin{align}
 R(\bg) =& D_1(\bg) D_2(\bg) D_3(\bg),\\
 W(\bg) =& \left\{
 D_{\beta\beta}(\bg) D_{\gamma\gamma}(\bg) - [ D_{\beta\gamma}(\bg)]^2
\right\} \beta^{-2}. 
\end{align}

The collective wave function $\Psi_{\alpha I M}(\b,\g,\Omega)$ is specified by the total
angular momentum $I$, its projection onto the $z$-axis of the laboratory frame $M$, and
$\alpha$ distinguishing the states with the same $I$ and $M$.
As the collective potential is independent of the Euler angles $\Omega$,
the collective wave function is written in the form:
 
\begin{align}
 \Psi_{\alpha IM}(\bg,\Omega) = 
 \sum_{K={\rm even}}\Phi_{\alpha IK}(\bg)\langle\Omega|IMK\rangle, \label{eq:Psi}
\end{align}
where
\begin{align}
\langle \Omega | IMK \rangle = 
\sqrt{\frac{2I+1}{16\pi^2 (1+\delta_{k0})}} [\Dc^{I}_{MK}(\Omega) + (-)^I \Dc^I_{MK}(\Omega)].
\end{align}
Here, $\Dc^I_{MK}$ is Wigner's rotation matrix and $K$ is the projection of the angular momentum
onto the $z$-axis in the body-fixed frame.
The summation over $K$ is taken from 0 to $I$ for even $I$ and from 2 to $I-1$ for odd $I$. 

The vibrational wave functions in the body-fixed frame, $\Phi_{\alpha IK}(\bg)$,  
are normalized as
\begin{align}
 \int d\beta d\gamma  |\Phi_{\alpha I}(\bg)|^2 |G(\bg)|^{\frac{1}{2}} = 1, 
\end{align}
where  
\begin{align}
 |\Phi_{\alpha I}(\bg)|^2 \equiv \sum_{K={\rm even}} |\Phi_{\alpha IK}(\bg)|^2,
\end{align}
and the volume element $G(\bg)$ is given by
\begin{align} \label{eq:metric}
 G(\bg) = 4\beta^8 W(\bg)R(\bg) \sin^2 3\gamma.
\end{align}
The symmetries and boundary conditions of the collective Hamiltonian and 
wave function are discussed in Ref. \cite{Kumar1967}.

\subsection{Constrained HFB + Local QRPA method}
In this subsection, we summarize the method of determining the
inertial masses and collective potential.
This method can be regarded as an approximation of 
the two-dimensional (2D) version of the ASCC method.
In this method,  we solve the local QRPA equations at each point 
on the $(\beta, \gamma)$ plane 
on top the CHFB state, and therefore we call it the CHFB+LQRPA method.
(See Ref.  \cite{Hinohara2010} for details.)

First, we solve the CHFB equation given by
  
\begin{align}
 \delta \bra{\phi(\bg)} \Hhat_{\rm CHFB}(\bg) \ket{\phi(\bg)} = 0,  \label{eq:CHFB} 
\end{align}
\begin{align}
 \Hhat_{\rm CHFB}(\bg) = \Hhat - \sum_{\tau}\lambda^{(\tau)}(\bg)\Nt^{(\tau)} 
- \sum_{m = 0, 2} \mu_{m}(\bg) \Dhatp_{2m}  \label{eq:H_CHFB} 
\end{align}
with four constraints
\begin{align}
 \bra{\phi(\bg)} \Nhat^{(\tau)} \ket{\phi(\bg)} = N^{(\tau)}_0,  
\quad (\tau = n, p) \\
 \bra{\phi(\bg)} \Dhatp_{2m} \ket{\phi(\bg)} = \Dp_{2m}, \quad (m = 0, 2)
\end{align}
where $\Dhatp_{2m}$ denotes Hermitian quadrupole operators, 
$\Dhat_{20}$ and $(\Dhat_{22} + \Dhat_{2-2})/2$ (for  $m = 0$ and  2, respectively).
We define the quadrupole deformation variables ($\beta, \gamma$) 
in terms of the expectation values of the quadrupole operators: 
\begin{align}
 \beta\cos\gamma &=  \eta D^{(+)}_{20},  \label{eq:definition1} \\
 \frac{1}{\sqrt{2}} \beta\sin\gamma &= \eta D^{(+)}_{22},
 \label{eq:definition2} 
\end{align}
where $\eta$ is a scaling factor (to be discussed in Section 2-4). 

Second, we solve the LQRPA equations on top of the CHFB states obtained above,

\begin{align}
 \delta & \bra{\phi(\bg)} [ \Hhat_{\rm CHFB}(\bg), \Qhat^i(\bg) ] \nonumber \\
 &- \frac{1}{i}\Phat_i(\bg) \ket{\phi(\bg)}  = 0, \label{eq:LQRPA1}\\
 \delta & \bra{\phi(\bg)} [ \Hhat_{\rm CHFB}(\bg), \frac{1}{i} \Phat_i(\bg)]
 \nonumber \\
 &- C_i(\bg) \Qhat^i(\bg) \ket{\phi(\bg)} = 0,  \quad\quad (i=1, 2). 
\label{eq:LQRPA2}
\end{align}
Here the infinitesimal generators, 
$\Qhat^i(\bg)$ and $\Phat_i(\bg)$, 
are local operators defined, for a given set of quadrupole deformation variables $(\bg)$, 
with respect to the CHFB state $\ket{\phi(\bg)}$.  
The quantity $C_i(\bg)$ is related to 
the eigenfrequency $\omega_i(\bg)$ of the local normal mode  
through $\omega_i^2(\bg)=C_i(\bg)$. 
Note that these equations are valid also for regions with negative curvature 
($C_i(\bg)<0$) where $\omega_i(\bg)$ takes an imaginary value. 
For selecting two collective modes from among many LQRPA modes,    
we use the criterion formulated in Ref. \cite{Hinohara2010}.  

Using the time derivatives of $D^{(+)}_{2m}$, 
the collective vibrational energy (\ref{eq:Tvib}) is written in a form of
\begin{equation}
T_{\rm vib}=\frac{1}{2}M_{00}(\dot D_{20}^{(+)})^2+M_{02}\dot D_{20}^{(+)}\dot D_{22}^{(+)}+\frac{1}{2}M_{22}(\dot D_{22}^{(+)})^2.
\label{eq:TvibD}
\end{equation}
In the 2D ASCC method, the vibrational part of the collective Hamiltonian is given by
\beq
T_{\rm vib}=\frac{1}{2}\sum_{i}\dot q_i^2
\eeq
under the condition that there exists a collective coordinate system ${(q_1,q_2)}$ where
the vibrational masses and stiffness tensors can be diagonalized globally.
Assuming one-to-one correspondence between ${(q_1,q_2)}$ and $(D^{(+)}_{20},D^{(+)}_{22})$, 
the vibrational masses in Eq.~(\ref{eq:TvibD})  are written as
\begin{align}
 M_{mm'}(\bg) = \sum_{i=1,2} \frac{\del q^i}{\del
 \Dp_{2m}} \frac{\del q^i}{\del \Dp_{2m'}}. \label{eq:M_mm}
\end{align} 
In this approximate version of the 2D ASCC method, we equate the momentum operator 
in the ASCC method with the one obtained by the LQRPA equations,  
and then the partial derivatives are easily evaluated as
\begin{align}
 \frac{\del \Dp_{20}}{\del q^i} =& \frac{\del}{\del q^i}\bra{\phi(\bg)}
 \Dhatp_{20} \ket{\phi(\bg)} \notag\\
  =& \bra{\phi(\bg)} [\Dhatp_{20}, \frac{1}{i}\Phat_i 
 (\bg)] \ket{\phi(\bg)}, \\
 \frac{\del \Dp_{22}}{\del q^i} =& \frac{\del}{\del q^i}\bra{\phi(\bg)}
 \Dhatp_{22} \ket{\phi(\bg)} \notag\\
  =& \bra{\phi(\bg)} [\Dhatp_{22}, \frac{1}{i}\Phat_i 
 (\bg)] \ket{\phi(\bg)}, 
\end{align}
without need of numerical derivatives. 


The rotational moments of inertia are calculated by solving the LQRPA equation 
\cite{Hinohara2010} for rotation on top of each CHFB state:

\begin{align}
 \delta \bra{\phi(\bg)} [\Hhat_{\rm CHFB}, \hat{\Psi}_k(\bg)]
- \frac{1}{i} (\Jc_k)^{-1} \Ihat_k \ket{\phi(\bg)} &= 0, \\ 
\bra{\phi(\bg)} [\hat{\Psi}_k(\bg), \Ihat_{k'}]\ket{\phi(\bg)} &= i \delta_{kk'},
\end{align}
where $\hat{\Psi}_k(\bg)$ and $\Ihat_k$ represent the rotational angle
and the angular momentum operators with respect to the principal axes
associated with the CHFB state $\ket{\phi(\bg)}$.
This is an extension of the Thouless-Valatin equation \cite{Thouless1962} 
for the HFB equilibrium state to non-equilibrium CHFB states.  
We call  $\Jc_k(\bg)$ and $D_k(\bg)$ determined by the above equations 
and  Eq.~(\ref{eq:Jc}) 
`LQRPA moments of inertia' and `LQRPA rotational masses', respectively.

Last, we solve the collective Schr\"odinger equation 
with the 5D quadrupole collective Hamiltonian 
whose collective potential and inertial masses are calculated as above.

\subsection{Calculation of electric quadrupole ($E2$) transitions and moments} 
\label{subsec:Electric}

The $E2$ operator in the intrinsic frame is defined as
\beq
\hat D^{(E2)} _{m}=\sum_{\tau=n, p}e^{(\tau)}_{\rm eff}\hat D^{(\tau)}_{2m},
\eeq
where $\hat D^{(\tau)}_{2m}$ are the quadrupole operator for protons or neutrons, 
and thus $\sum_{\tau}\hat D^{(\tau)}_{2m}=\hat D_{2m}$.
The $E2$ operator in the laboratory frame is given by
\beq
\hat D^{\prime (E2)}_{m}=\sum_{m'} \mathcal{D}^2_{mm'}(\Omega)\hat D^{(E2)}_{m'} 
\eeq
with Wigner's $\mathcal{D}$ functions.
The reduced $E2$ transition probability $B(E2)$ 
and the spectroscopic quadrupole moment  $Q$ are given by  
\begin{align}
  B(E2;\alpha I \rightarrow \alpha'I')
=\left(2I+1\right)^{-1}|\brared{\alpha I}\hat D'^{(E2)} \ketred{\alpha' I'}|^2 \label{eq:BE2}\\
  Q(\alpha I)=\sqrt{\frac{16\pi}{5}}\bra{\alpha,I,M=I}\hat D'^{(E2)}_0\ket{\alpha, I, M=I}
\end{align}

The reduced matrix element in Eq.~(\ref{eq:BE2}) is easily evaluated through  the relation,
\begin{align}
\bra{\alpha,I,M=I}\hat D'^{(E2)}_0\ket{\alpha ', I', M'=I}
=\begin{pmatrix}
  I & 2 & I' \\
 -I & 0 & I
\end{pmatrix}
\brared{\alpha I}\hat D'^{(E2)} \ketred{\alpha' I'}. 
\end{align}
Substituting Eq. (\ref{eq:Psi}) into  $\ket{\alpha, I, M}$, we obtain 
\begin{align}
&\brared{\alpha I} \hat D'^{(E2)} \ketred{\alpha ' I'} \notag\\
=& \sqrt{(2I+1)(2I'+1)}(-)^I\sum_{K}
\left\{
\begin{pmatrix}
  I & 2 & I' \\
 -K & 0 & K
\end{pmatrix}\right.
\bra{\Phi_{\alpha,I,K}}\hat D^{(E2)}_{0+}\ket{\Phi_{\alpha', I',K}} \notag\\
+&\sqrt{(1+\delta_{K0})}
\left[
\begin{pmatrix}
  I & 2 & I' \\
 -K-2 & 2 & K
\end{pmatrix}\right.
\bra{\Phi_{\alpha,I,K+2}}\hat D^{(E2)}_{2+}\ket{\Phi_{\alpha', I',K}} \notag\\
+&\left.\left.(-)^{I+I'}
\begin{pmatrix}
  I & 2 & I' \\
 K & 2 & -K-2
\end{pmatrix}
\bra{\Phi_{\alpha,I,K}}\hat D^{(E2)}_{2+}\ket{\Phi_{\alpha', I',K+2}} \right]\right\},
\label{eq:redmat}
\end{align}
with $\hat D^{(E2)}_{m+}=(\hat D^{(E2)}_{m}+\hat D^{(E2)}_{-m})/2$.

The intrinsic matrix elements appearing in the above formula are evaluated as 
\begin{equation}
\bra{\Phi_{\alpha,I,K}}\hat D^{(E2)}_{m+}\ket{\Phi_{\alpha', I',K'}}
=\int d\beta d\gamma |G|^{\frac{1}{2}} \Phi^*_{\alpha IK}(\beta,\gamma)
D^{(E2)}_{m+}(\beta,\gamma) \Phi_{\alpha' I'K'}(\beta,\gamma), 
\end{equation}
where 
\begin{equation}
D^{(E2)}_{m+}(\beta,\gamma)=\bra{\phi(\beta,\gamma)}\hat D^{(E2)}_{m+}
\ket{\phi(\beta,\gamma)}.
\end{equation}

We also calculate the electric monopole transition matrix elements.  
It is evaluated, to the lowest order in $\beta$, by \cite{Kumar1975}.

 \begin{align}
 \rho(E0;i \rightarrow f)&=\frac{3}{4}\frac{Z}{\pi} \bra{i}
 \beta^2 \ket{f}\delta_{I_i I_f} \notag\\ 
    &=\frac{3}{4}\frac{Z}{\pi}\delta_{I_i I_f} \sum_{K_i,K_f}
\int d\g d\b |G|^{\frac{1}{2}} \Phi^*_{\alpha_i I_iK_i}(\beta,\gamma)\beta^2  
\Phi_{\alpha_f I_fK_f}(\beta,\gamma). \label{eq:rhoE0}
 \end{align}

\subsection{Details of numerical calculation}
\label{subsec:Details}

In this work, we adopt a version of the P+Q interaction which includes the quadrupole 
pairing interaction as well as the monopole pairing interaction.
We take two major shells as the active model space for protons and neutrons.
The interaction parameters are determined as follows:
for $^{72}$Kr, we use the same values of the monopole pairing interaction strength $G^{(\tau)}_0$ 
and the quadrupole particle-hole interaction strength $\chi$ 
as used in the previous ASCC work \cite{Hinohara2008}, 
which were determined  such that the magnitude of the quadrupole deformation 
\be~and monopole pairing gaps
at the oblate and prolate HFB minima obtained in the Skyrme-HFB calculation 
\cite{Yamagami2001} were approximately reproduced.
These values are scaled for $^{74}$Kr and $^{76}$Kr 
assuming a simple mass number dependence 
\cite{Baranger1968}: 
$G_0^{(\tau)}\sim A^{-1}$ and $\chi'\equiv\chi b^4 \sim A^{-\frac{5}{3}}$.
We determine the quadrupole-pairing strength 
following the Sakamoto-Kishimoto prescription \cite{Sakamoto1990}.
As in the conventional treatment of the P+Q model, we omit the Fock term.
Therefore, we use the abbreviation HB instead of HFB in the following.   
The CHB + LQRPA equations are solved at each mesh point on the
$(\b,\g)$ plane. This two-dimensional mesh consists of the
60 $\times$ 60 points specified by
\begin{align}
 \beta_i  = (i - 0.5) \times 0.01, \quad (i = 1, \cdots 60), \\
 \gamma_j = (j - 0.5) \times 1^\circ, \quad (j = 1, \cdots 60), 
\end{align}
with which we can perform a parallel computation easily for each mesh point.  
The effective charges are given by $e^{(n)}_{\rm eff}= \delta e_{\rm pol}$ for neutrons and 
$e^{(p)}_{\rm eff}= e+\delta e_{\rm pol}$ for protons.
In this paper, we use the polarization charge $\delta e_{\rm pol}= 0.834e$, 
which is determined such that the calculated value of 
$B(E2;2^+_1 \rightarrow 0_1^+)$ reproduces the experimental data for \kr{74} \cite{Clement2007}. 

\section{Results of calculation and discussion}
\label{sec:result}

\begin{figure}[htb]
\begin{center}
\subfigure[$^{72}$Kr]{\includegraphics[height=0.35\textwidth,keepaspectratio,clip,trim=60 0 160 0]{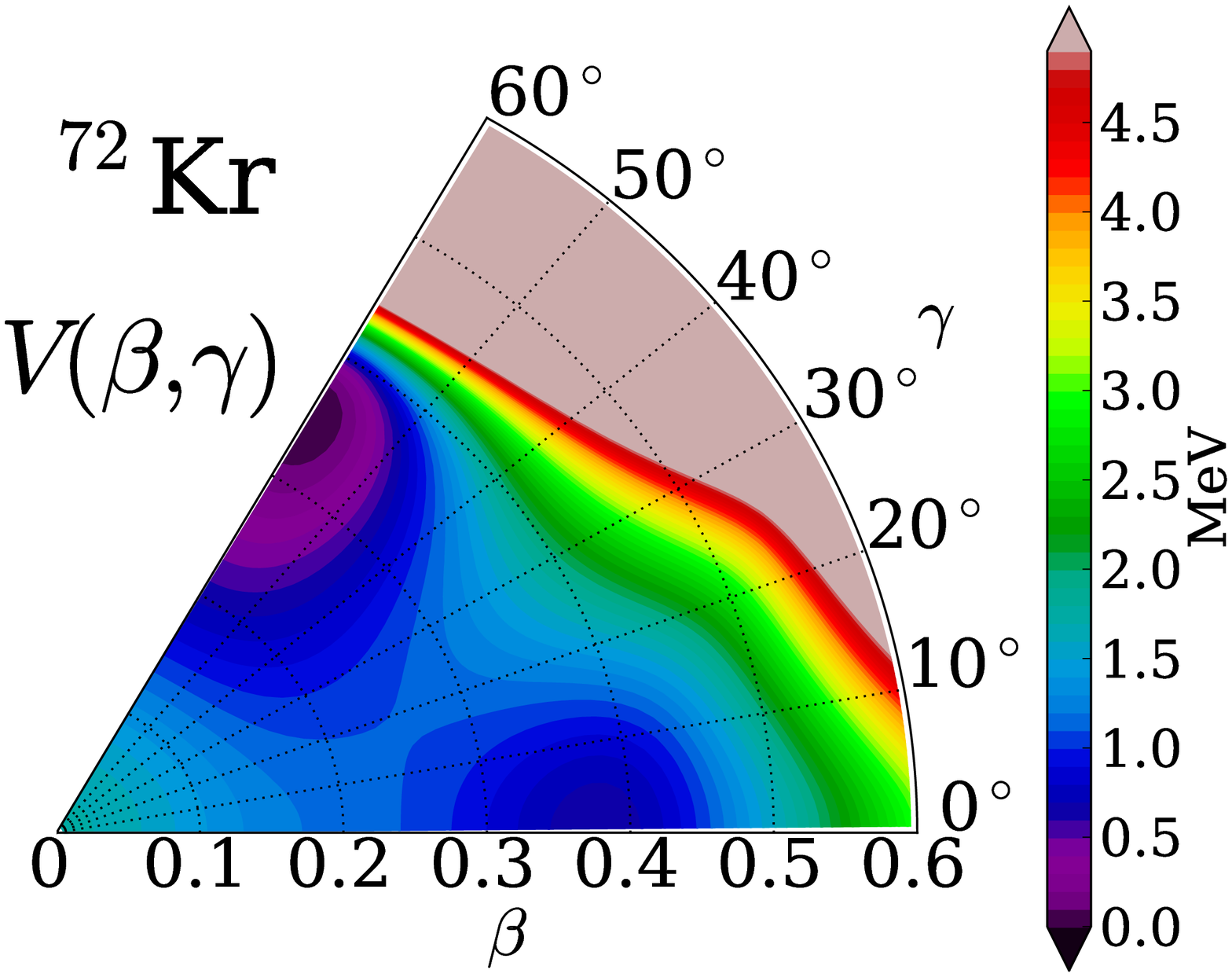}} 
\subfigure[$^{74}$Kr]{\includegraphics[height=0.35\textwidth,keepaspectratio,clip,trim=60 0 160 0]{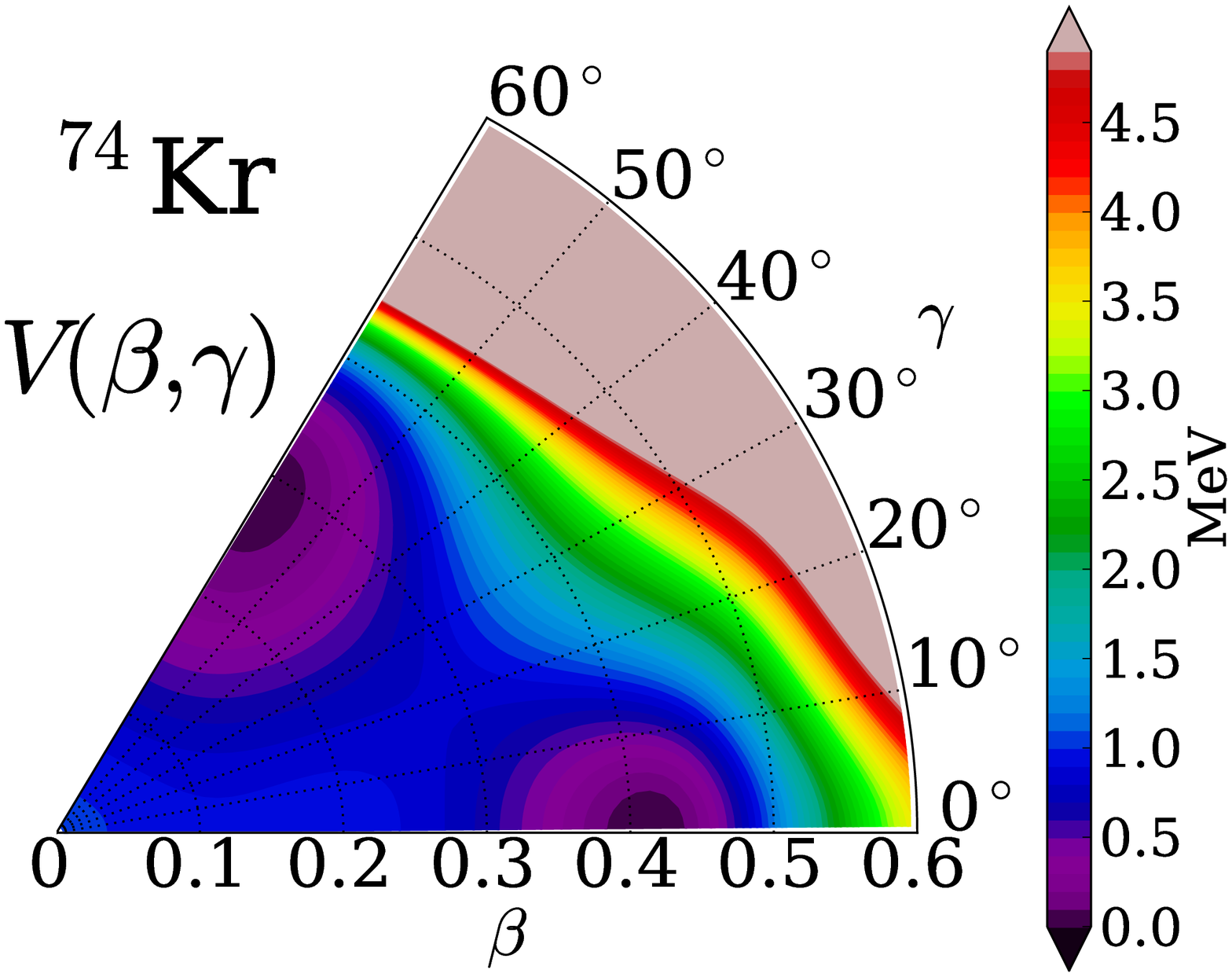}} 
\subfigure[$^{76}$Kr]{\includegraphics[height=0.35\textwidth,keepaspectratio,clip,trim=60 0 60 0]{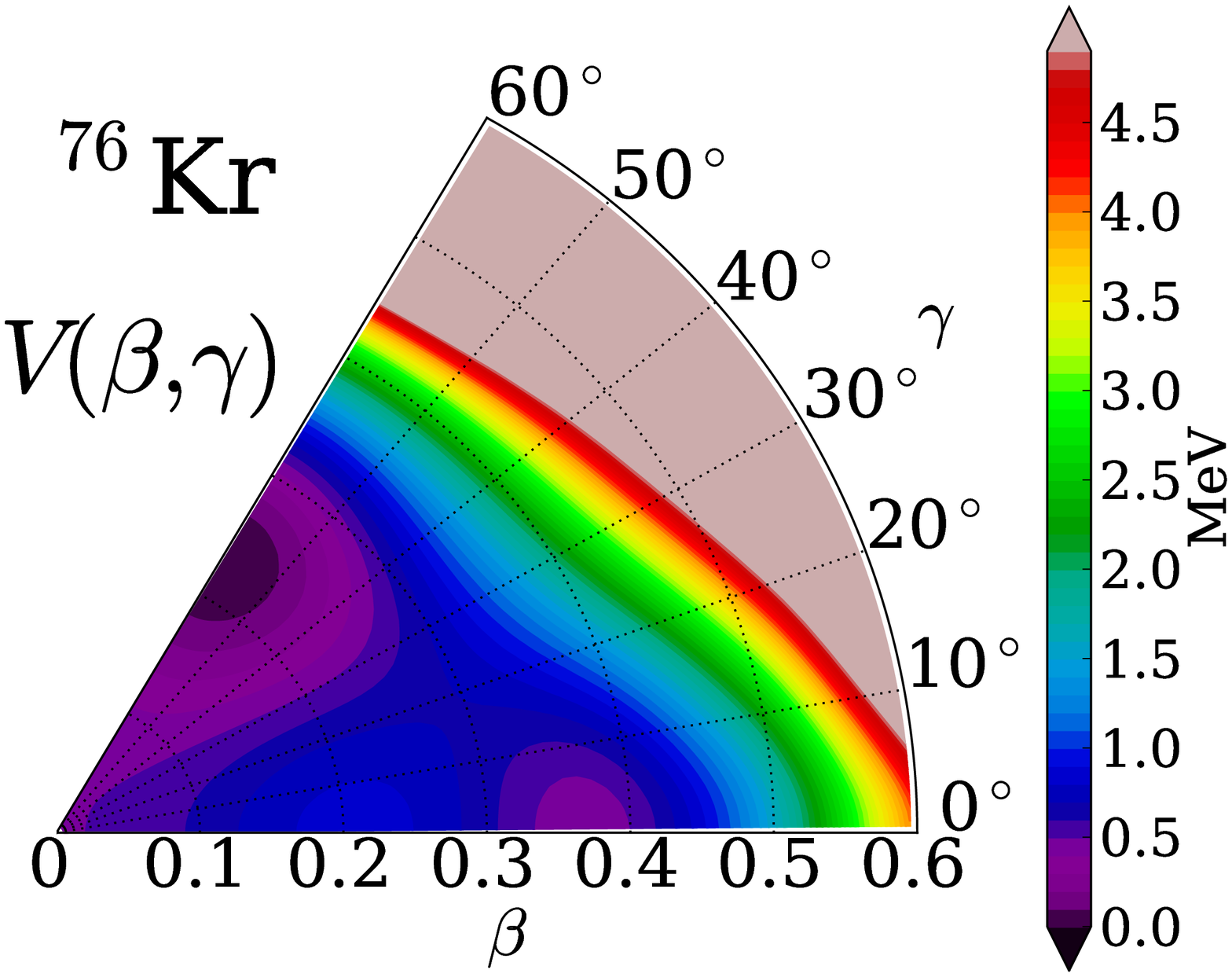}} 
\end{center}
\caption{Collective potential energy surfaces $V(\beta,\gamma)$ for $^{72,74,76}$Kr. 
The regions higher than 5 MeV (measured from the HFB minima) 
are colored rosy-brown. }
\label{fig:potential}
\end{figure}

 In  this section, we present the result of the numerical calculation
 for $^{72}$Kr, $^{74}$Kr, and  $^{76}$Kr and discuss properties of low-lying 
collective modes of excitation in these nuclei.
 Figure~\ref{fig:potential} shows the collective potential calculated for $^{72-76}$Kr.
 All of the potential have two local minima: one is oblate and the other is prolate.
 The spherical shape is a local maximum in any case.
 In \kr{72}, the absolute minimum is oblate and the prolate minimum is 
approximately 600 keV higher.
 In contrast, the prolate minimum is lower than the oblate one in \kr{74}, 
which suggest there may occur
 a shape transition from oblate to prolate between these isotopes.
 The experimental data \cite{Clement2007} 
of the spectroscopic quadrupole moments for \kr{74} and \kr{76}
 indicate that they are prolate in the ground band.
 One may expect that the absolute minimum is oblate in \kr{72} 
and it becomes prolate for $A \geq 74$.  
 However, it is not as simple as one might expect:
 the oblate minimum becomes the absolute minimum again in \kr{76}. 
 This does not necessarily mean that our calculation fails in the reproduction of the 
 ground band shape in \kr{76}. 
As we shall discuss below, we have to take into account 
dynamical effects beyond the mean field.

\begin{figure}[htb]
\begin{center}
\subfigure[$D_{\b\b}(\bg)$]{\includegraphics[height=0.37\textwidth,keepaspectratio,clip,trim=60 0 160 0]
{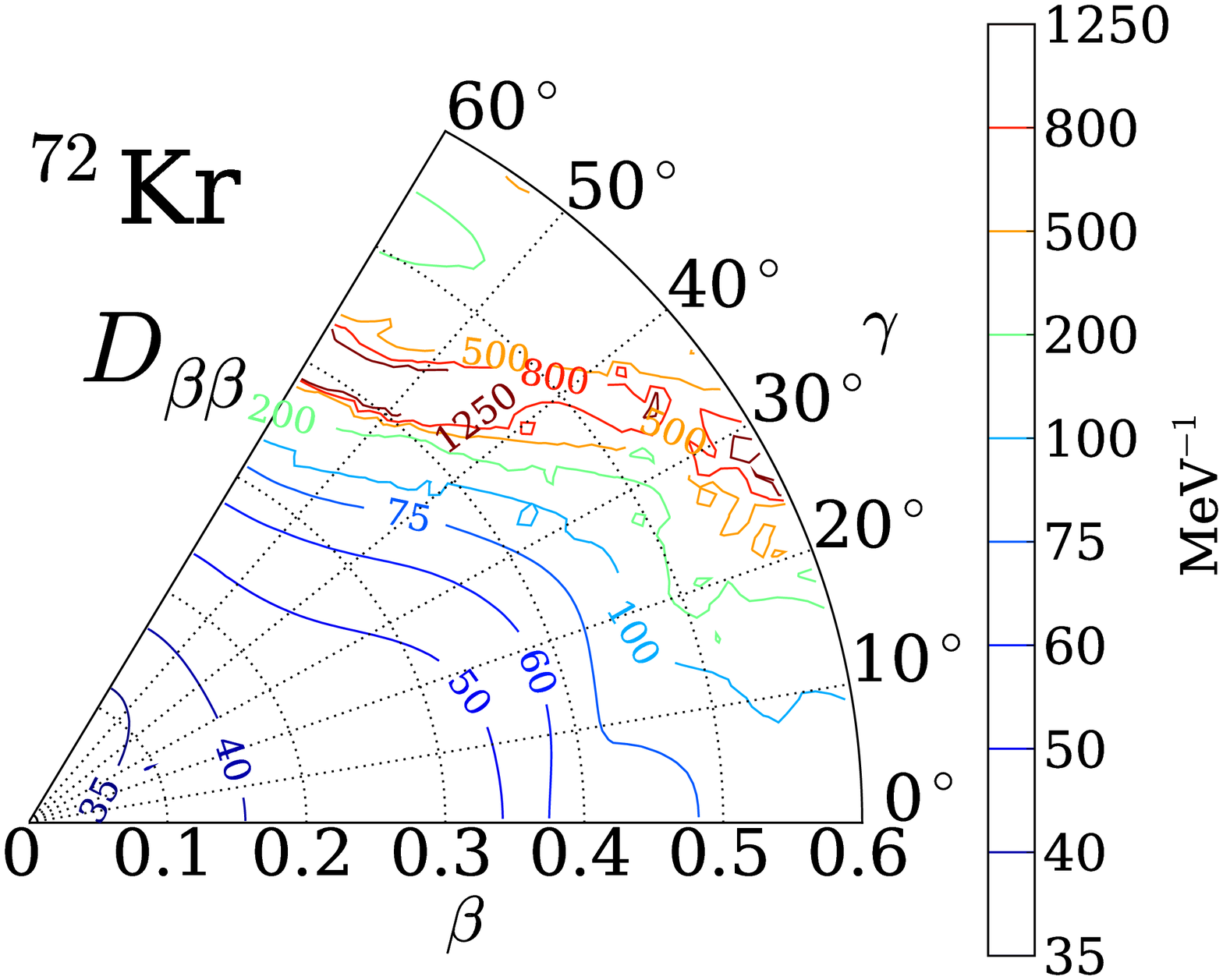}} 
\subfigure[$D_{\b\g}(\bg)/\beta$]{\includegraphics[height=0.37\textwidth,keepaspectratio,clip,trim=60 0 160 0]
{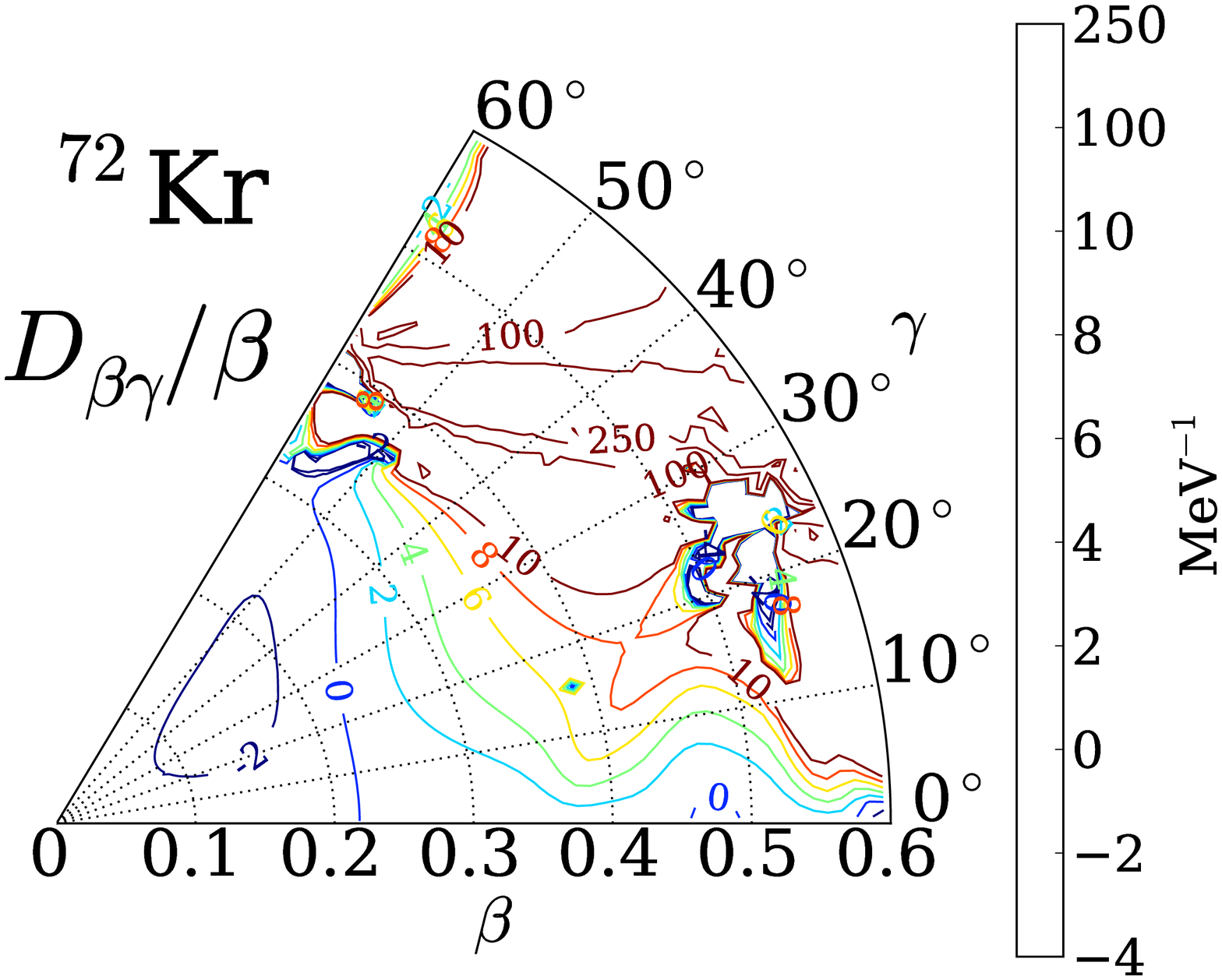}} 
\subfigure[$D_{\g\g}(\bg)/\beta^2$]{\includegraphics[height=0.37\textwidth,keepaspectratio,clip,trim=60 0 160 0]
{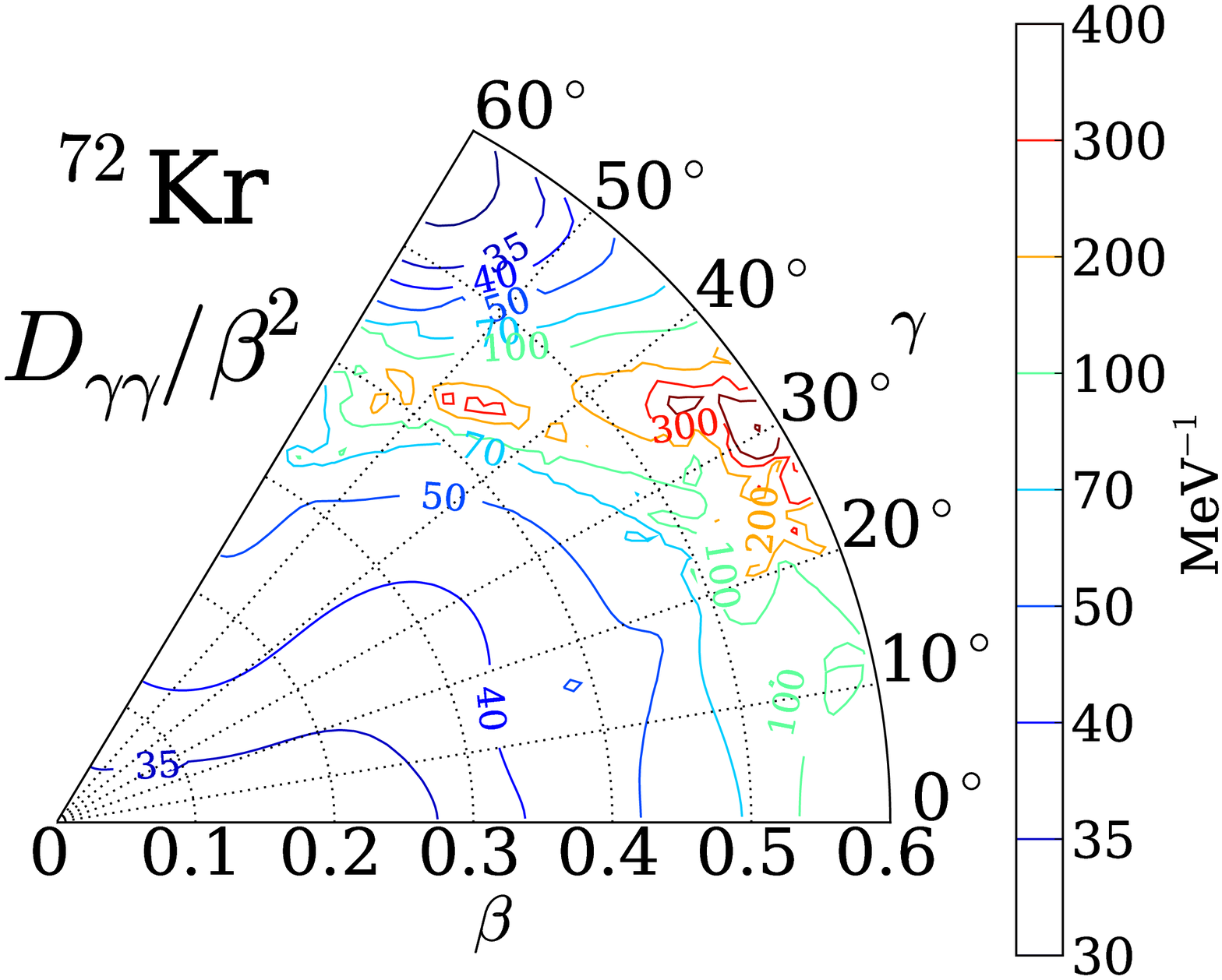}} 
\end{center}
\caption{Vibrational inertial masses, 
$D_{\beta\beta}(\bg)$, $D_{\beta\gamma}(\bg)/\beta$  
and $D_{\gamma\gamma}(\bg)/\beta^2$, in unit of MeV$^{-1}$ 
calculated for $^{72}$Kr.}
\label{fig:vibM72Kr}
\end{figure}

\begin{figure}[htb]
\begin{center}
\subfigure[$D_{\b\b}/D_{\b\b}^{\rm (IB)}$]{\includegraphics[width=0.4\textwidth,keepaspectratio,clip,trim=0 0 160 0]
{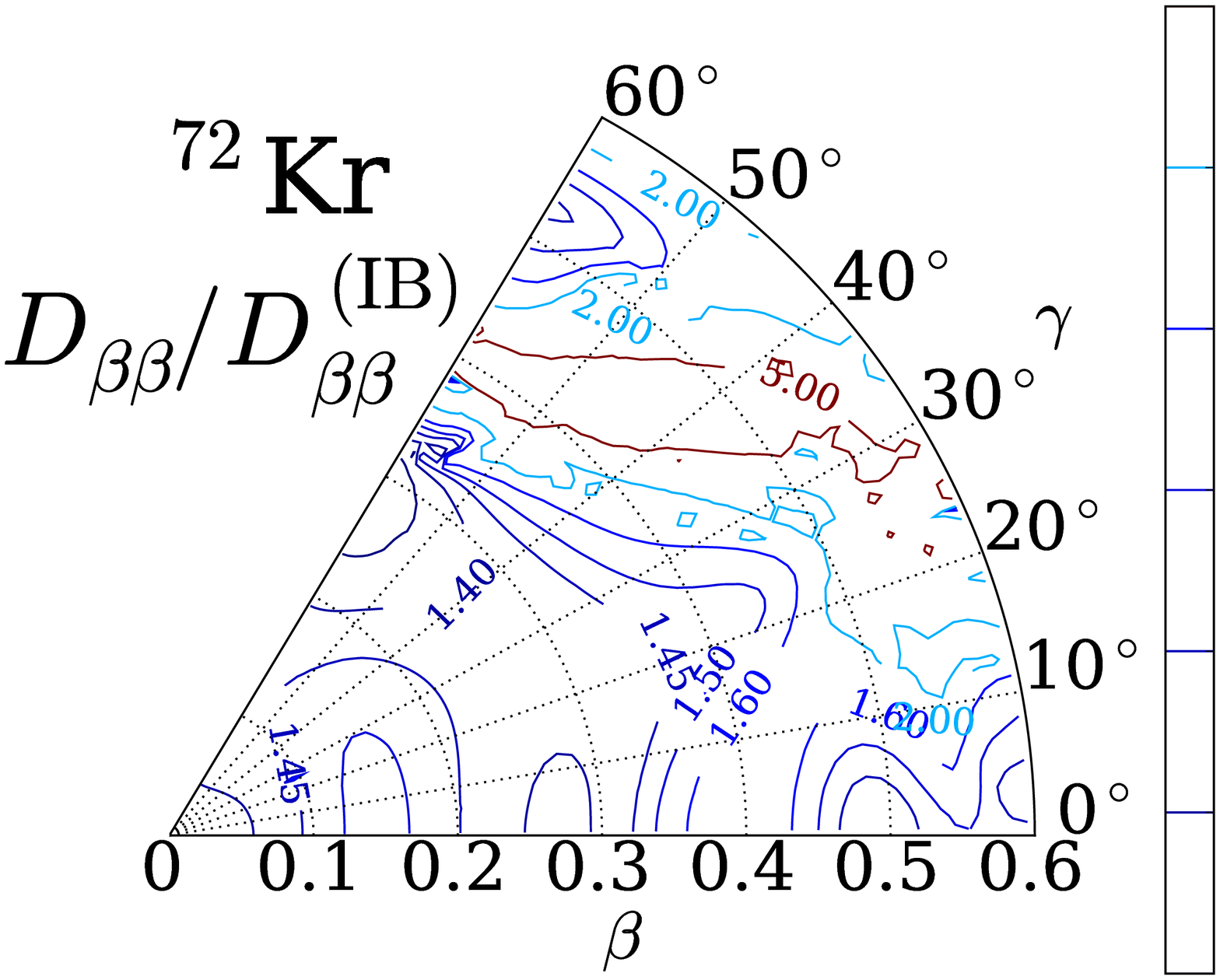}} 
\subfigure[$D_{\g\g}/D_{\g\g}^{\rm (IB)}$]{\includegraphics[width=0.4\textwidth,keepaspectratio,clip,trim=0 0 160 0]
{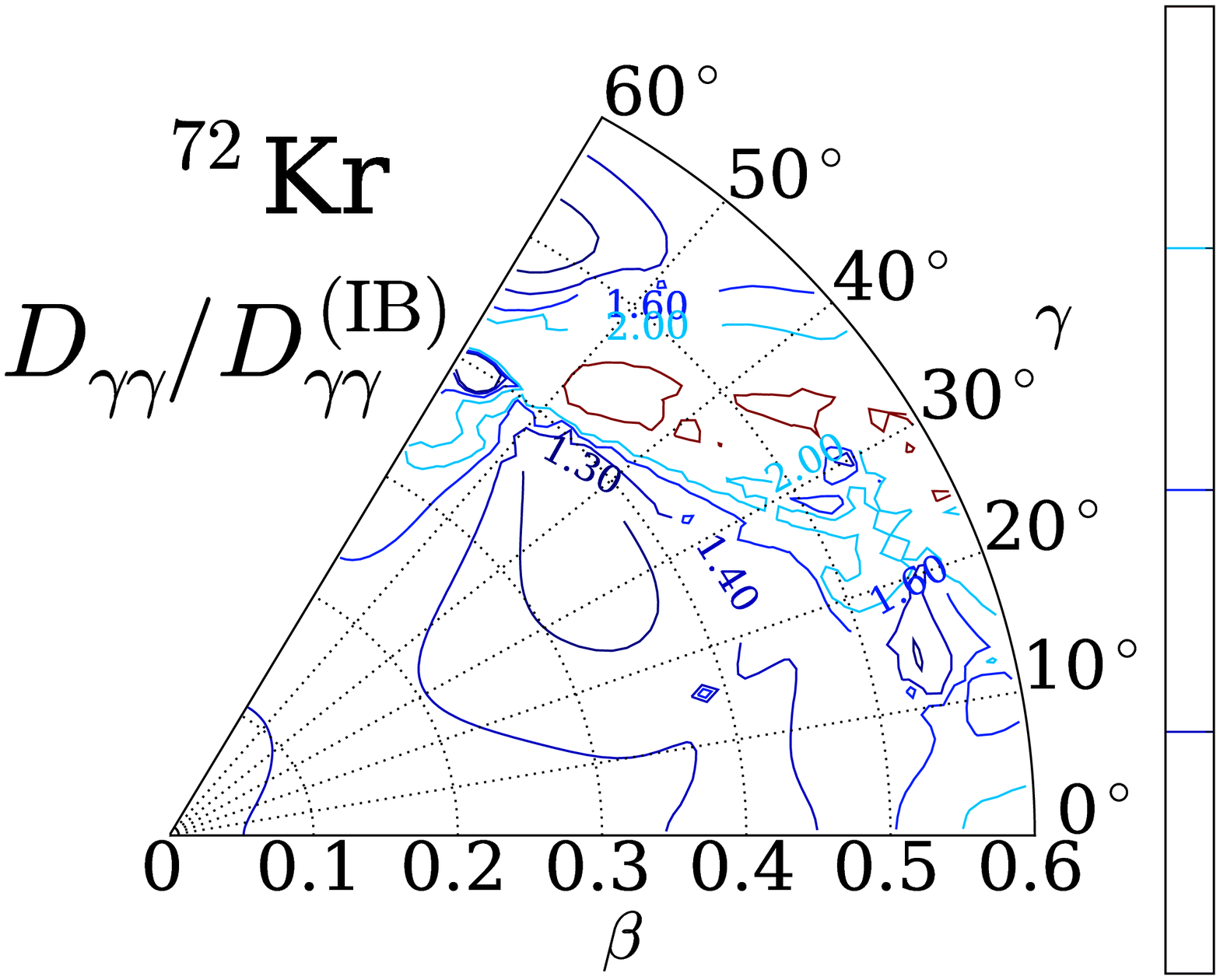}} 
\end{center}
\caption{Ratios of the LQRPA vibrational masses to the IB cranking masses, 
$D_{\beta\beta} / D^{({\rm IB})}_{\beta\beta}$ and 
$D_{\gamma\gamma} / D^{({\rm IB})}_{\gamma\gamma}$, 
calculated for $^{72}$Kr.}
\label{fig:ratiovib}
\end{figure}

 We show in Fig.~\ref{fig:vibM72Kr} the $\b - \g$ dependence of the vibrational masses for \kr{72}.
 One can see their deviation from a constant.
 Although they have a complicated structure at large deformation compared 
to that at small deformation, 
 this region hardly contributes to the calculated results as the potential is high.
 We also depict the ratios of the LQRPA vibrational masses to the IB vibrational masses 
in Fig.~\ref{fig:ratiovib}.

\begin{figure}[htb]
\begin{center}
\subfigure[$\Jc_{1}(\bg)$]{\includegraphics[height=0.33\textwidth,keepaspectratio,clip,trim=20 0 160 0]{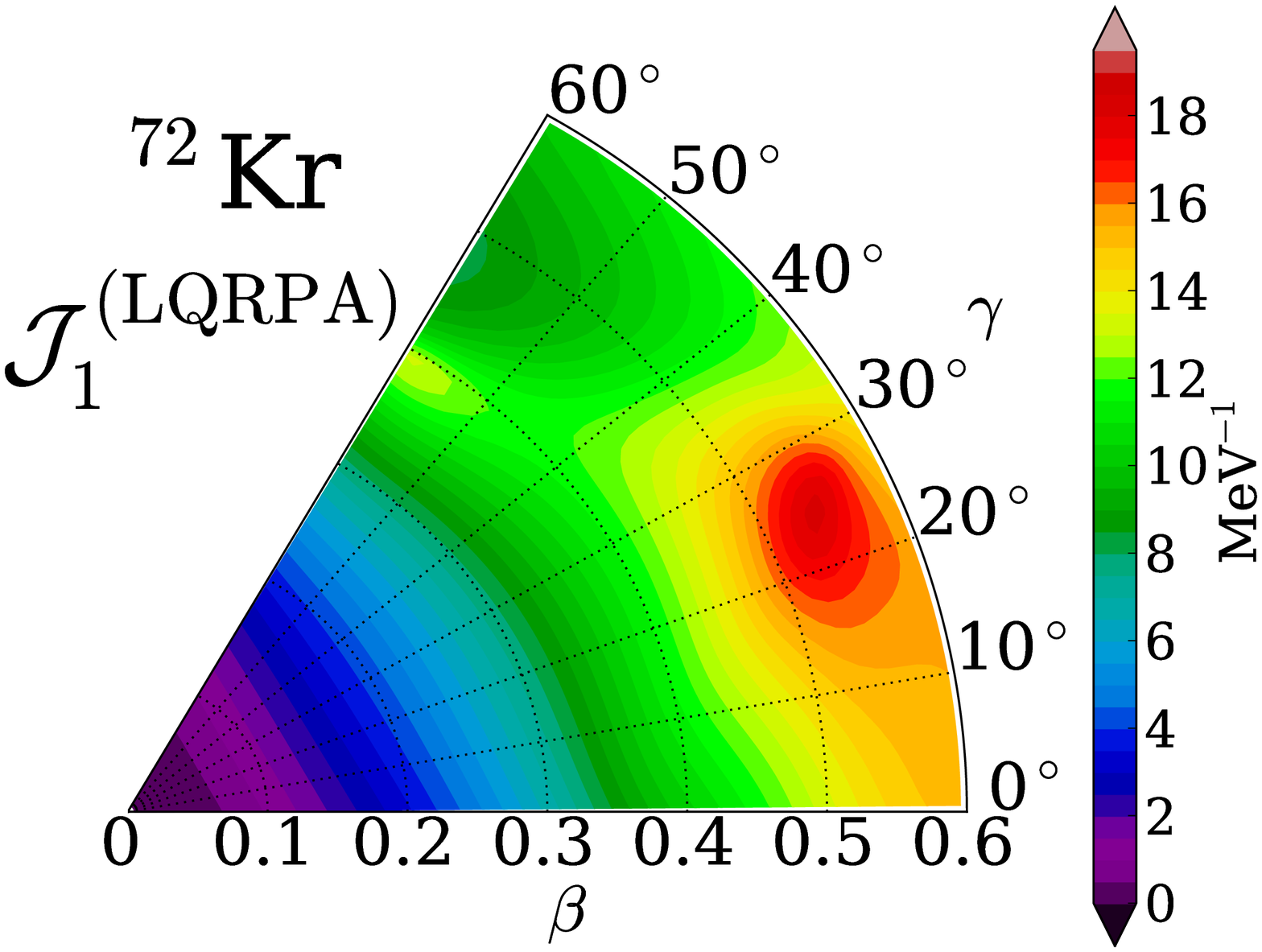}} 
\subfigure[$\Jc_{2}(\bg)$]{\includegraphics[height=0.33\textwidth,keepaspectratio,clip,trim=20 0 160 0]{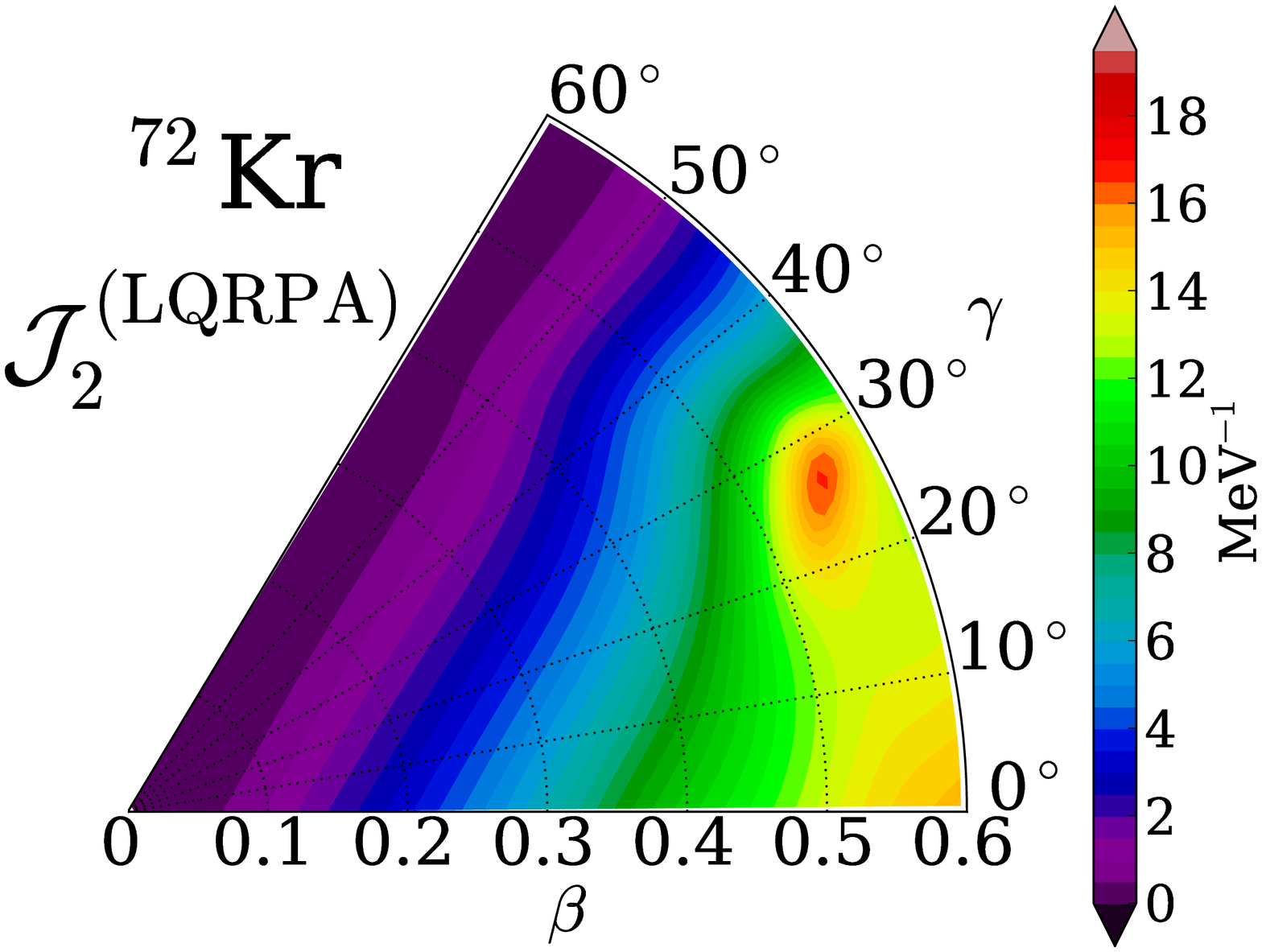}} 
\subfigure[$\Jc_{3}(\bg)$]{\includegraphics[height=0.33\textwidth,keepaspectratio,clip,trim=20 0 80 0]{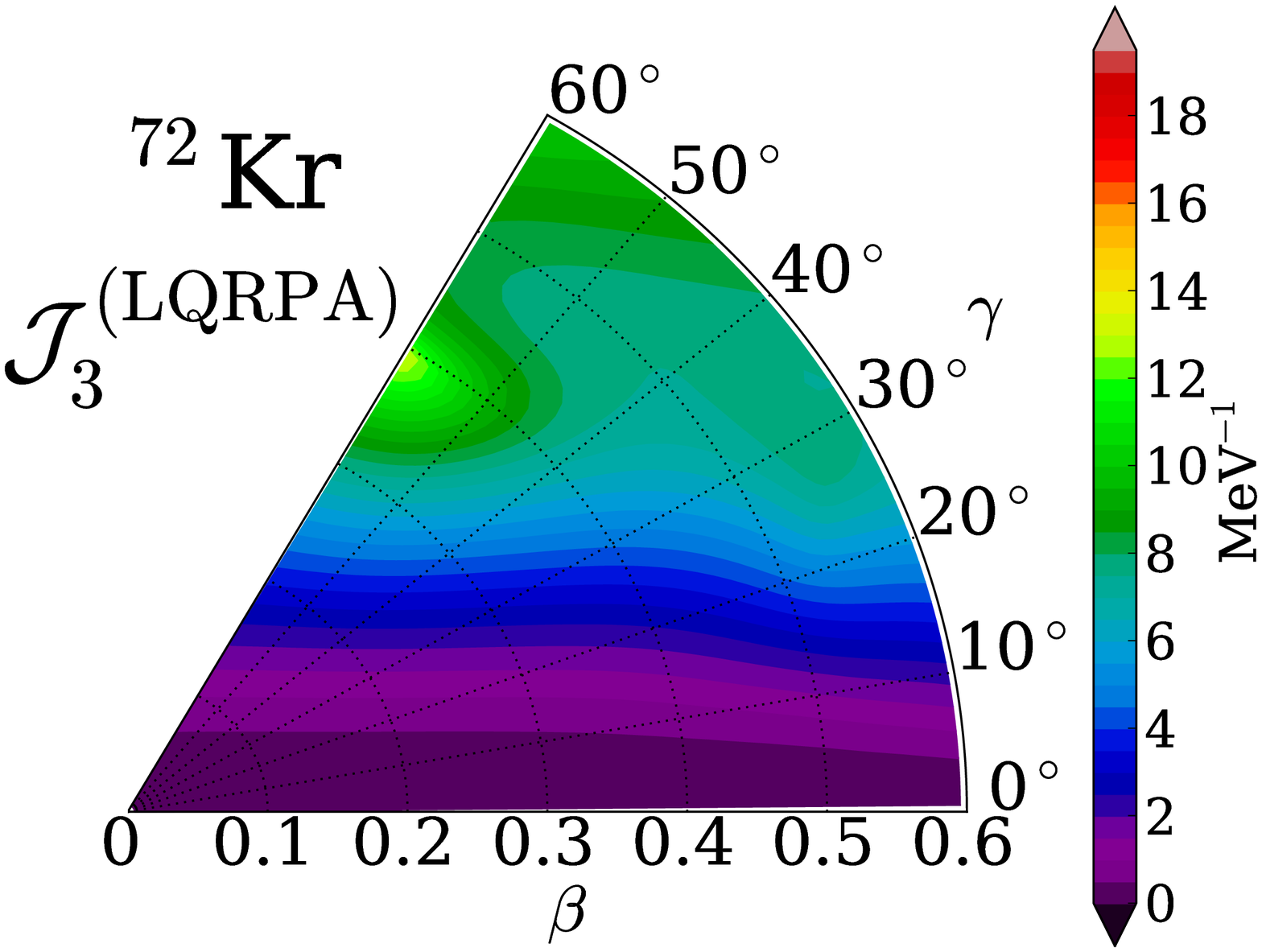}} 
\end{center}
\caption{LQRPA rotational moments of inertia in unit of MeV$^{-1}$, calculated for $^{72}$Kr.}
\label{fig:MoI72Kr}
\end{figure}

\begin{figure}[htb]
\begin{center}
\subfigure[$\Jc_1/\Jc_1^{\rm (IB)}$]
{\includegraphics[height=0.33\textwidth,keepaspectratio,clip,trim=30 0 160 0]{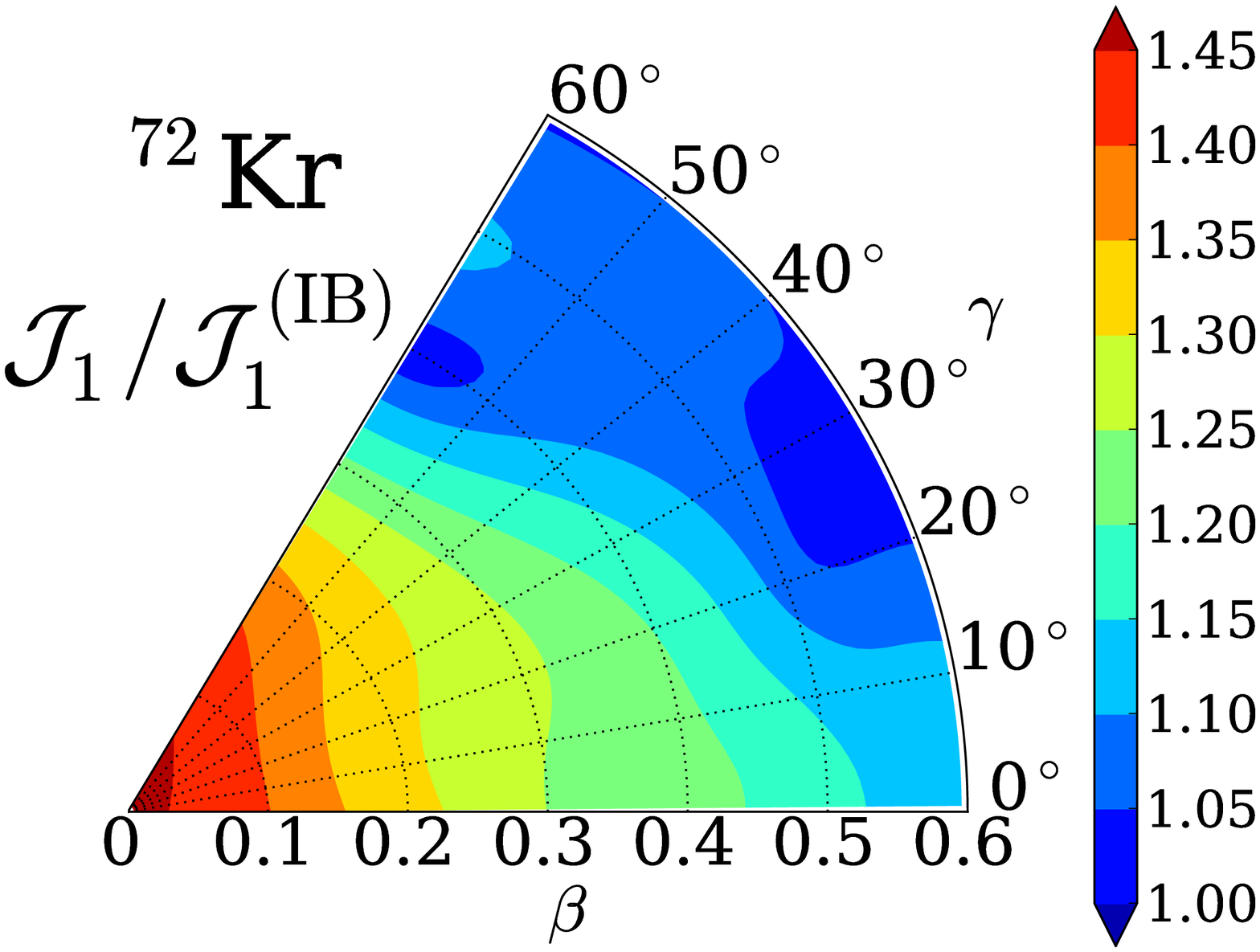}} 
\subfigure[$\Jc_{2}/\Jc_2^{\rm (IB)}$]
{\includegraphics[height=0.33\textwidth,keepaspectratio,clip,trim=30 0 160 0]{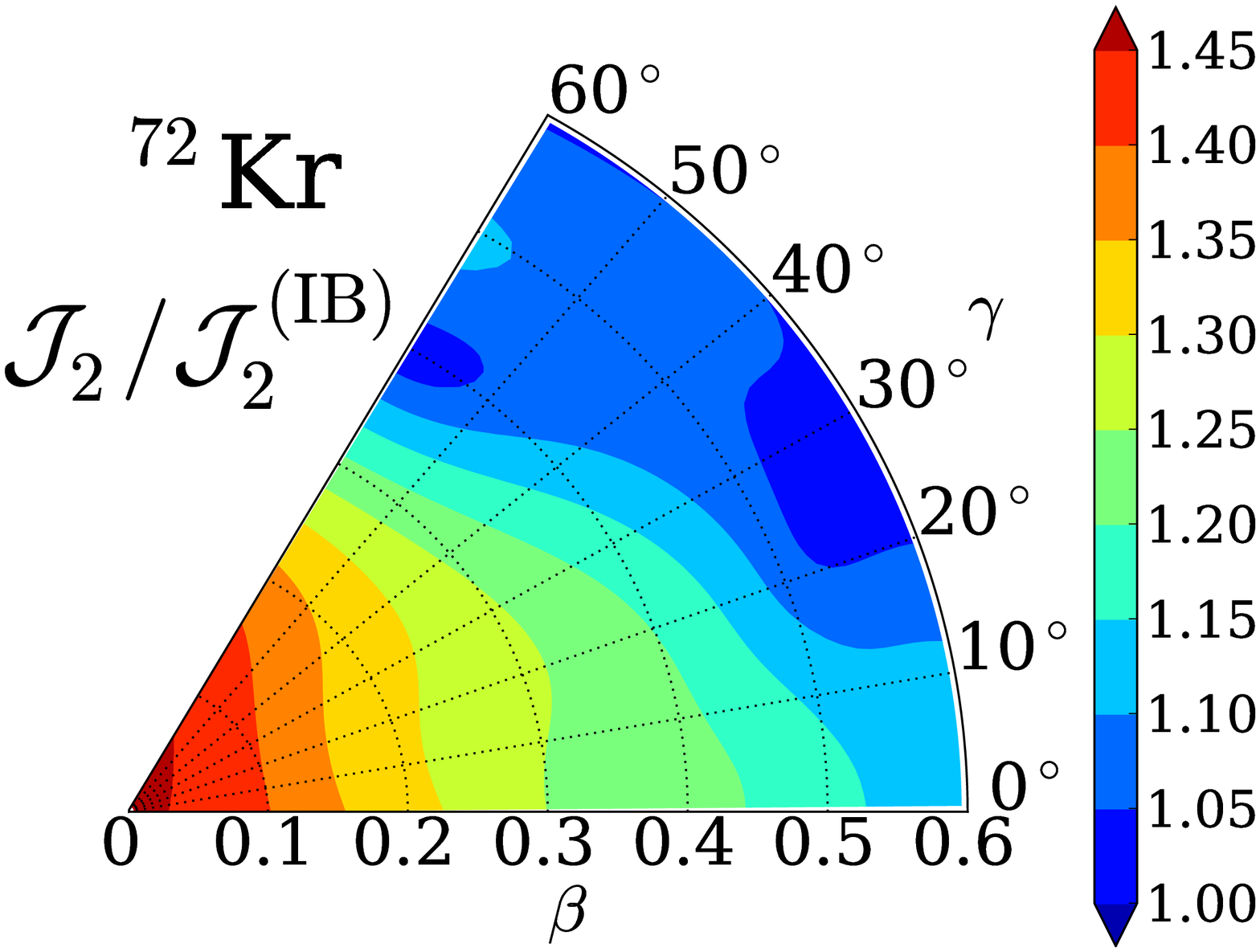}}  
\subfigure[$\Jc_{3}/\Jc_3^{\rm (IB)}$]
{\includegraphics[height=0.33\textwidth,keepaspectratio,clip,trim=30 0 70 0]{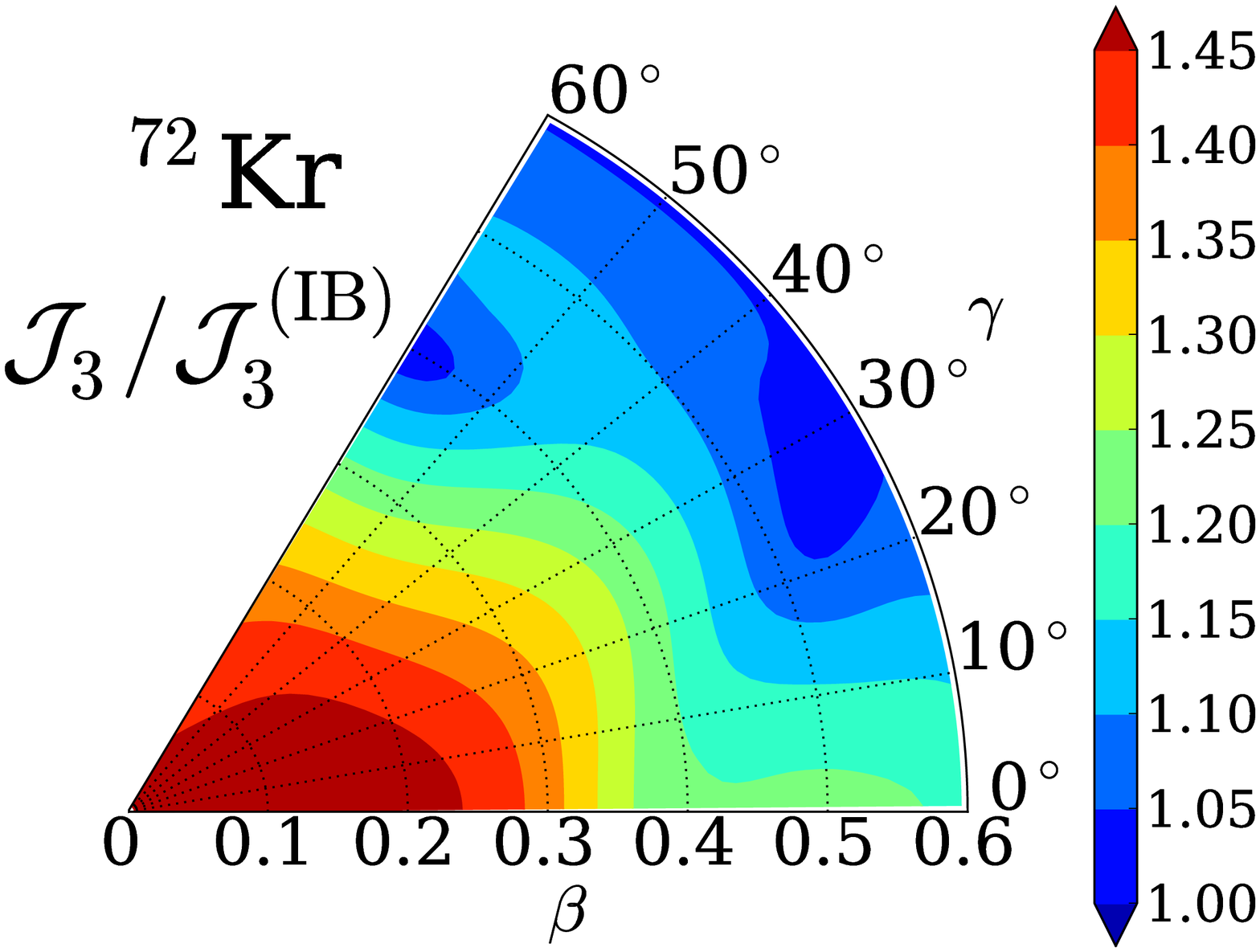}}  
\end{center}
\caption{Ratios of the LQRPA rotational moments of inertia 
to the IB cranking moments of inertia, calculated for $^{72}$Kr.}
\label{fig:ratioMoI}
\end{figure}

 We plot the LQRPA moments of inertia for \kr{72} in Fig.~\ref{fig:MoI72Kr}.
 We can see deviation from the irrotational moments of inertia.
 The ratios of the LQRPA moments of inertia to the Inglis-Belyaev ones are plotted
 in Fig.~\ref{fig:ratioMoI}.
 All the ratios shown in Figs.~\ref{fig:ratiovib} and ~\ref{fig:ratioMoI} are larger than 1 on the entire ($\bg$) plane and depends 
on \be~and \ga.
 This indicates not only that the time-odd components of the mean field 
(which enhance the inertial masses) should be taken into account
 but also that a simple remedy used in Refs. \cite{Niksic2009,Li2009,Li2010}, i. e., 
a simple multiplication of a constant factor to the IB cranking moments of inertia, 
is insufficient.

 \subsection{$^{72}$Kr}
\begin{figure}[h]
\begin{center}
\includegraphics[width=0.4\textwidth,angle=-90,keepaspectratio,clip]{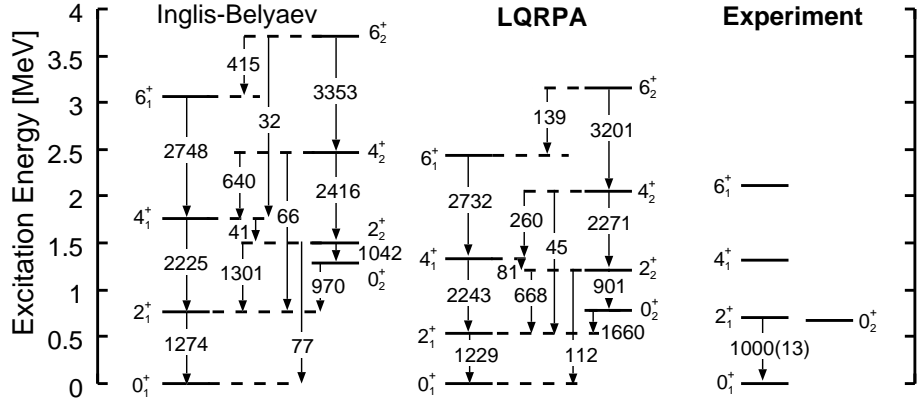}
\end{center}
\caption{Excitation spectra and $B(E2)$ values calculated for $^{72}$Kr 
by means of the CHB+LQRPA method (denoted LQRPA) and 
experimental data \cite{Bouchez2003,Gade2005,Fischer2003}.  
For comparison, results calculated using the IB cranking masses 
(denoted Inglis-Belyaev) are also shown. 
Only $B(E2)$'s larger than 1 Weisskopf unit  
are shown in units of $e^2{\rm fm}^4$.}
\label{fig:spectra72}
\end{figure}

\begin{figure}[htb]
\begin{center}
\begin{tabular}{ll}
\includegraphics[height=0.28\textwidth,keepaspectratio,clip,trim=72 0 160 0]
{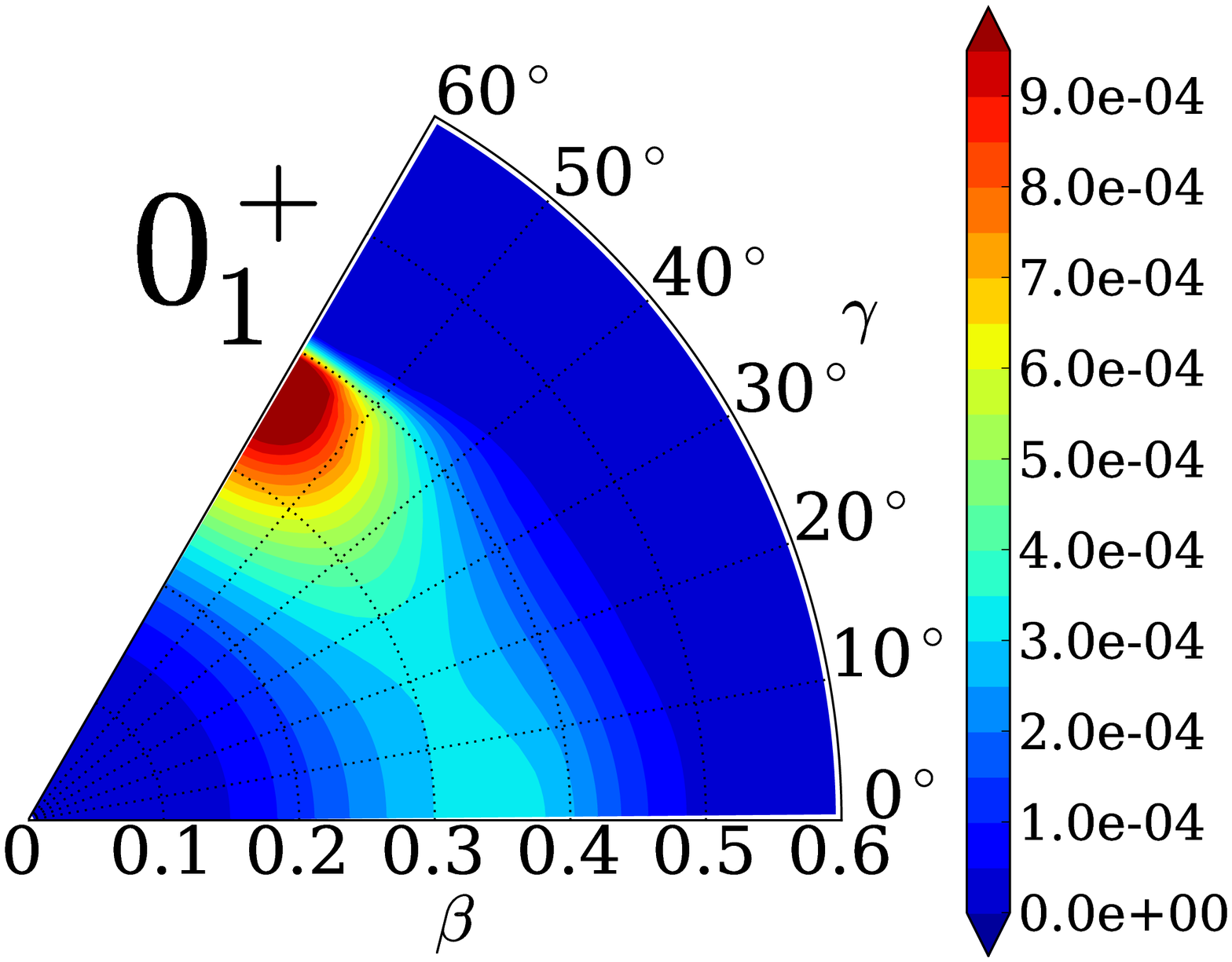} 
\includegraphics[height=0.28\textwidth,keepaspectratio,clip,trim=72 0 160 0]
{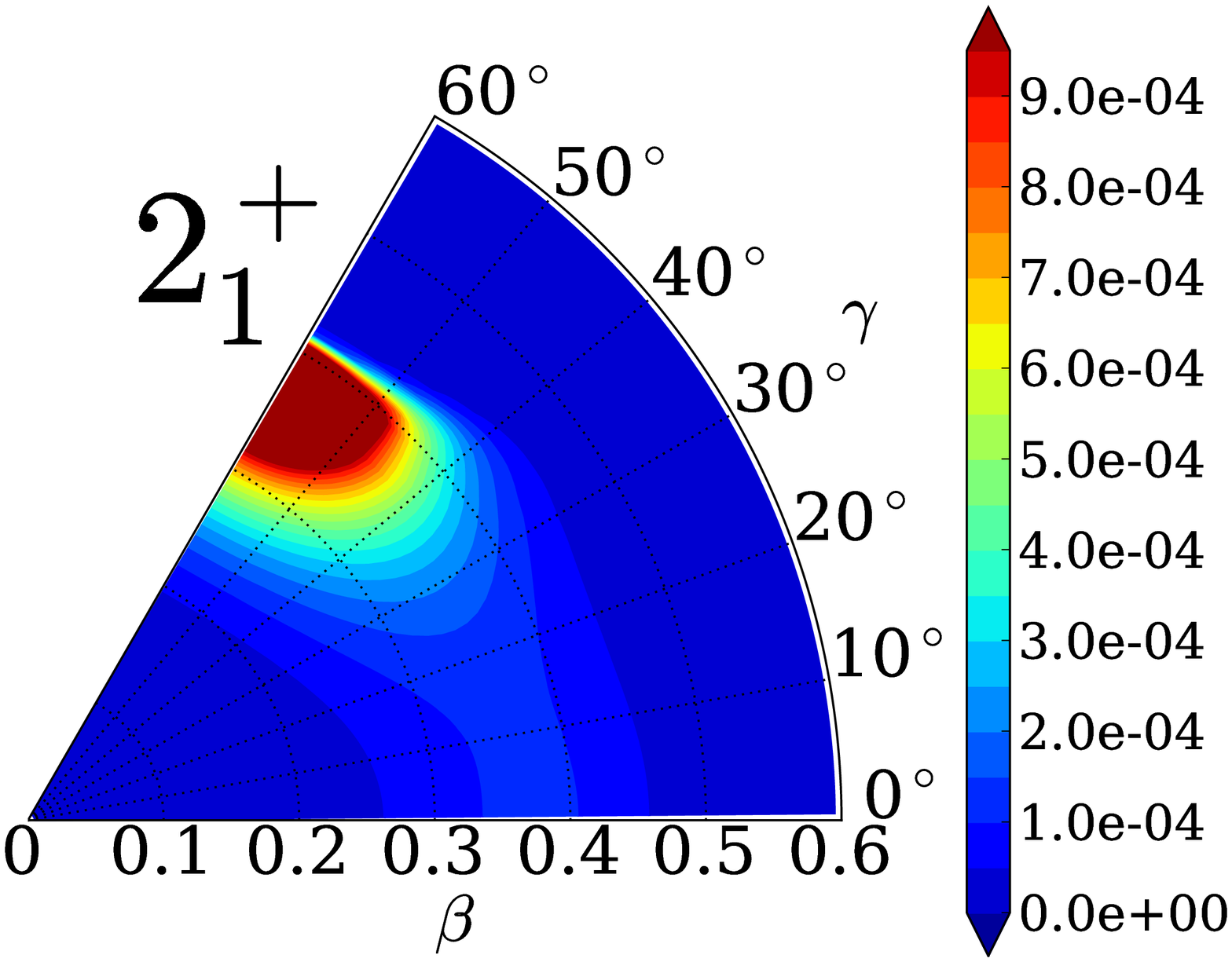} 
\includegraphics[height=0.28\textwidth,keepaspectratio,clip,trim=72 0 160 0]
{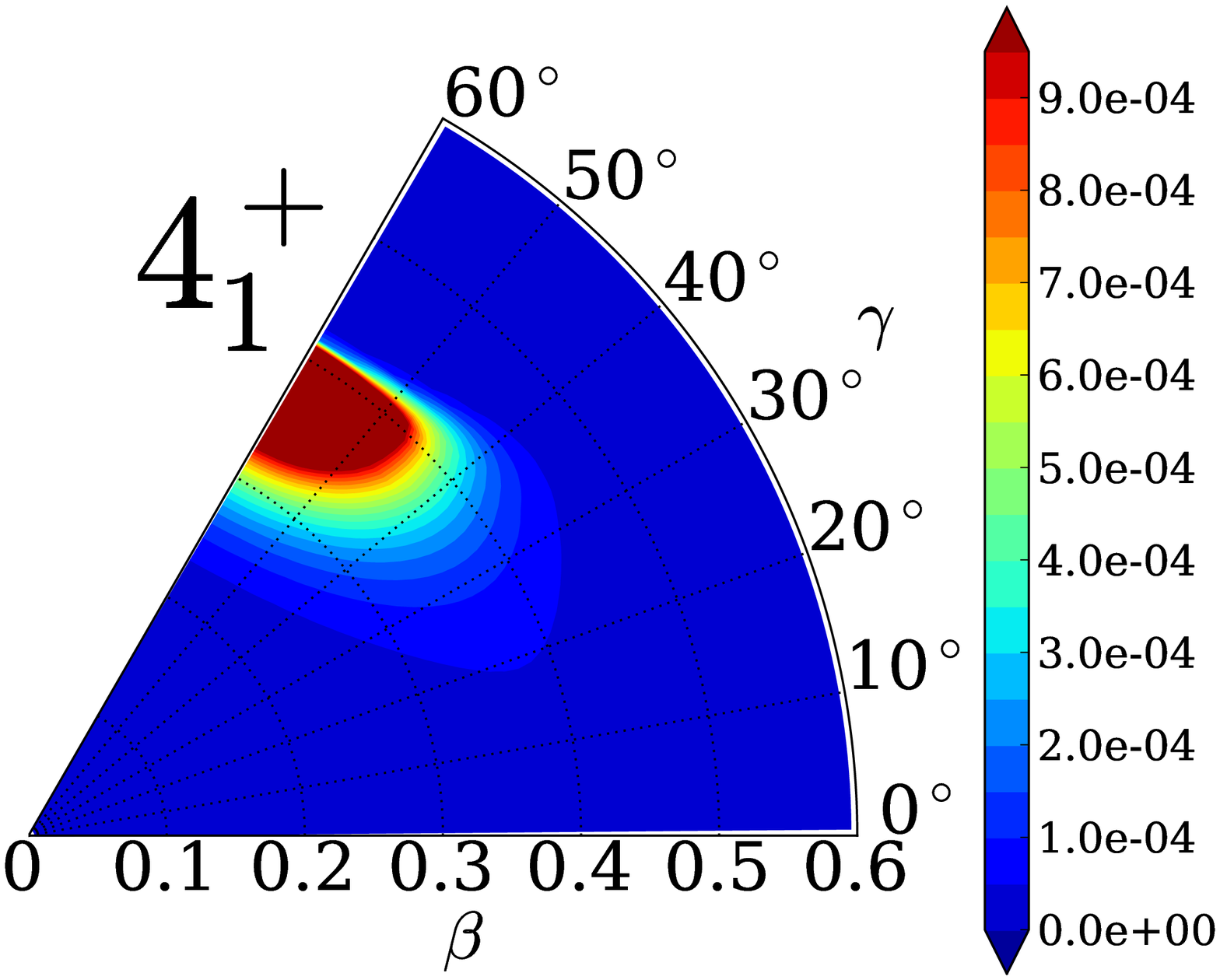} 
\includegraphics[height=0.28\textwidth,keepaspectratio,clip,trim=72 0 0 0]
{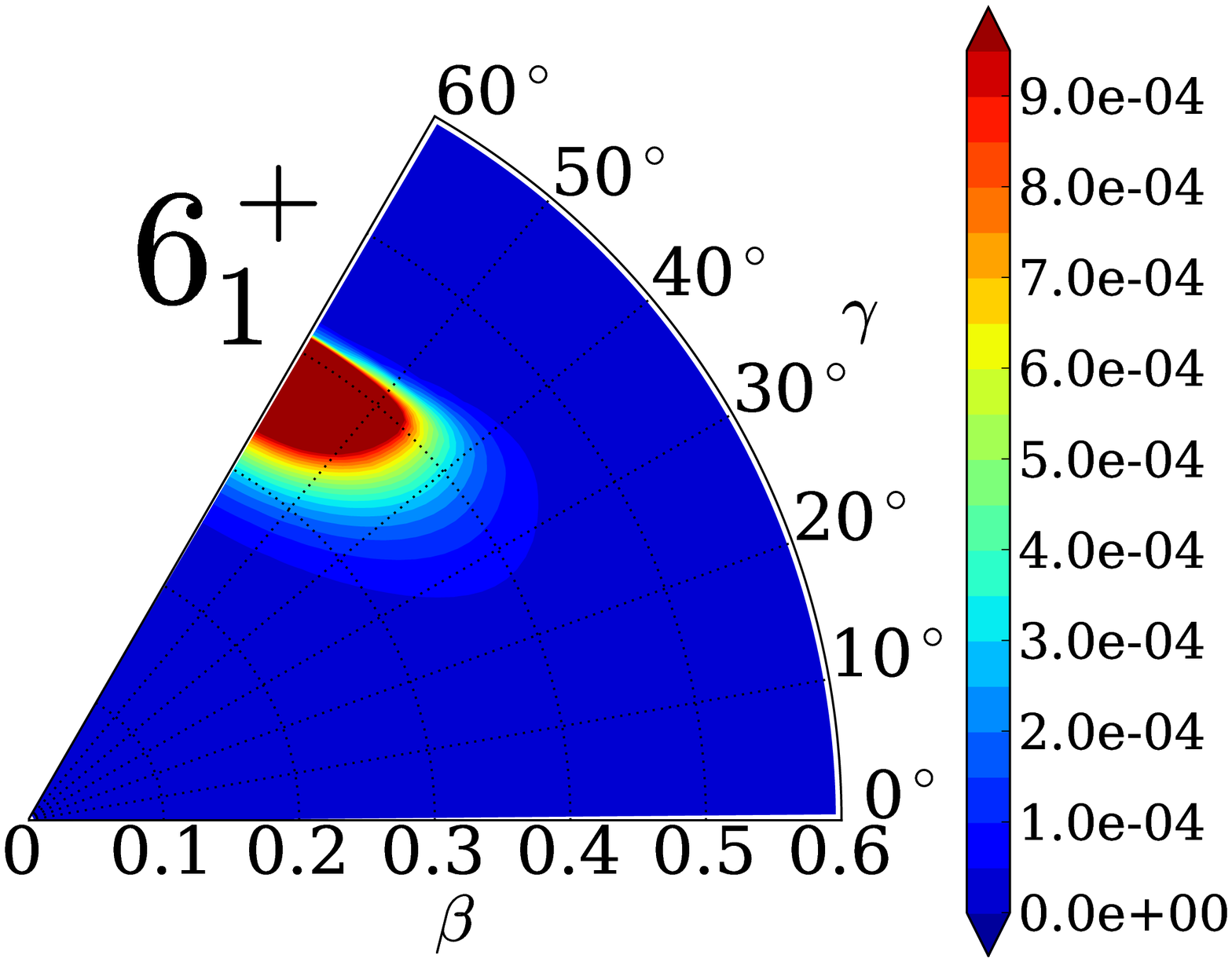}\\ 
\includegraphics[height=0.28\textwidth,keepaspectratio,clip,trim=72 0 160 0]
{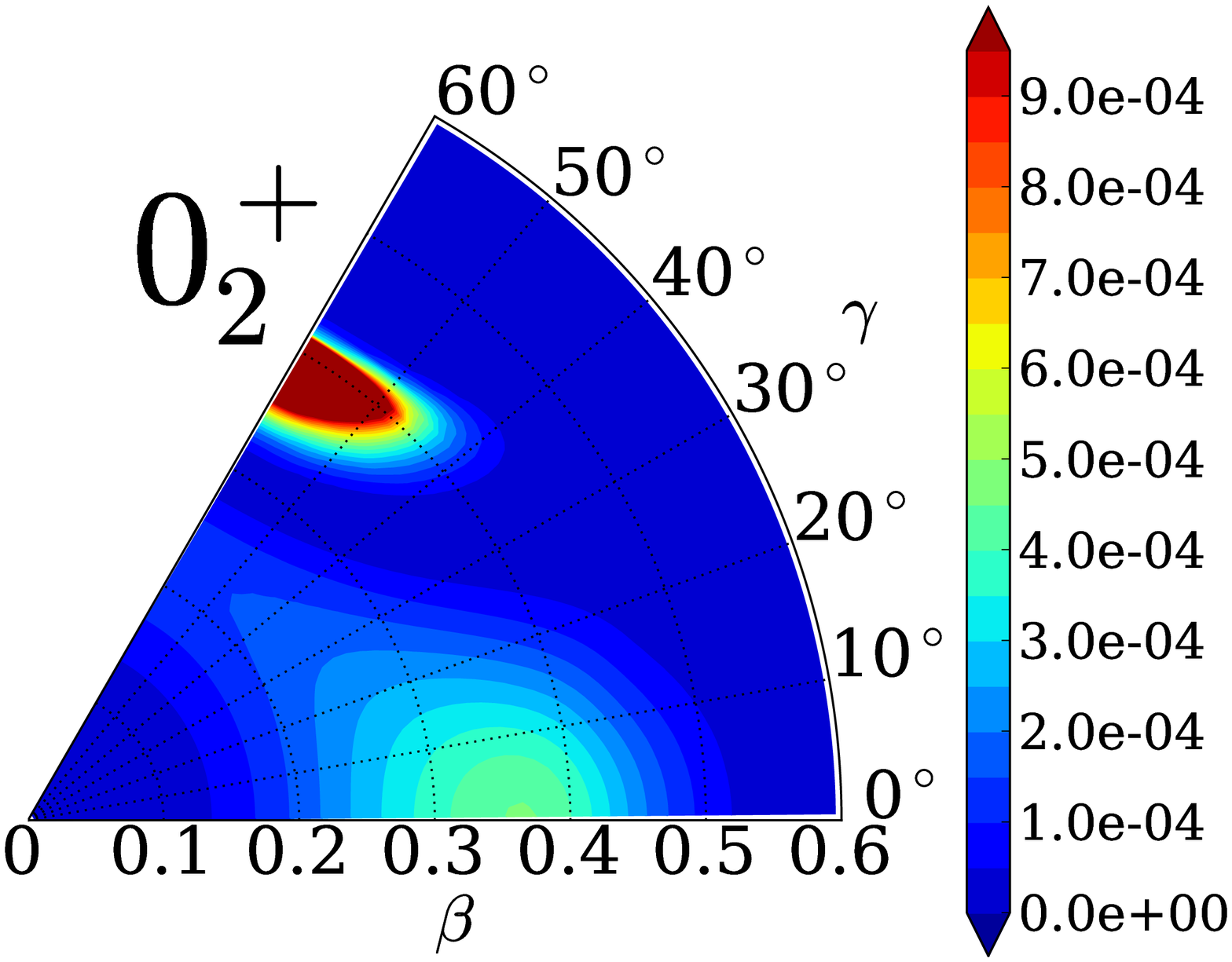} 
\includegraphics[height=0.28\textwidth,keepaspectratio,clip,trim=72 0 160 0]
{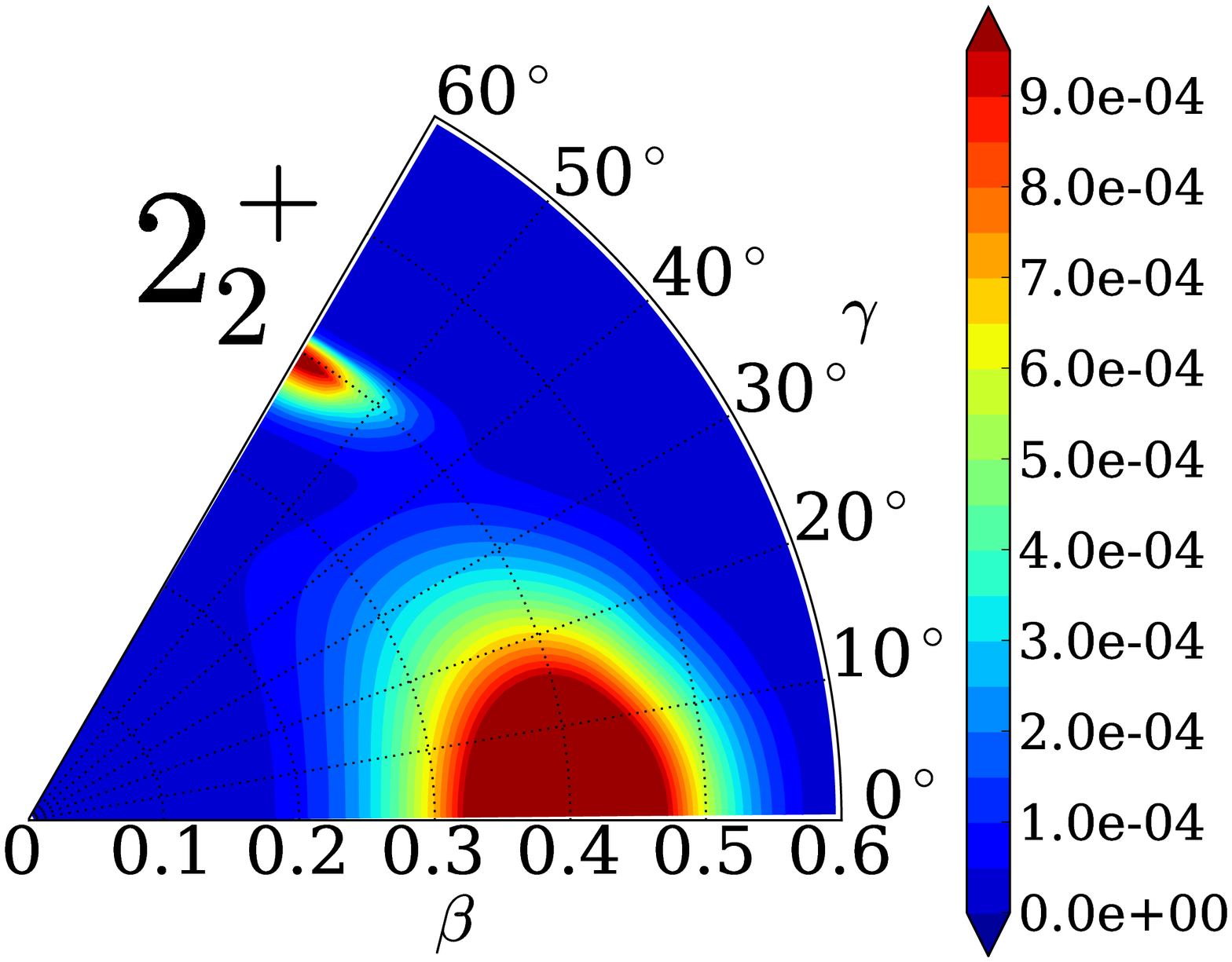} 
\includegraphics[height=0.28\textwidth,keepaspectratio,clip,trim=72 0 160 0]
{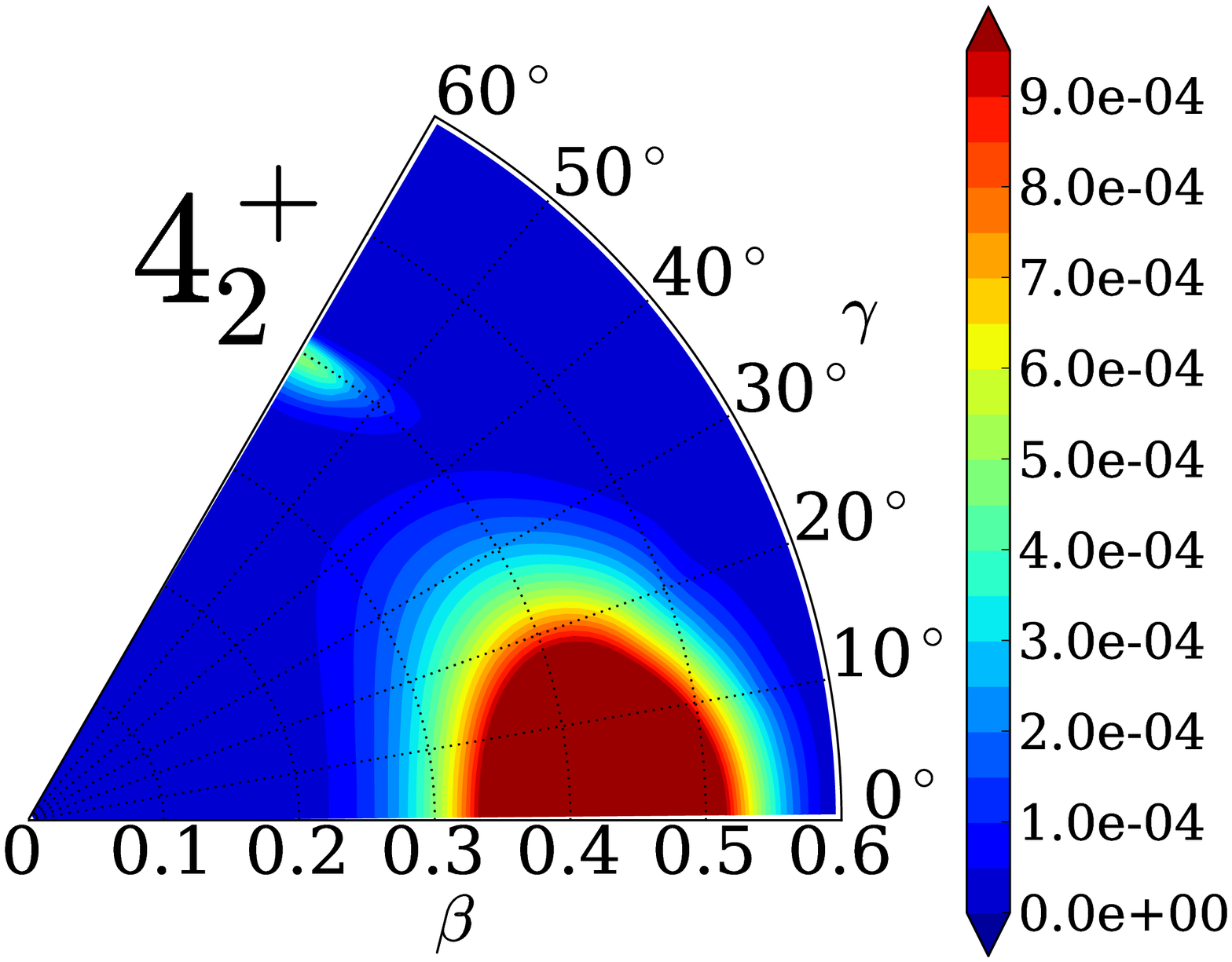} 
\includegraphics[height=0.28\textwidth,keepaspectratio,clip,trim=72 0 0 0]
{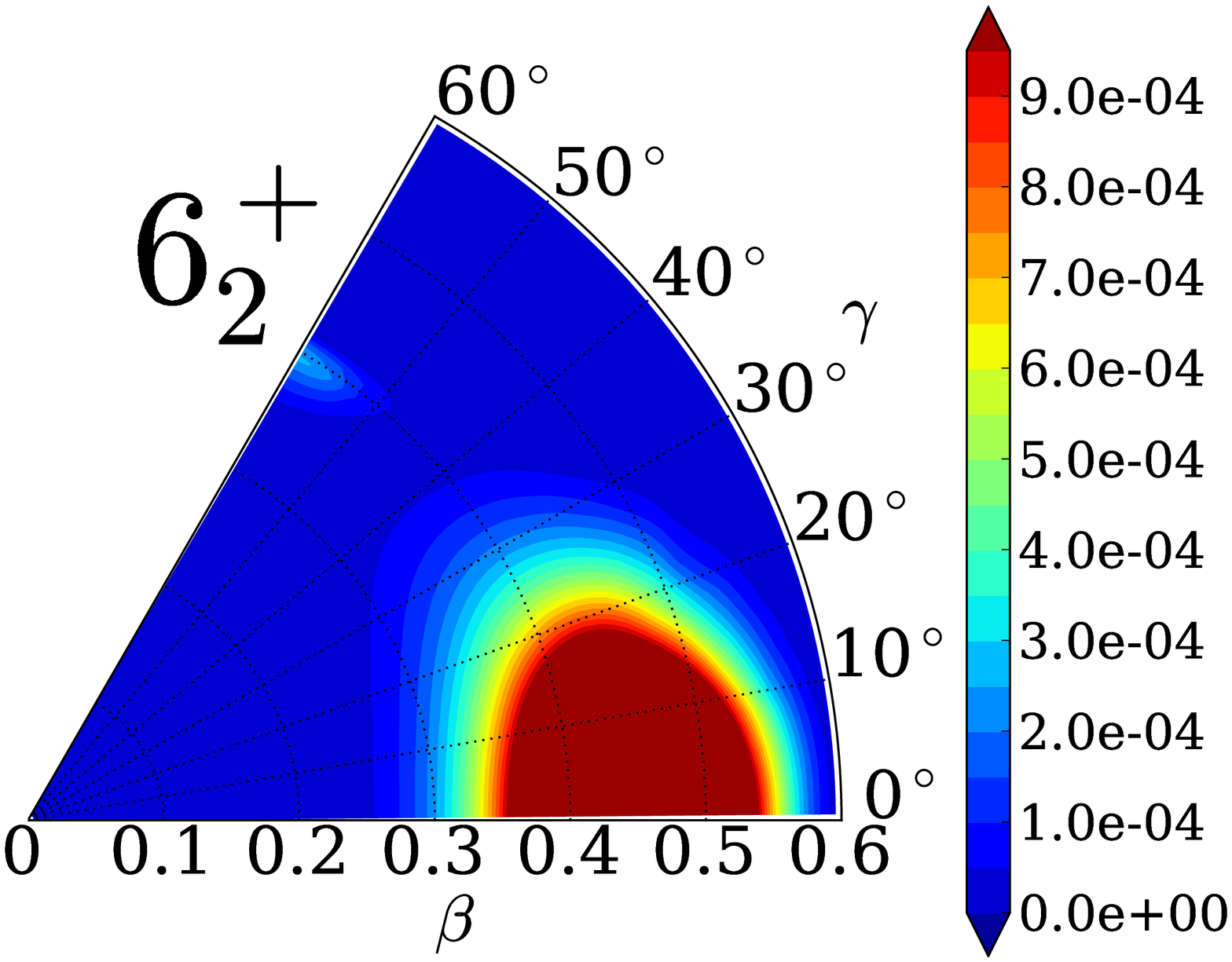}
\end{tabular}
\end{center}
\caption{Vibrational wave functions squared,  
$\beta^4|\Phi_{\alpha I}(\beta,\gamma)|^2$, for $^{72}$Kr.}
\label{fig:wf72}
\end{figure}

We show the excitation spectra and $B(E2)$ values calculated for \kr{72} 
in Fig.~\ref{fig:spectra72} 
together with the experimental data.   
The result of calculation for spectroscopic quadrupole moments 
are shown in Fig.~\ref{fig:Q} together with those for \kr{74, 76}.
In Fig.~\ref{fig:wf72} the collective wave function squared,  
$\b^4\sum_K|\Phi_{\alpha IK}(\b,\g)|^2$, are plotted with a factor $\b^4$ multiplied.
The $\b^4$ factor carries the dominant \be~dependence of 
the volume element $G^{1/2}(\b,\g)$. 
Note that, if the whole volume element was multiplied, all the state would look triaxial 
because of the factor $\sin 3\g$, which vanishes on the oblate and prolate lines. 
 In Fig.~\ref{fig:spectra72}, the result obtained using the IB cranking masses 
instead of the LQRPA collective masses are also shown for comparison.   

 As expected from Figs.~\ref{fig:ratiovib} and \ref{fig:ratioMoI}, 
 the excitation energies obtained with the IB cranking masses are higher than 
those obtained with the LQRPA collective masses.
 Moreover, the excitation energies obtained with the LQRPA collective masses are 
in better agreement with the experimental data except for the $2_1^+$ state.
 The observed excitation energy of the $0_2^+$ state locates close to the $2_1^+$ state. 
It is seen that the $0_2^+$ energy obtained with the LQRPA masses is 
in much better agreement with the experimental data than that with the IB cranking masses.

 The collective wave function of the ground state has a peak around the oblate potential minimum.
 It has a tail to the prolate region, however.
 As angular momentum increases, the localization of the collective wave function develops.
 The collective wave function of the $0_2^+$ state consists of two component:
 one is a sharp peak on the oblate side and the other is a component 
spreading around the prolate region somewhat broadly.
 It has a node in the \be~direction on the oblate side but the \ga-vibrational component is strongly mixed.
 In contrast to the $0_2^+$ state,
the effects of the $\beta$ vibrational excitation are weak for the other yrare states.   
The vibrational wave function has two peaks at $I=2$.
 The oblate peak shrinks rapidly but the prolate one grows with angular momentum 
 due to the orthogonality to the yrast states.
 These behaviors of the vibrational wave function are consistent 
with the results of the (1+3) dimensional ASCC calculation \cite{Hinohara2008, Hinohara2009}.
 There, it is suggested that the rotational effect may assist the localization of the
 collective wave function.

The $E2$ transition strengths clearly indicate that the shape-coexistence-like 
character becomes stronger with increasing angular momentum: 
the interband transitions between the initial and final states having equal angular momentum become weaker and weaker.

 The signs of the spectroscopic quadrupole moments shown in Fig.~\ref{fig:Q}(a)     
are positive for the yrast states indicating their oblate-like character,   
while they are negative for the yrare states indicating their prolate-like character.  
Their magnitudes increase as angular momentum increases, 
which reflects the growth of localization of the vibrational wave function.  

 \subsection{$^{74}$Kr}
\begin{figure}[h]
\begin{center}
\includegraphics[height=0.57\textheight,angle=90,keepaspectratio,clip]{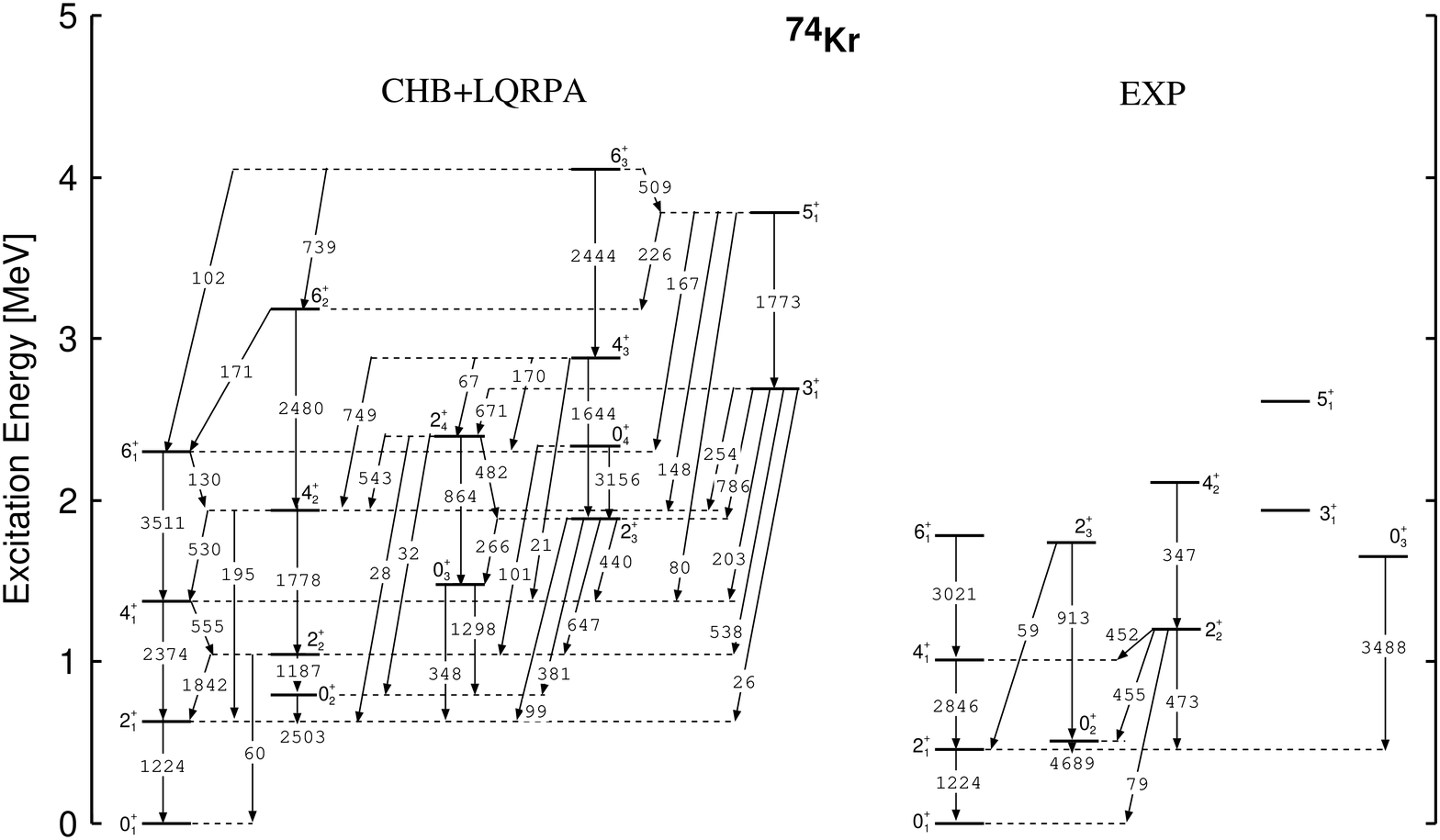}
\end{center}
\caption{Same as Fig.~\ref{fig:spectra72} but for $^{74}$Kr. 
Experimental data are taken from Ref.~\cite{Clement2007}
.}
\label{fig:spectra74}
\end{figure}

\begin{figure}[htb]
\begin{center}
\begin{tabular}{llll}
\includegraphics[height=0.3\textwidth,keepaspectratio,clip,trim=72 0 160 0]
{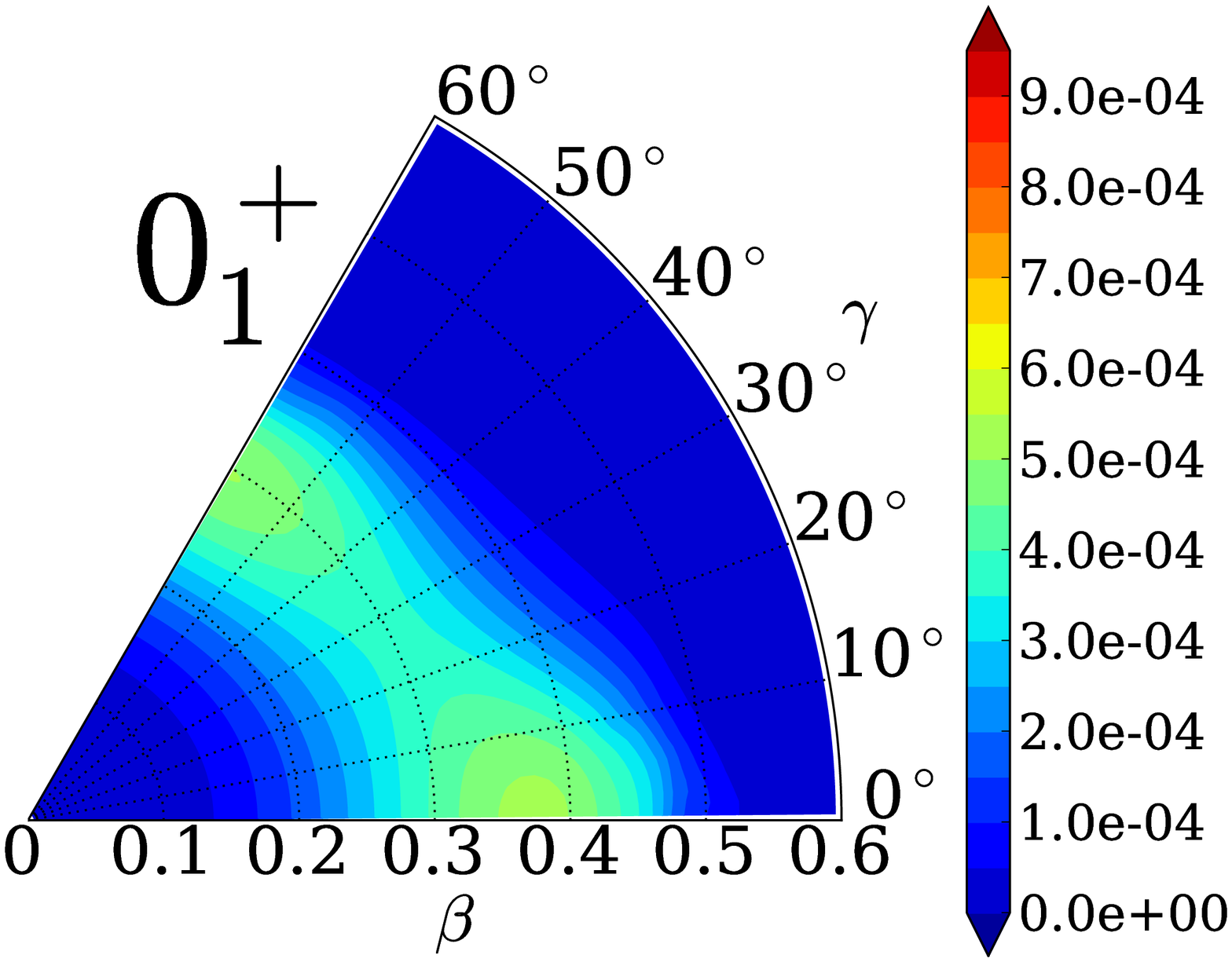}  
\includegraphics[height=0.3\textwidth,keepaspectratio,clip,trim=72 0 160 0]
{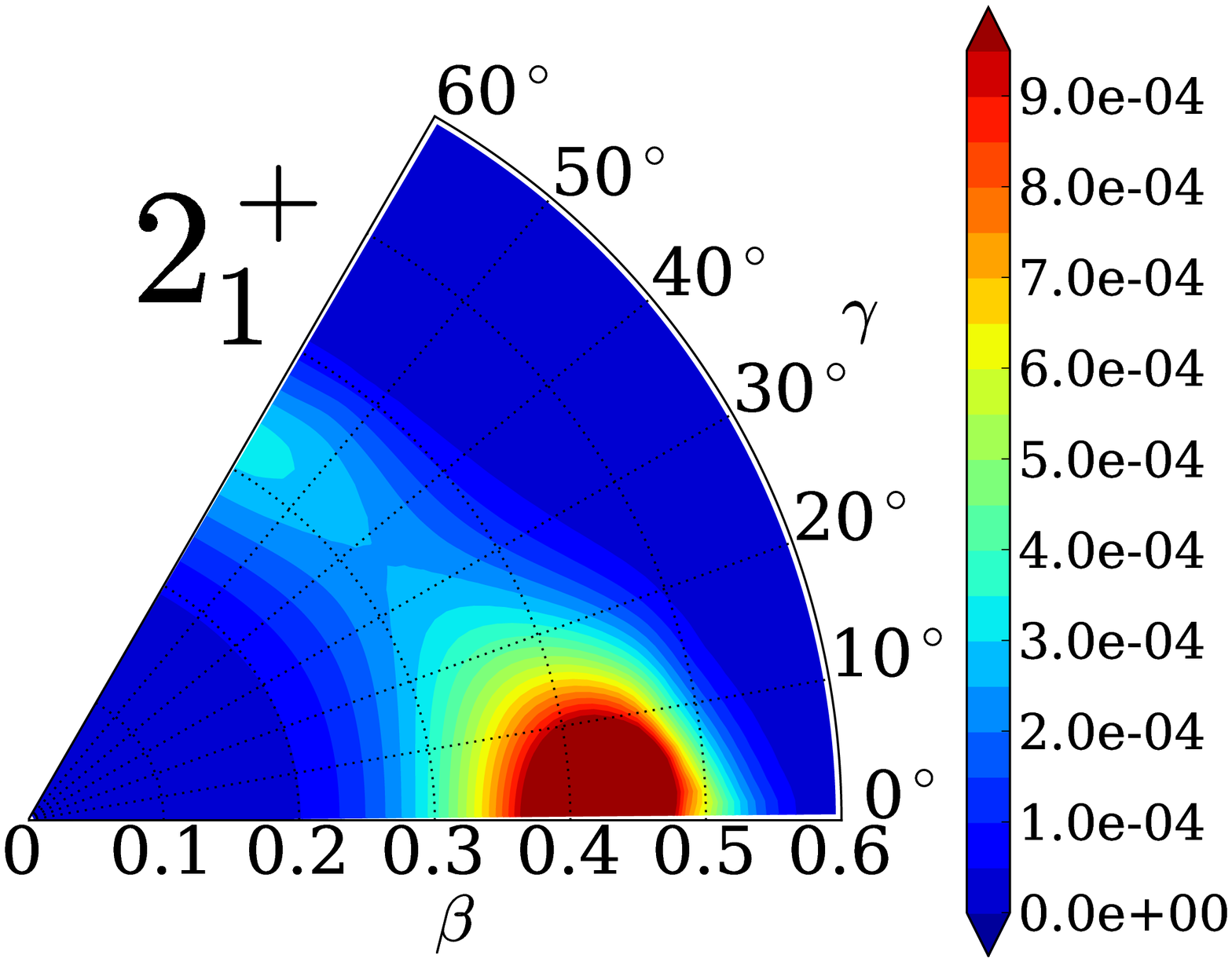} 
\includegraphics[height=0.3\textwidth,keepaspectratio,clip,trim=72 0 160 0]
{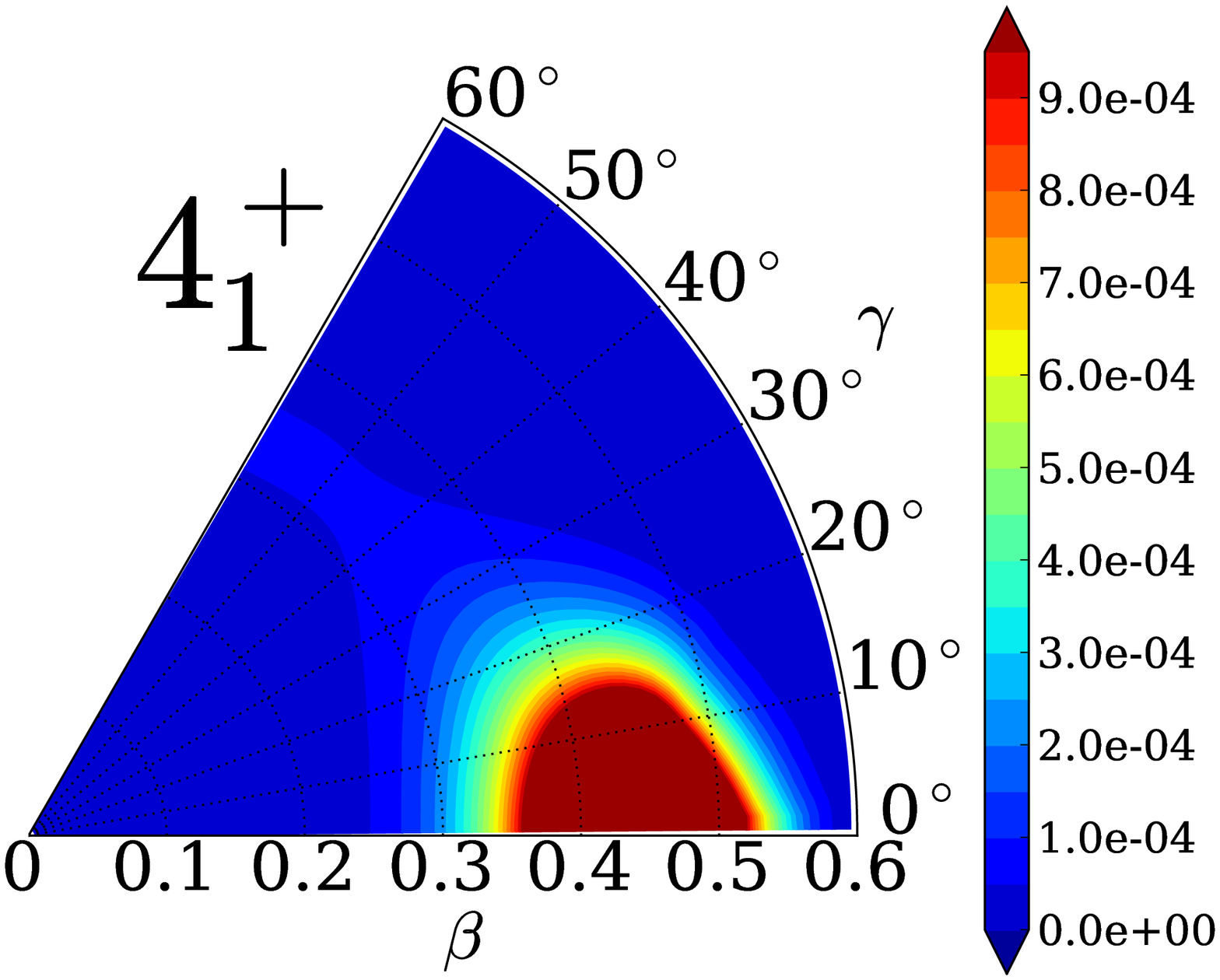}  
\includegraphics[height=0.3\textwidth,keepaspectratio,clip,trim=72 0 0 0]
{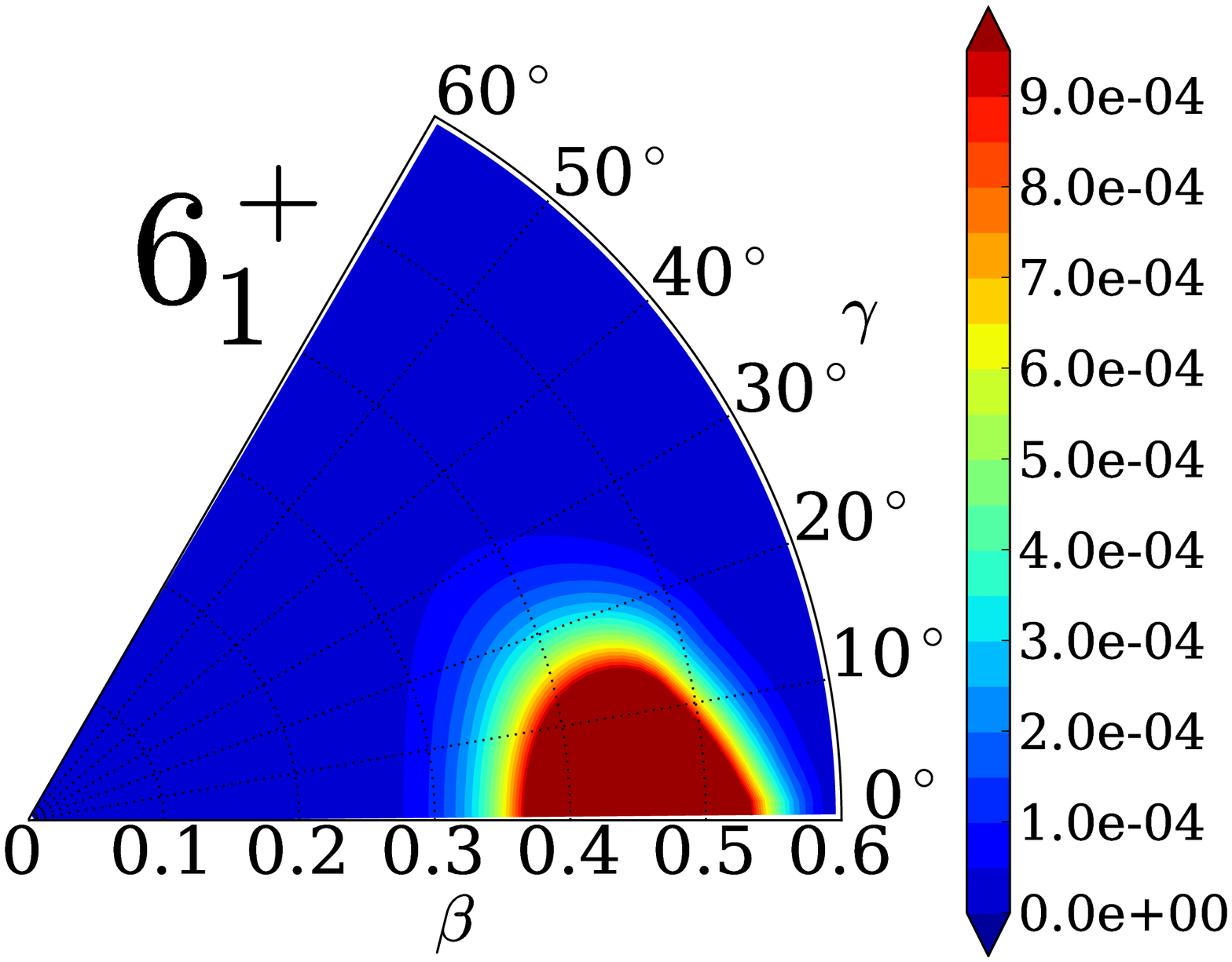} \\ 
\includegraphics[height=0.3\textwidth,keepaspectratio,clip,trim=72 0 160 0]
{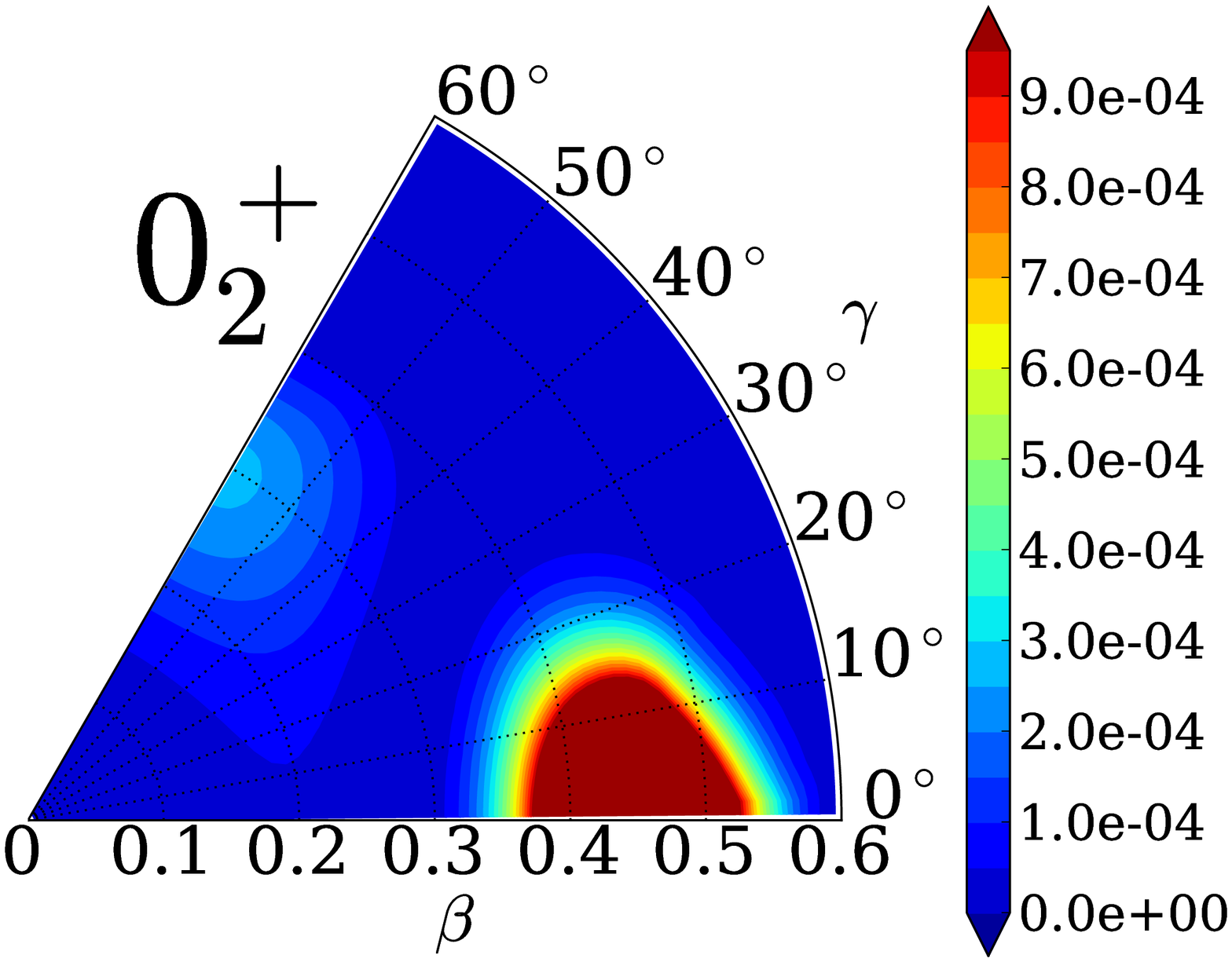}  
\includegraphics[height=0.3\textwidth,keepaspectratio,clip,trim=72 0 160 0]
{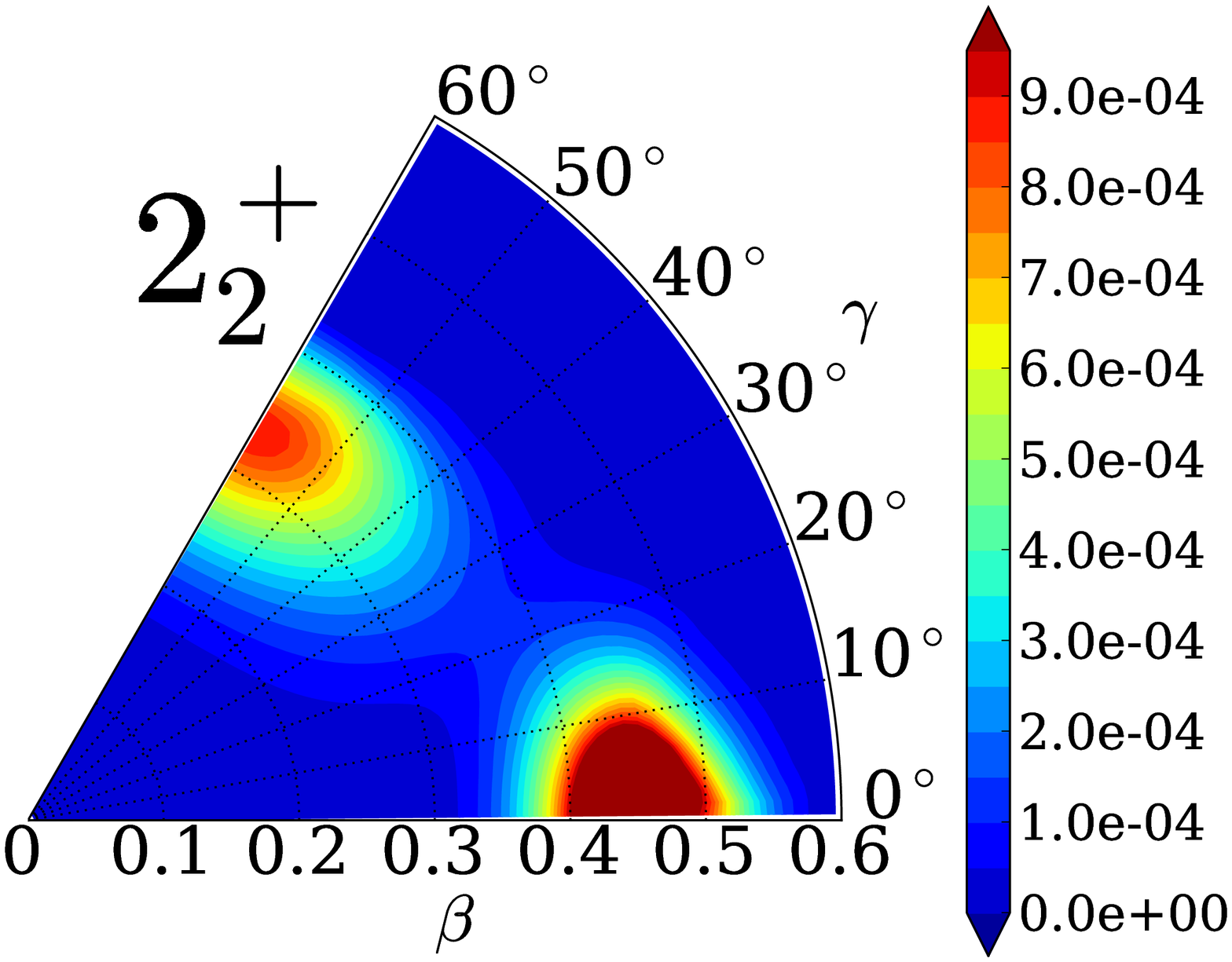}  
\includegraphics[height=0.3\textwidth,keepaspectratio,clip,trim=72 0 160 0]
{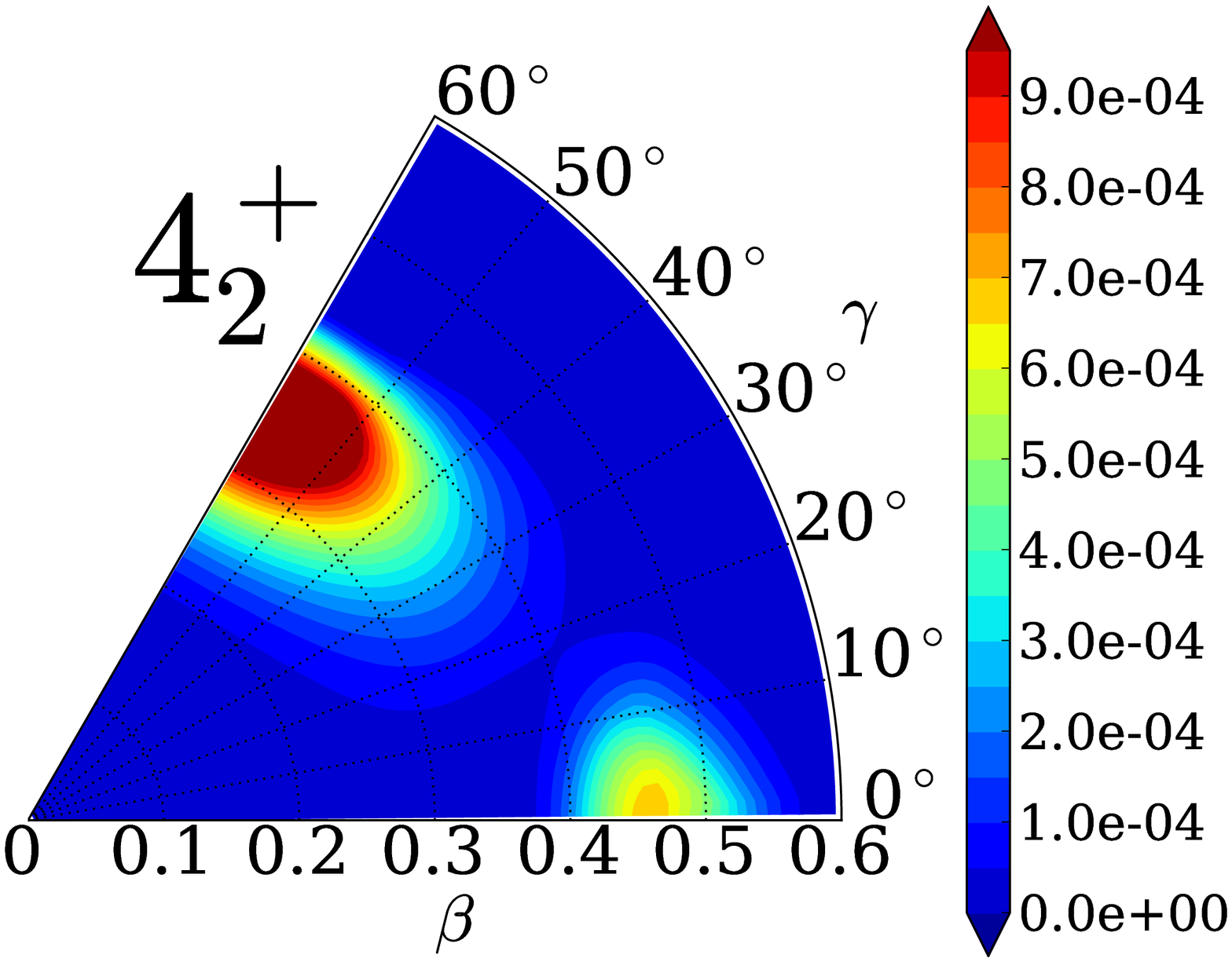}  
\includegraphics[height=0.3\textwidth,keepaspectratio,clip,trim=72 0 0 0]
{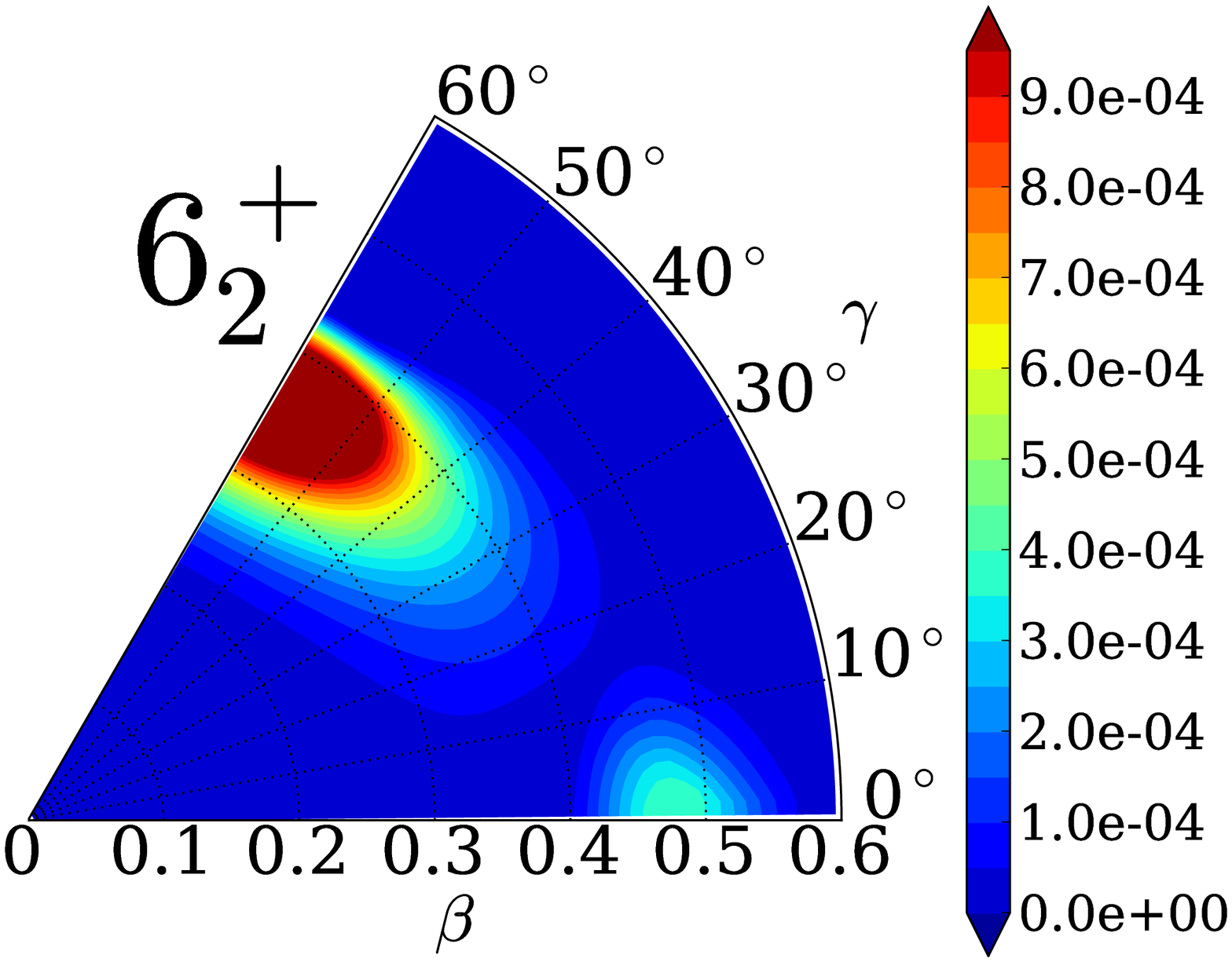} \\
\includegraphics[height=0.3\textwidth,keepaspectratio,clip,trim=72 0 160 0]
{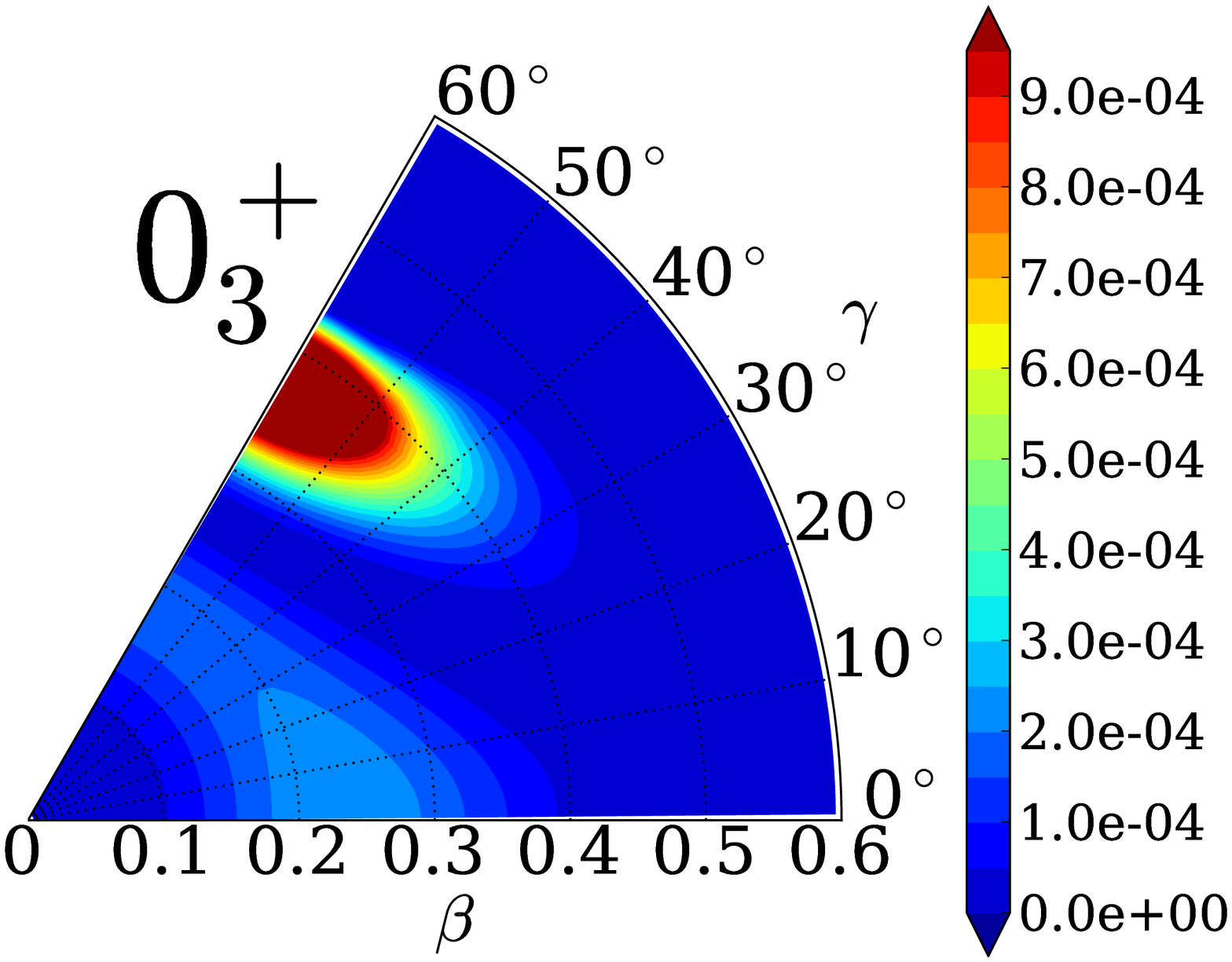}  
\includegraphics[height=0.3\textwidth,keepaspectratio,clip,trim=72 0 160 0]
{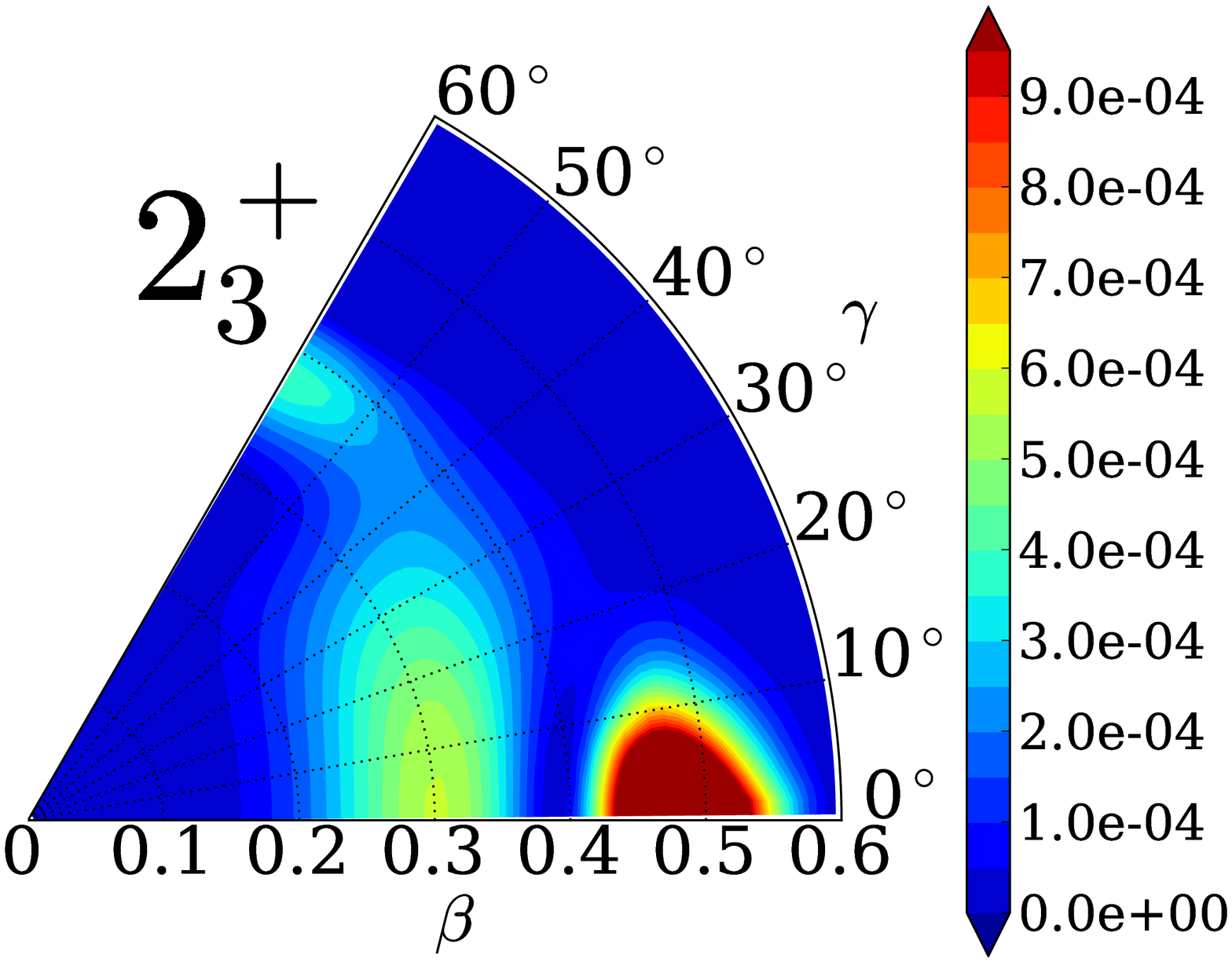}  
\includegraphics[height=0.3\textwidth,keepaspectratio,clip,trim=72 0 160 0]
{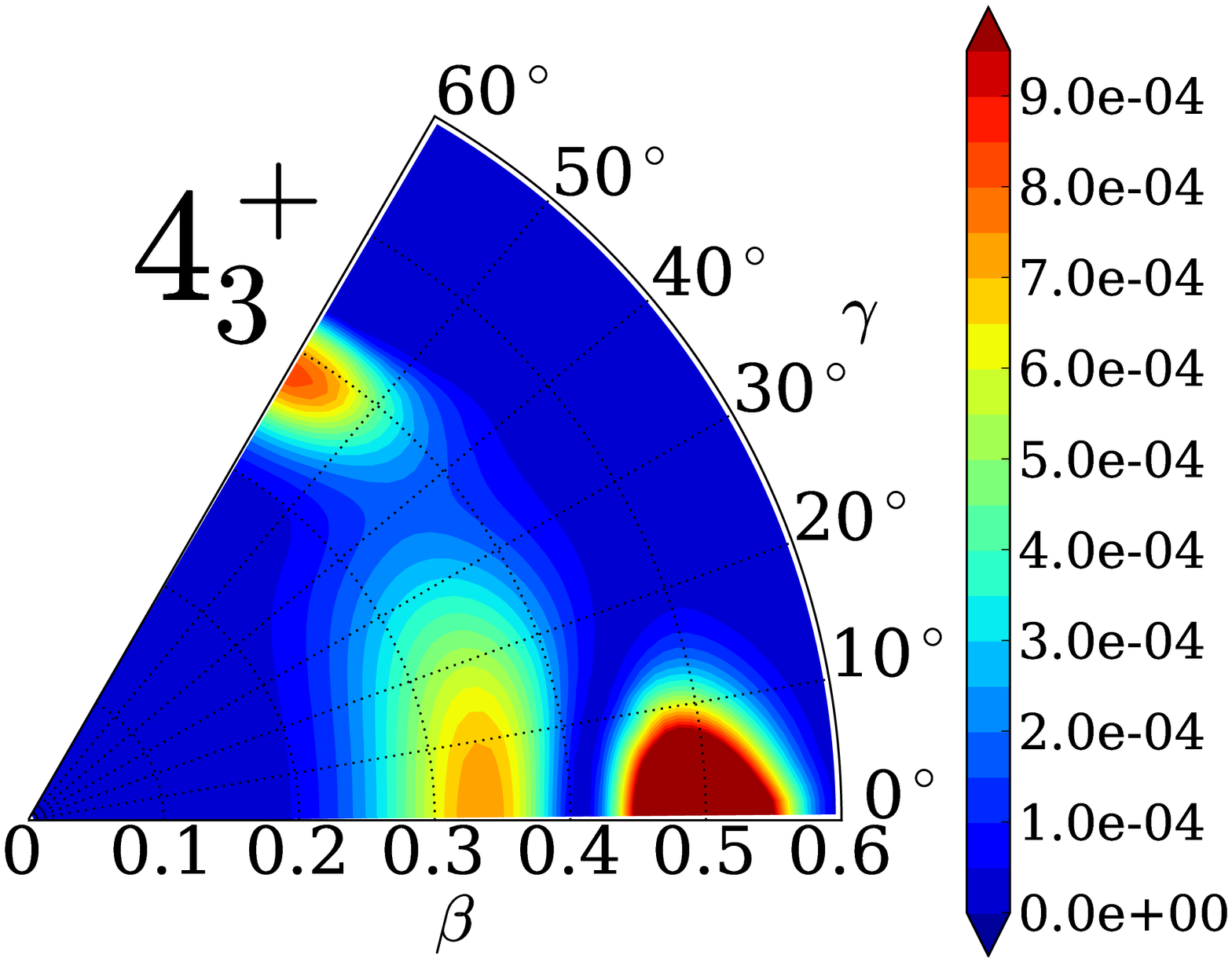}  
\includegraphics[height=0.3\textwidth,keepaspectratio,clip,trim=72 0 0 0]
{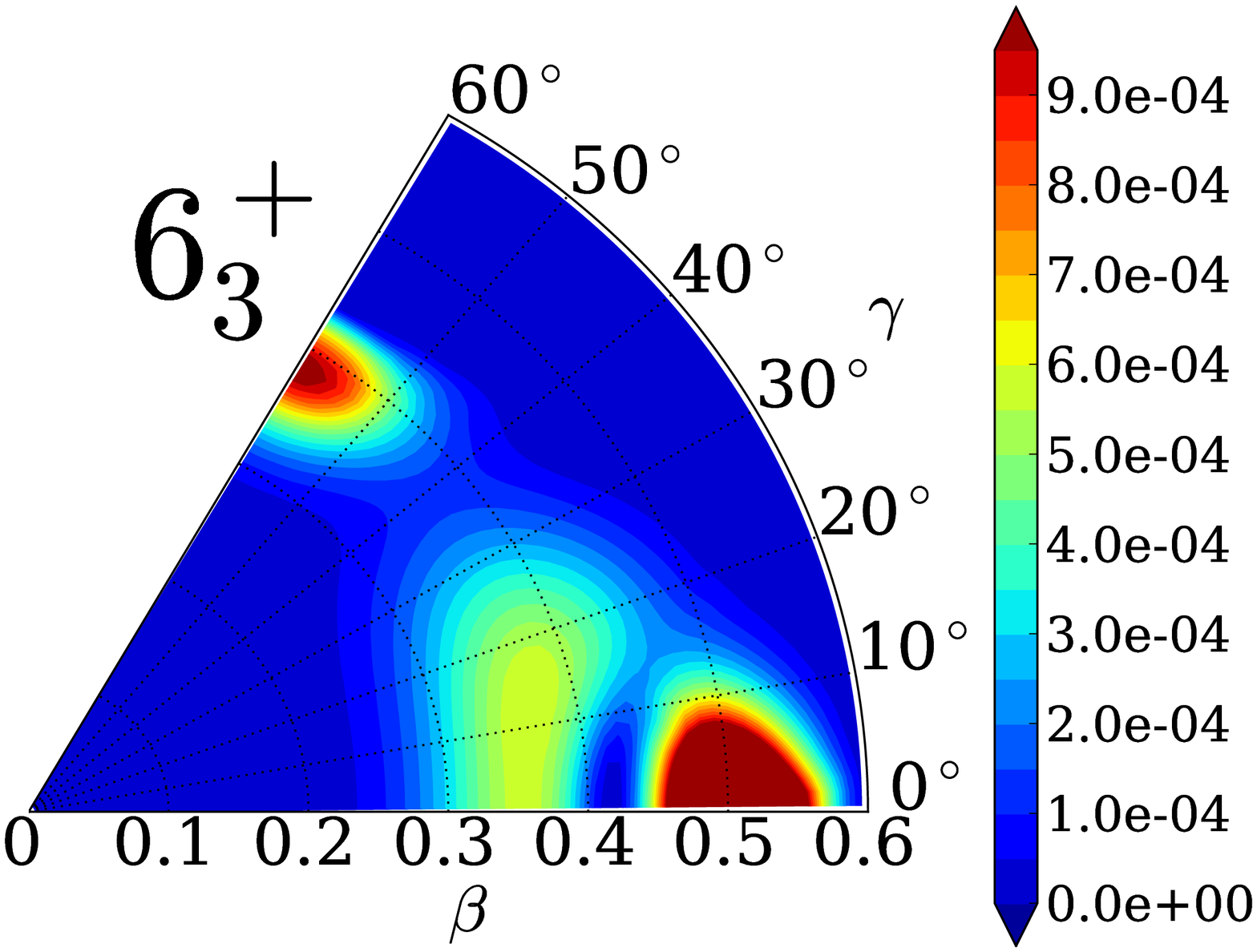} \\
\includegraphics[height=0.3\textwidth,keepaspectratio,clip,trim=72 0 160 0]
{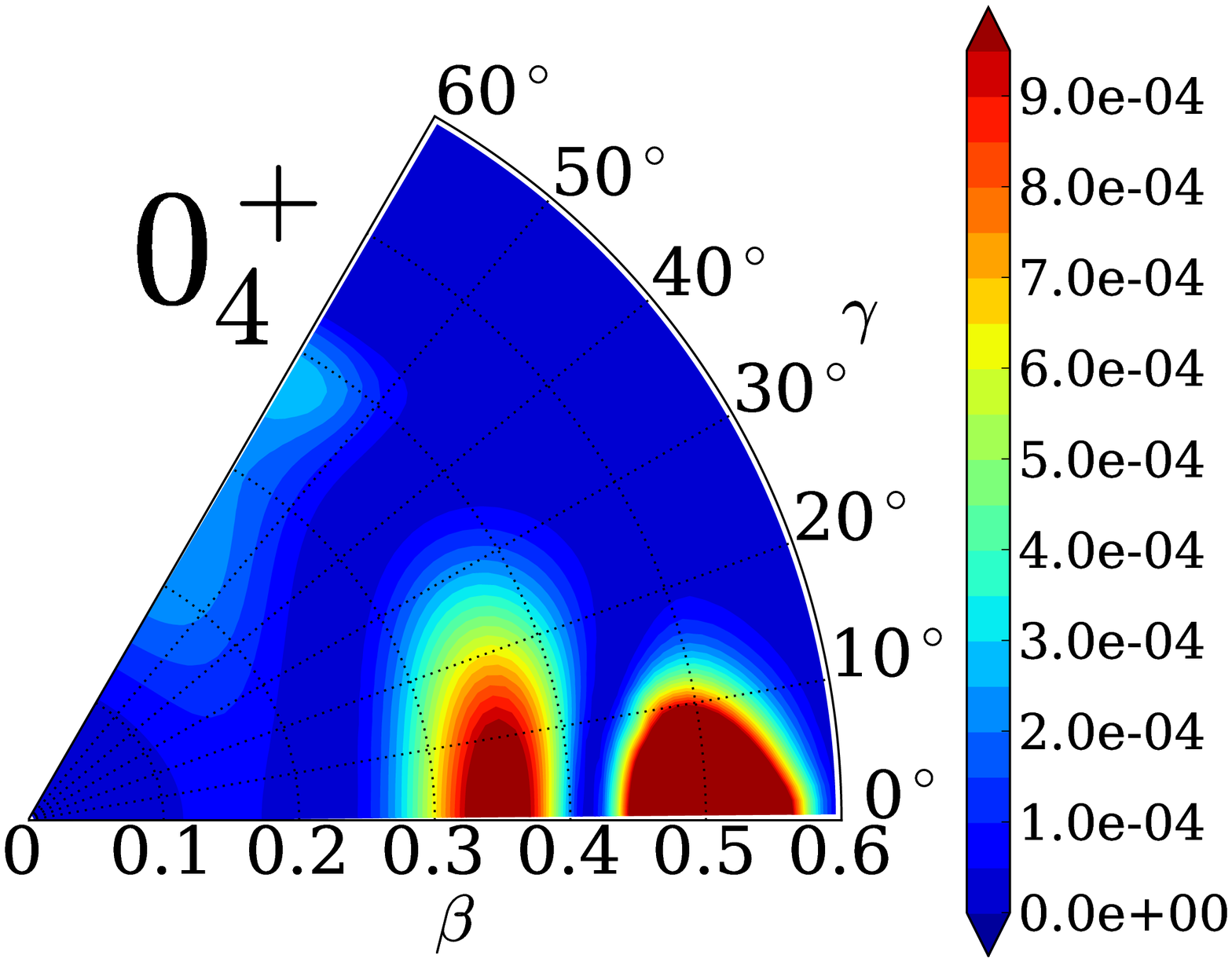}  
\includegraphics[height=0.3\textwidth,keepaspectratio,clip,trim=72 0 0 0]
{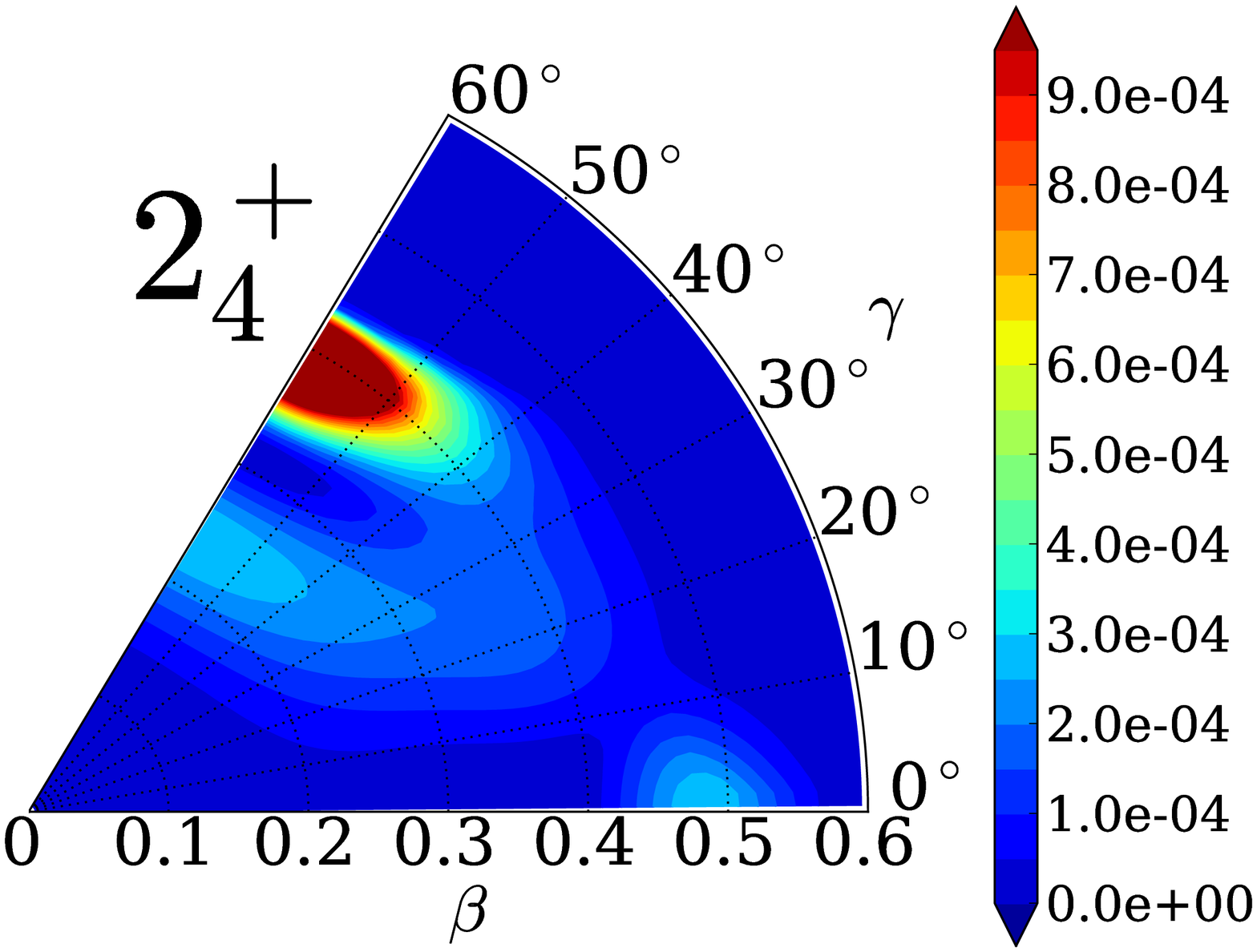}    
\end{tabular}
\end{center}
\caption{Vibrational wave functions squared, 
$\beta^4|\Phi_{\alpha I}(\beta,\gamma)|^2$, for $^{74}$Kr.}
\label{fig:wf74}
\end{figure}

\begin{figure}[h]
\begin{center}
\includegraphics[height=0.3\textwidth,keepaspectratio,clip]{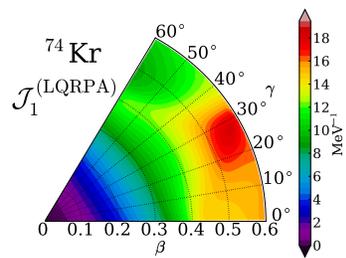}
\end{center}
\caption{The LQRPA moment of inertia $\Jc_1$ for rotation about the $x$-axis, 
calculated for \kr{74}. }
\label{fig:J174}
\end{figure}

The excitation spectrum and $B(E2)$ values calculated for \kr{74} 
are presented in Fig.~\ref {fig:spectra74} 
together with the experimental data,   
while the collective wave function squared are displayed in Fig.~\ref{fig:wf74}.    
It is seen that the collective wave function of the ground state spreads over 
the entire $\gamma$ region along the potential valley.
However, it localizes more and more on the prolate side with increasing angular momentum.
This behavior results from the $\beta$-$\gamma$ dependence of the rotational 
moments of inertia:
one can clearly see the oblate-prolate asymmetry for $\b \gtrsim 0.4$ 
in the moment of inertia $\Jc_1$ displayed in Fig.~\ref{fig:J174}.
Although the $0_2^+$ state has somewhat a \be-vibrational component, 
 the yrare collective wave function for $I \geq 2 $ has almost one-dimensional structure 
in the \ga~direction  with a constant \be.  
 While the collective wave function has a two-peak structure,
 the prolate peak shrinks with increasing angular momentum 
due to the orthogonality to the yrast state.


The third and fourth states for each angular momentum can be regarded as admixtures of  
\be-vibrational and \ga-vibrational components.
On one hand, the $0_3^+$ and $2_4^+$ states have a node in the \be~direction 
on the oblate side. On the other hand, $0_4^+$, $2_3^+$, $4_3^+$ and $6_3^+$ have a node 
on the prolate side.  


It is clearly seen in Fig.~8~
that the result of calculation reproduces well the significant enhancement of  
$B(E2; 0_2^+ \rightarrow 2_1^+)$, $B(E2; 2_2^+ \rightarrow 0_2^+)$ and 
$B(E2; 2_2^+ \rightarrow 2_1^+)$,   
found in experiments, 
as well as the large $B(E2)$ values within the ground band. 

The spectroscopic quadrupole moments for \kr{74}, shown in Fig.~\ref{fig:Q}(b), 
are also in excellent agreement with the experimental data, 
aside from a minor disagreement for the $2_3^+$ state.
In particular, the characteristic features, such as their signs 
and the increasing tendency of their magnitudes with angular momentum 
in the ground band, are well reproduced.

 \subsection{\kr{76}}
\begin{figure}[h]
\begin{center}
\includegraphics[height=0.57\textheight,angle=90,keepaspectratio,clip]{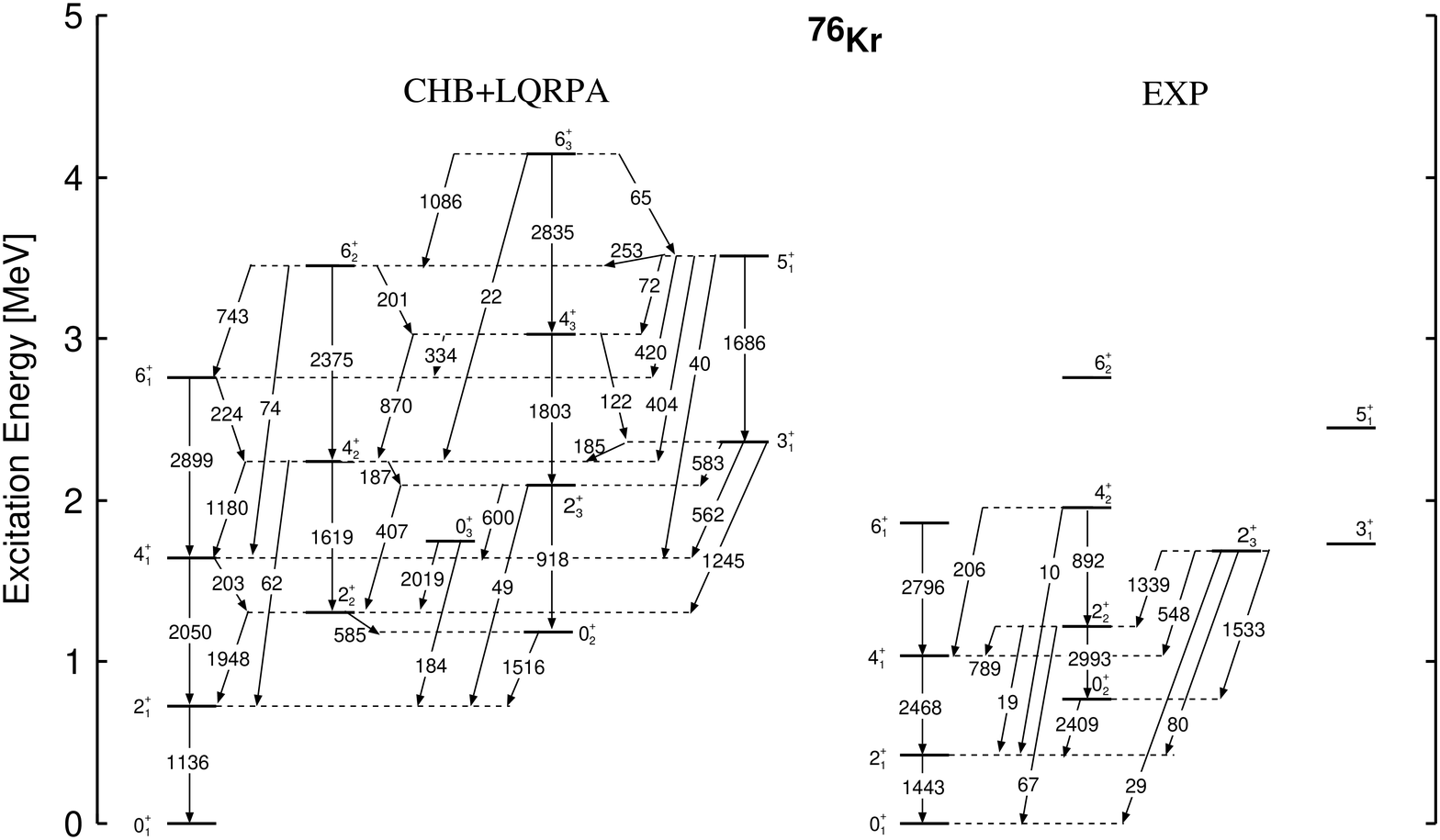}
\end{center}
\caption{Same as Fig.~\ref{fig:spectra72} but for $^{76}$Kr.
Experimental data are taken from Ref.~\cite{Clement2007}
.}
\label{fig:spectra76}
\end{figure}

\begin{figure}[htb]
\begin{center}
\begin{tabular}{lll}
\includegraphics[height=0.3\textwidth,keepaspectratio,clip,trim=72 0 160 0]
{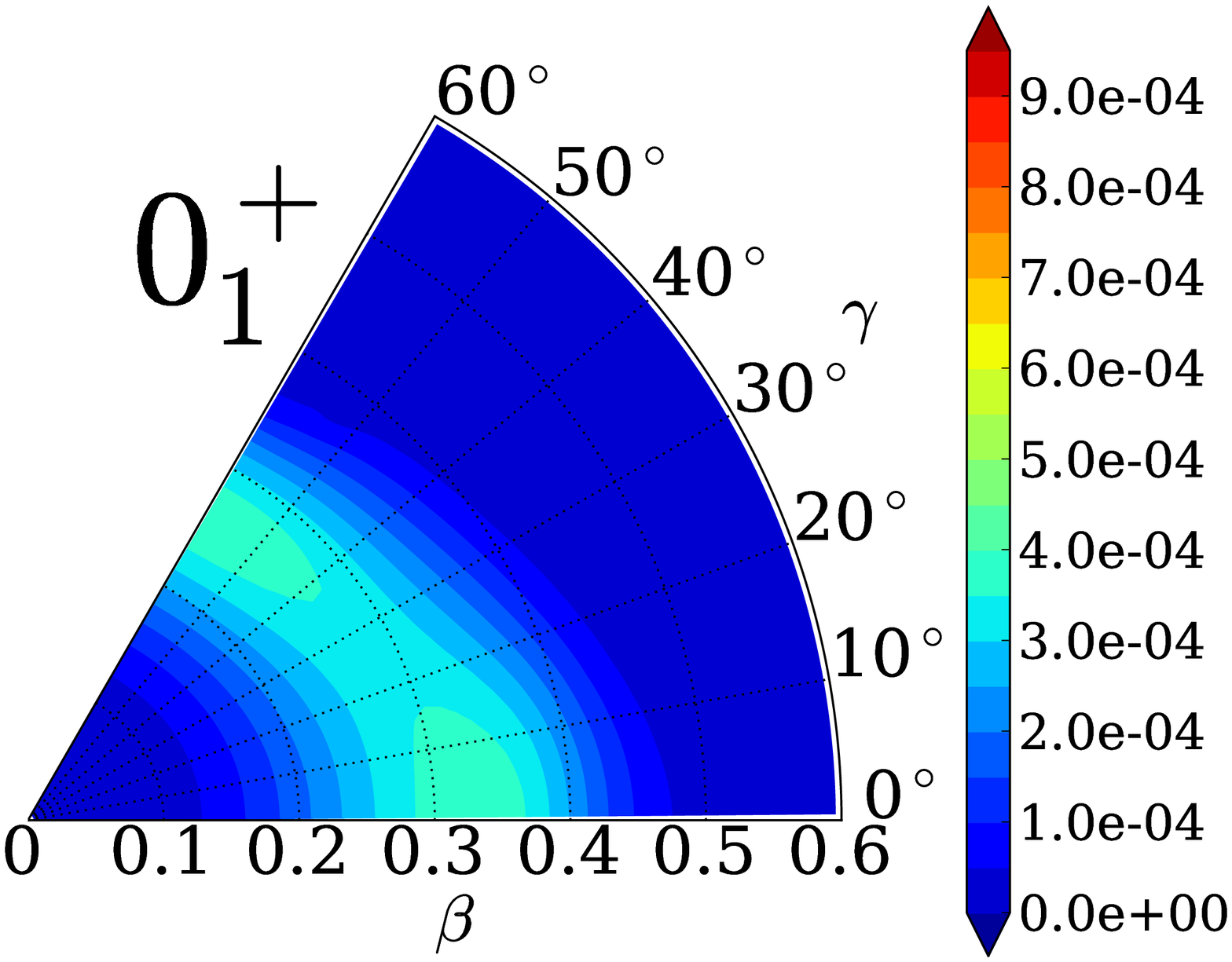} 
\includegraphics[height=0.3\textwidth,keepaspectratio,clip,trim=72 0 160 0]
{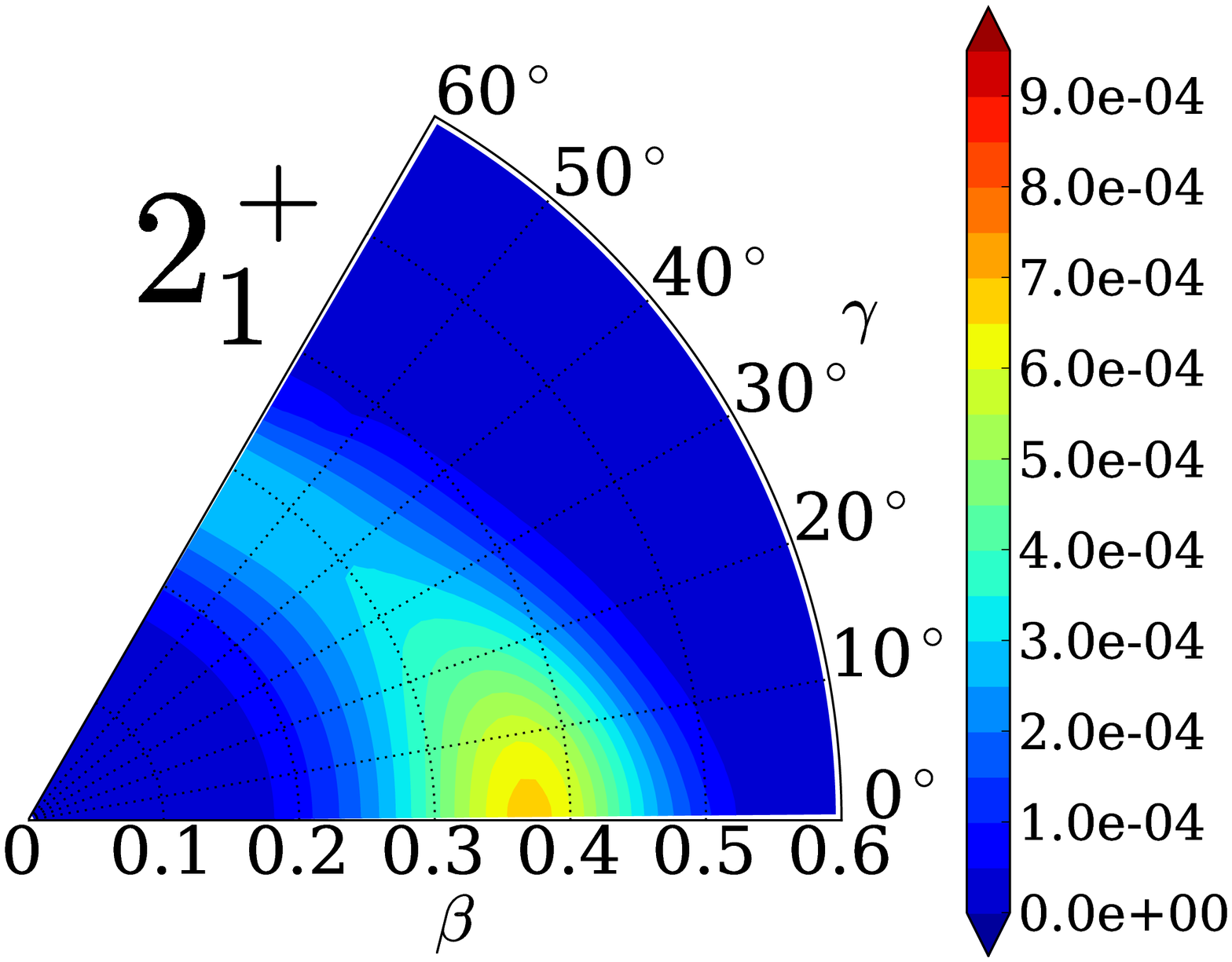}  
\includegraphics[height=0.3\textwidth,keepaspectratio,clip,trim=72 0 160 0]
{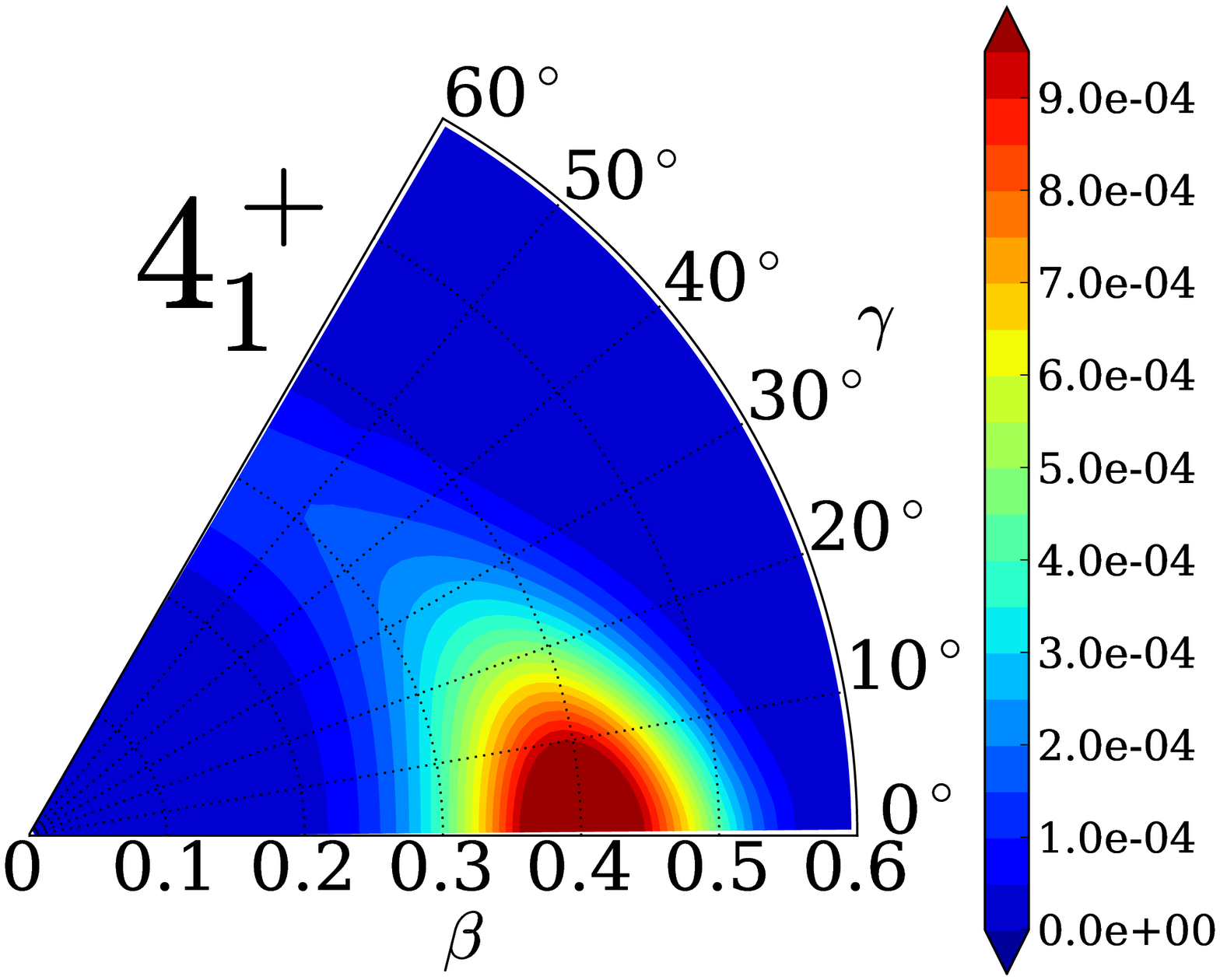} 
\includegraphics[height=0.3\textwidth,keepaspectratio,clip,trim=72 0 0 0] 
{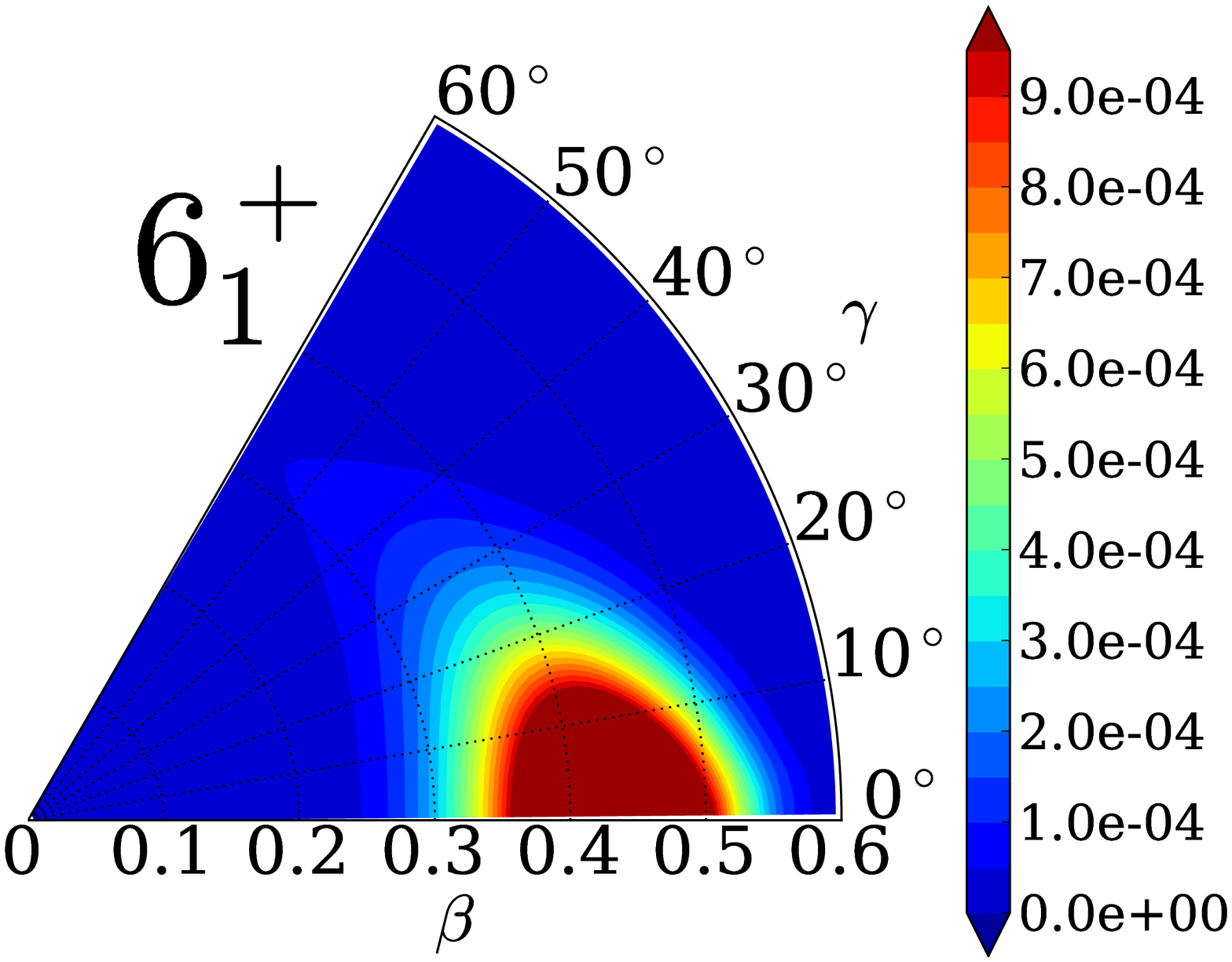} \\
\includegraphics[height=0.3\textwidth,keepaspectratio,clip,trim=72 0 160 0]
{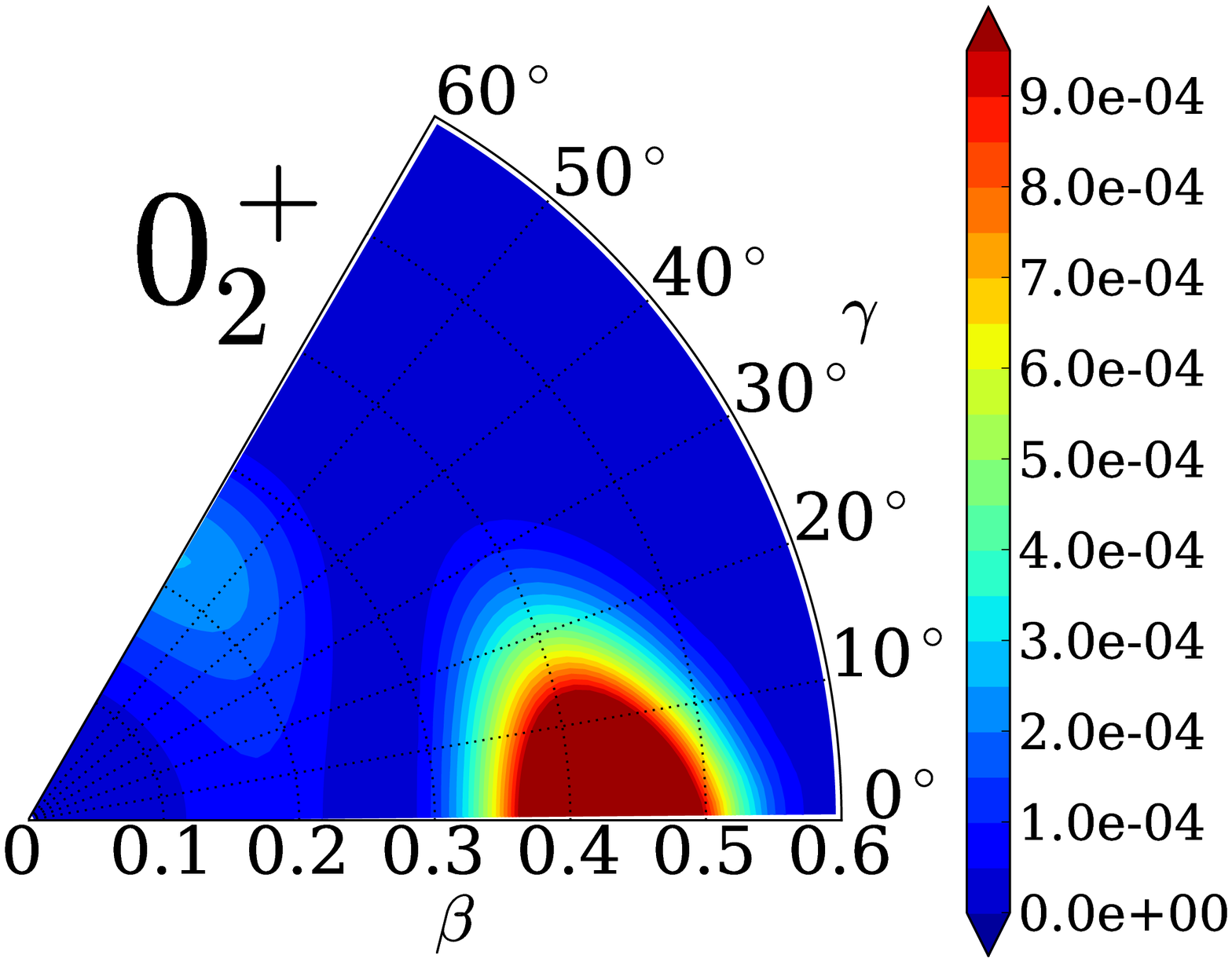} 
\includegraphics[height=0.3\textwidth,keepaspectratio,clip,trim=72 0 160 0]
{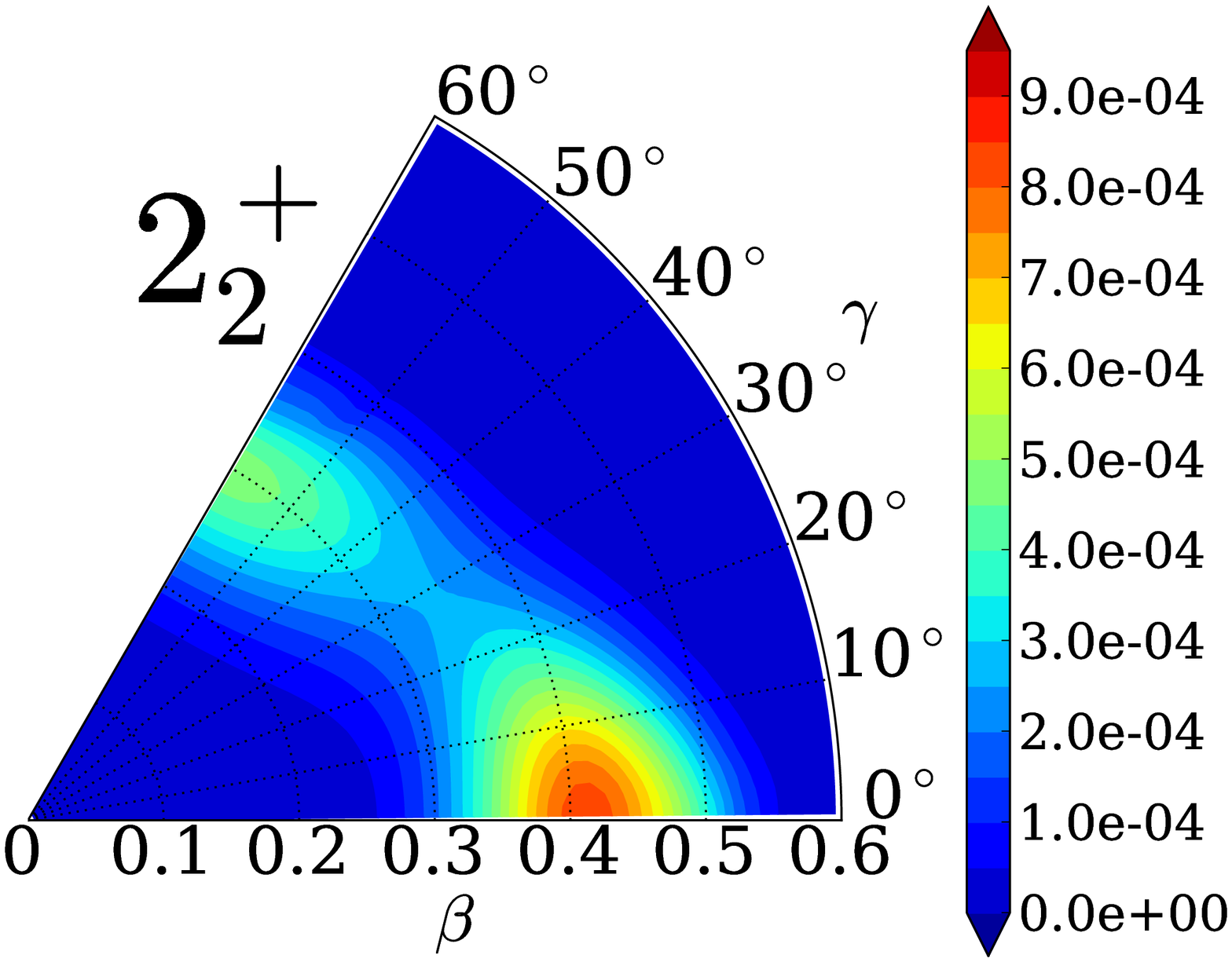} 
\includegraphics[height=0.3\textwidth,keepaspectratio,clip,trim=72 0 160 0]
{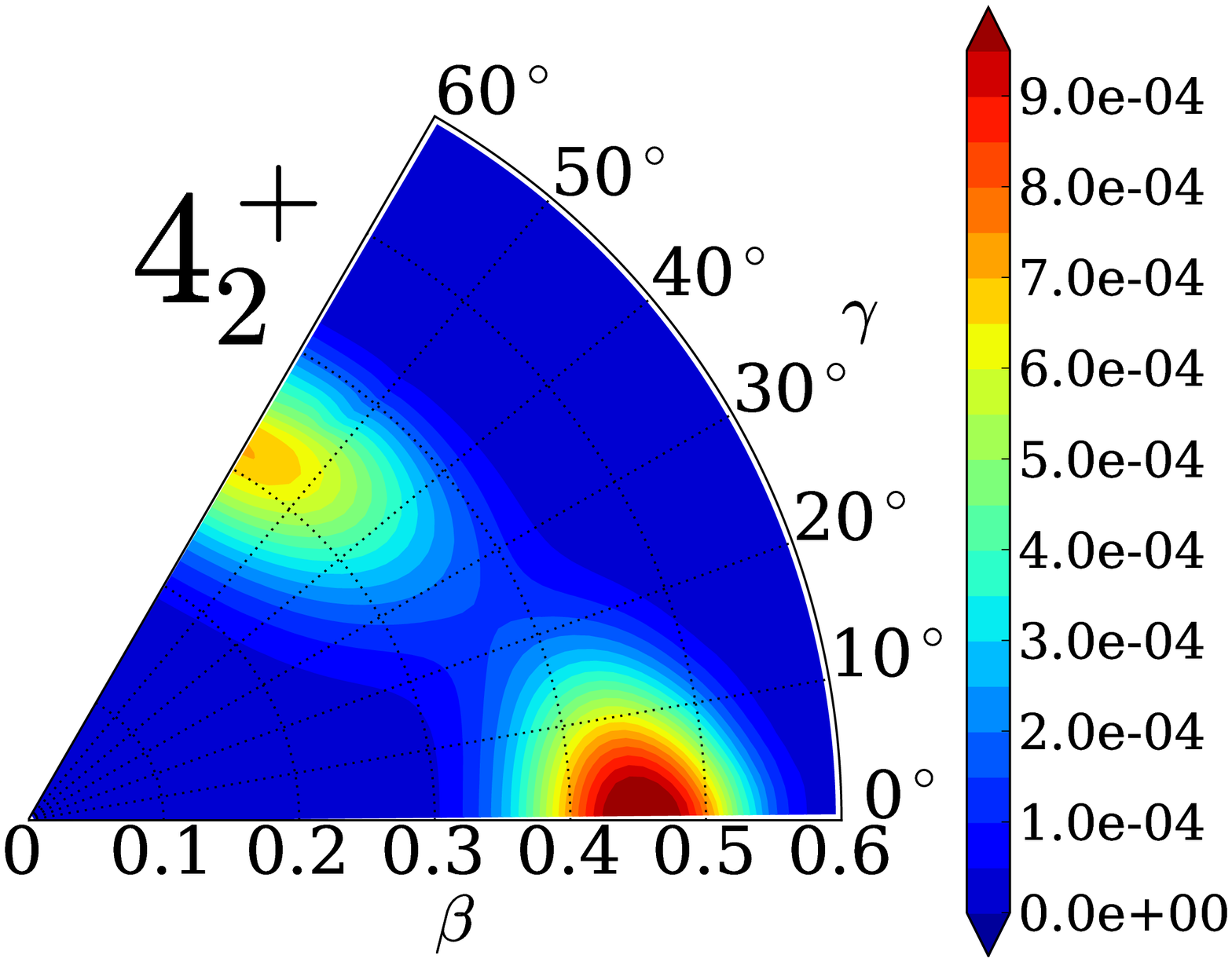} 
\includegraphics[height=0.3\textwidth,keepaspectratio,clip,trim=72 0 0 0]
{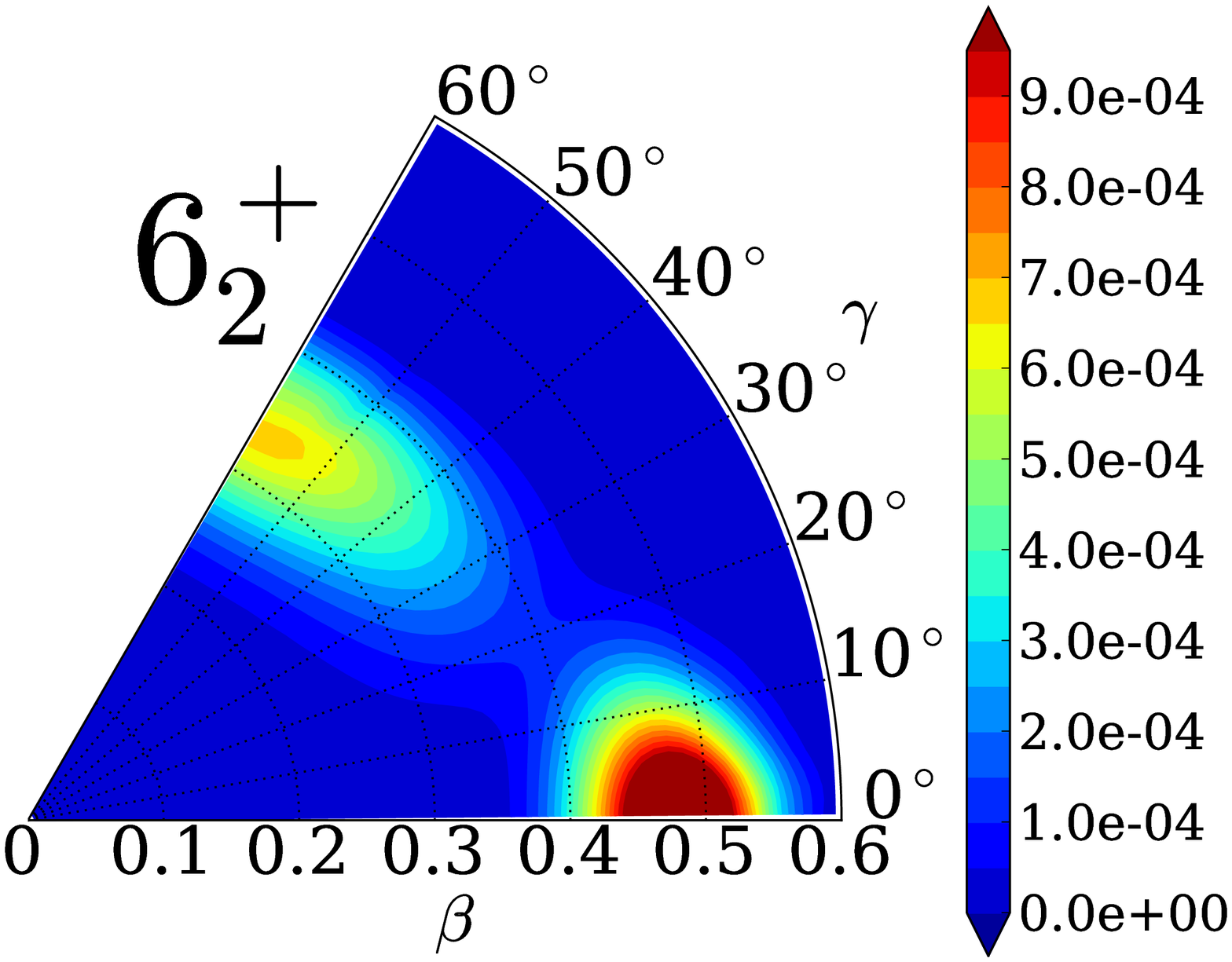} \\
\includegraphics[height=0.3\textwidth,keepaspectratio,clip,trim=72 0 160 0]
{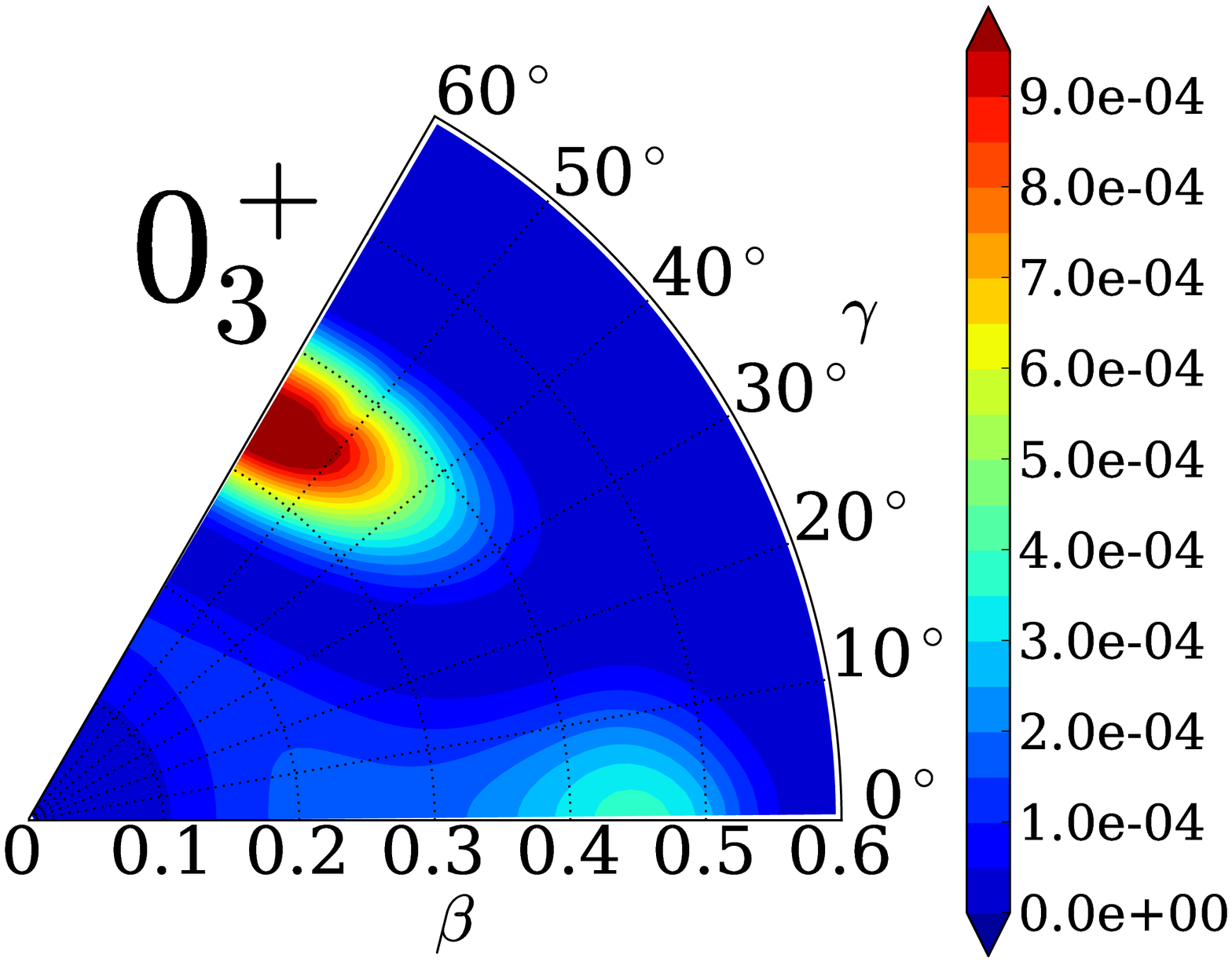} 
\includegraphics[height=0.3\textwidth,keepaspectratio,clip,trim=72 0 160 0]
{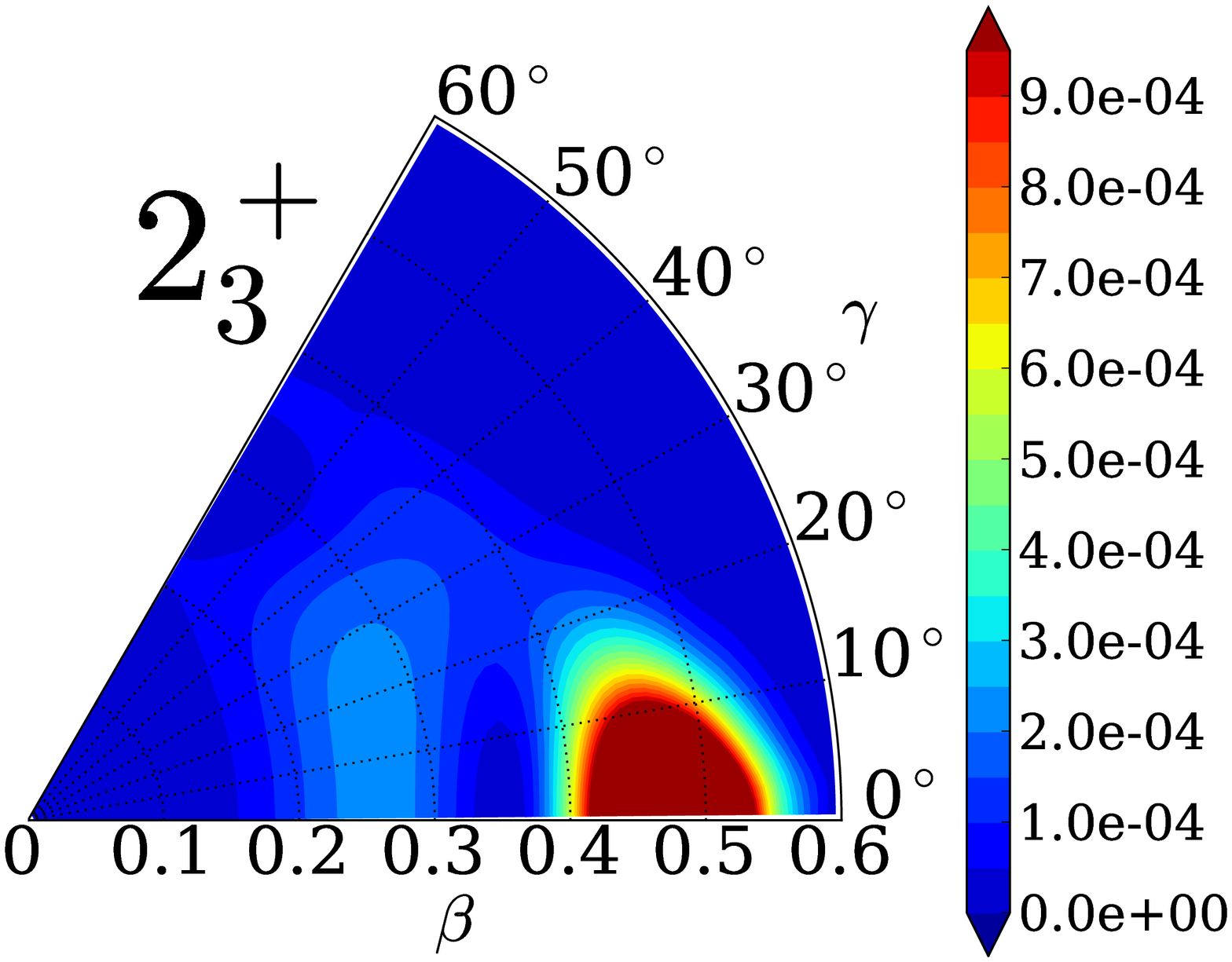} 
\includegraphics[height=0.3\textwidth,keepaspectratio,clip,trim=72 0 160 0]
{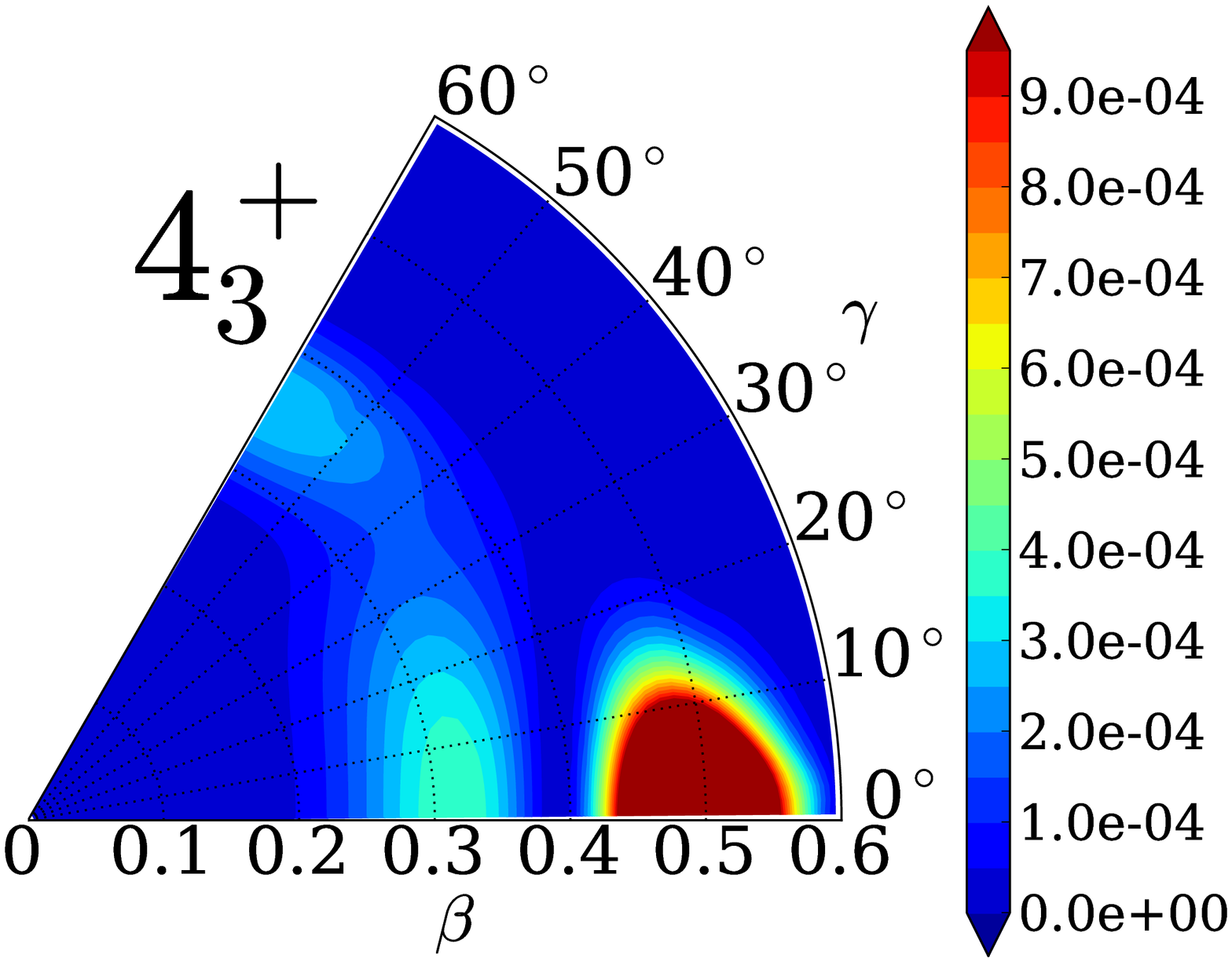} 
\includegraphics[height=0.3\textwidth,keepaspectratio,clip,trim=72 0 0 0]
{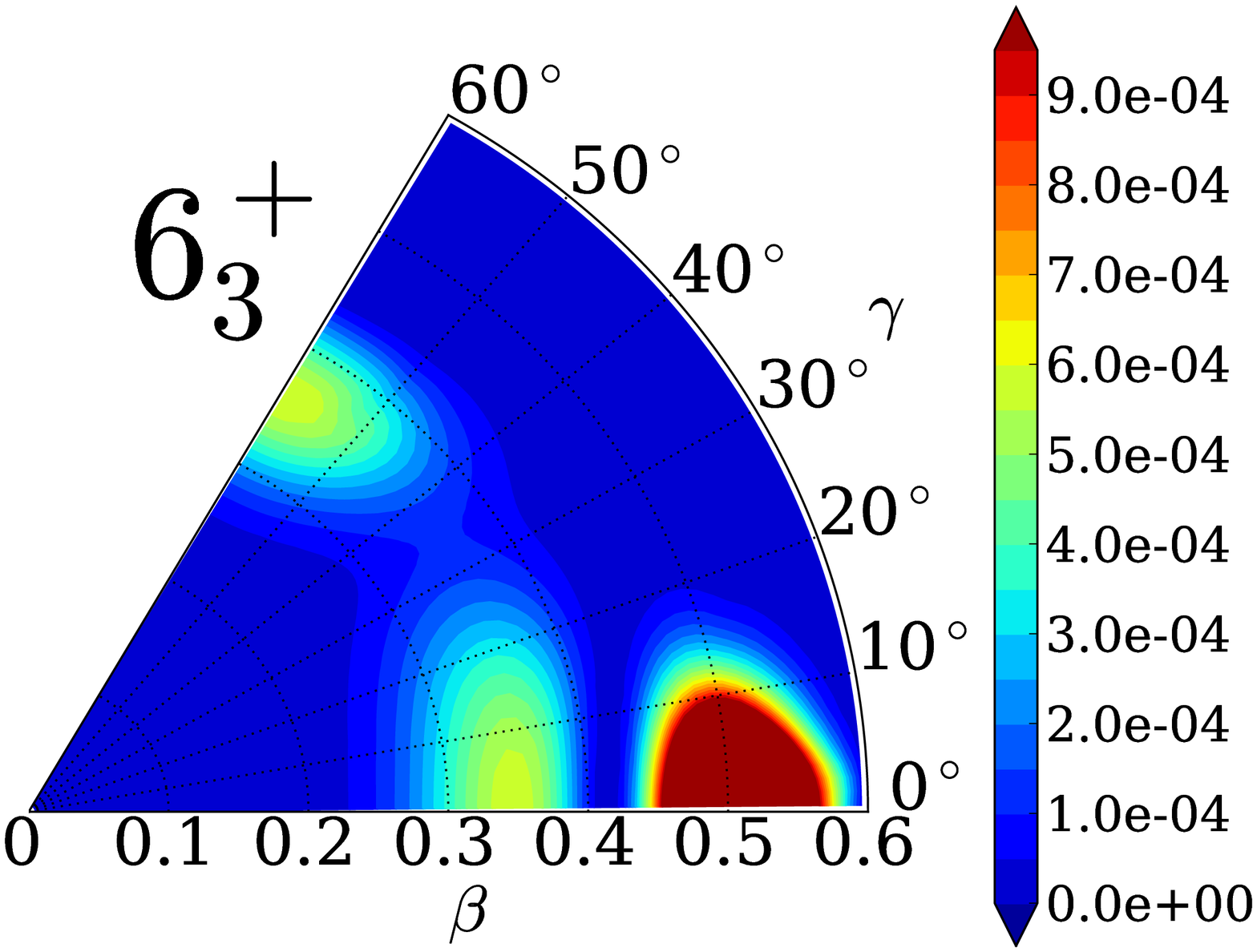}
\end{tabular} 
\end{center}
\caption{Vibrational wave functions squared, 
$\beta^4|\Phi_{\alpha I}(\beta,\gamma)|^2$, for $^{76}$Kr.}
\label{fig:wf76}
\end{figure}

\begin{figure}[htb]
\begin{center}
\subfigure[\kr{72}]{\includegraphics[width=0.4\textwidth,keepaspectratio,clip]{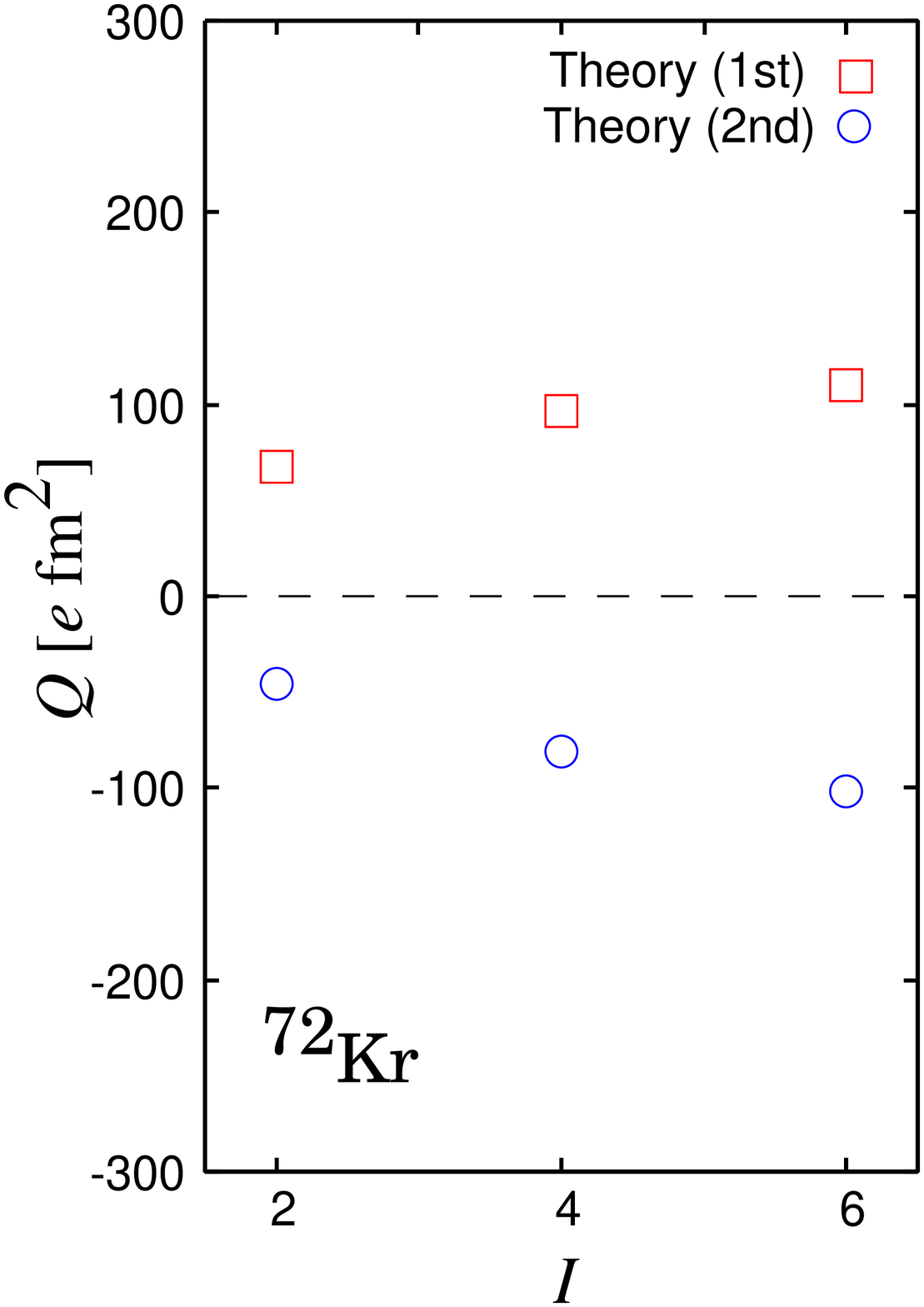}} 
\subfigure[\kr{74}]{\includegraphics[width=0.4\textwidth,keepaspectratio,clip]{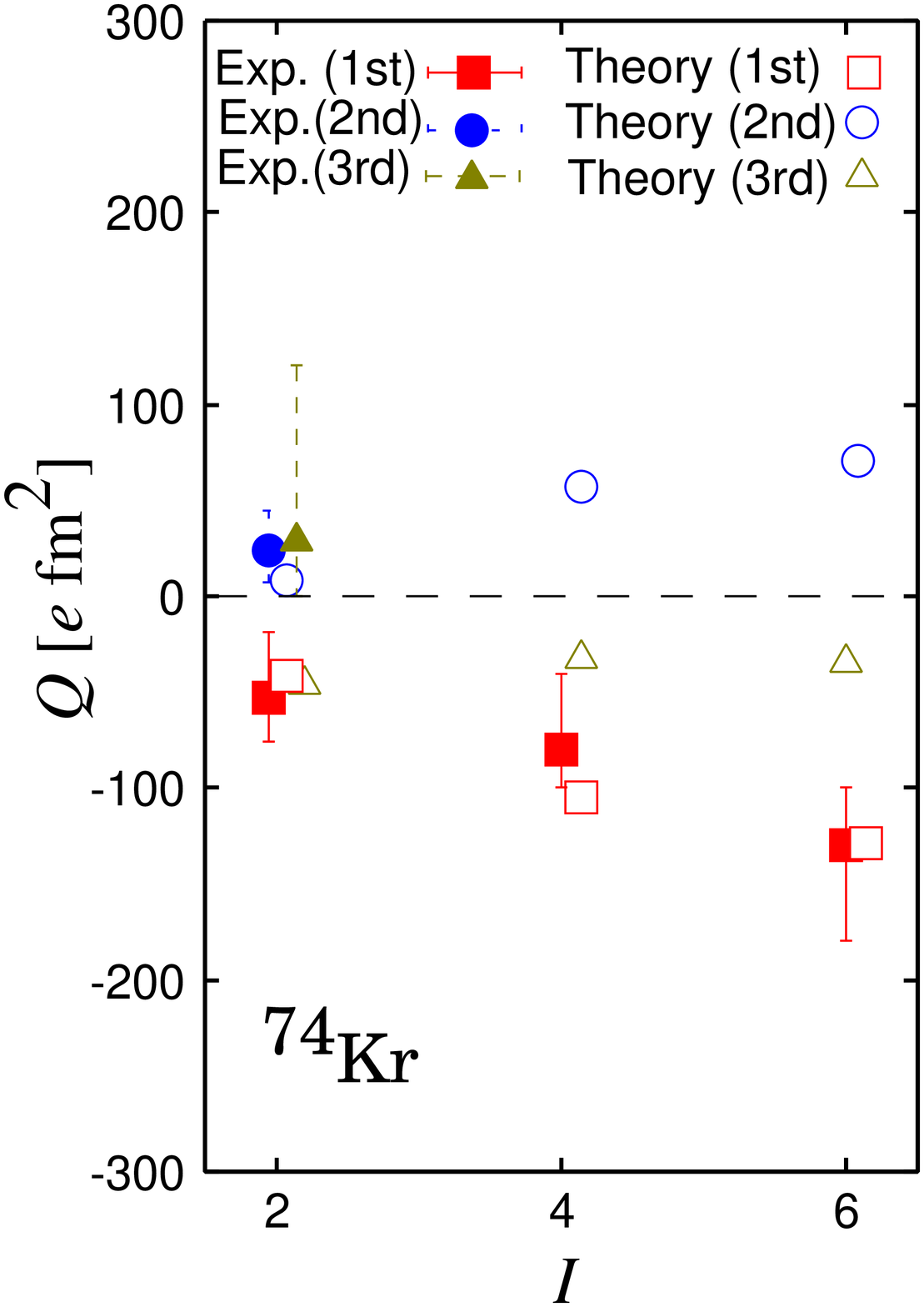}} \\  
\subfigure[\kr{76}]{\includegraphics[width=0.4\textwidth,keepaspectratio,clip]{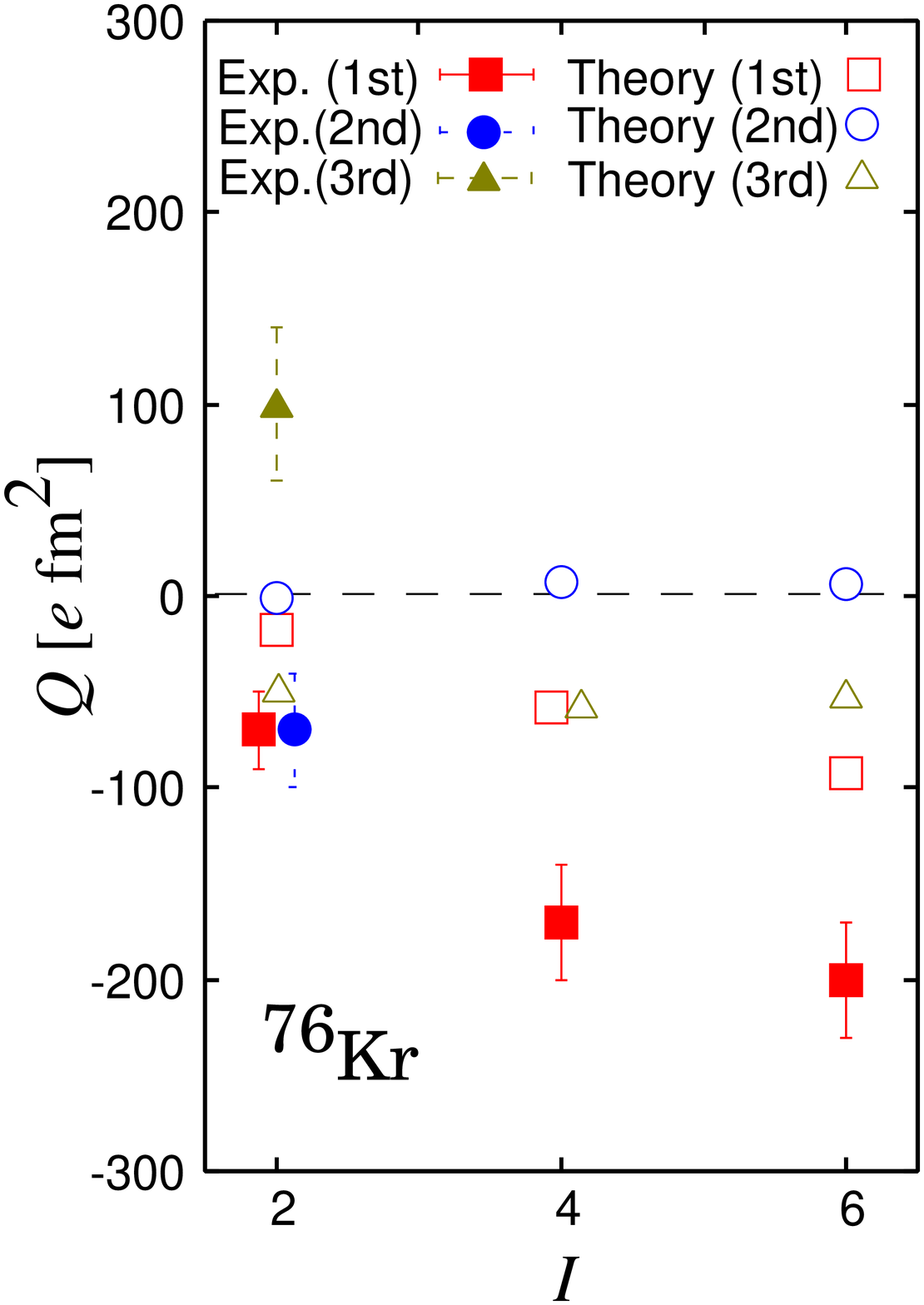}}  
\end{center}
\caption{Spectroscopic quadrupole moments in unit of $e~{\rm fm}^2$ of  
the first (square), second (circle) and third (triangle) states 
for each angular momentum in $^{72,74,76}$Kr.
Calculated values are shown by open symbols, 
while experimental data \cite{Clement2007} are indicated by filled symbols.}
\label{fig:Q}
\end{figure}

 The excitation spectrum and $B(E2)$ values calculated for \kr{76} 
are presented in Fig.~\ref {fig:spectra76} 
together with the experimental data,   
while the collective wave function squared are displayed in Fig.~\ref{fig:wf76}. 
 The behavior of the collective wave function for \kr{76} is qualitatively similar to that for \kr{74},
 except that the development of the localization of the yrare wave function is weaker and it has
 a more remarkable two-peak structure. 
The mechanism of appearance of the two-peak structure is discussed
 from a general viewpoint of oblate-prolate symmetry breaking in Ref.~\cite{Sato2010}. 

 The difference to be most noted between \kr{74} and \kr{76} is 
the location of the minimum of the collective potential. 
 One can see the importance of the rotational effect from Fig.~\ref{fig:wf76}.
 Although the potential has the oblate absolute minimum in \kr{76}, 
the vibrational wave functions of the $2_1^+$, $4_1^+$ and $6_1^+$ states 
still localize on the prolate side.
 It implies that the potential and rotational effects compete and 
the rotational effect dominates over the potential one with increasing angular momentum.
 This calculation serves as a good example suggesting that
 the nuclear shape cannot be determined only by 
the location of the potential minimum. 

The spectroscopic quadrupole moments, shown in Fig.~\ref{fig:Q}(c), 
qualitatively agree with the experimental data.
The sign and increase with angular momentum are well reproduced for the ground band.
Their absolute magnitudes are, however, too small compared with the experimental data. 
This seems to indicate that the calculated magnitudes of the quadrupole deformation 
and/or the localization of the collective wave functions in the $(\beta, \gamma)$ space 
are insufficient. 
It is interesting to note that the calculated quadrupole moment for the $2_3^+$ state 
better agrees with the experimental data for the $2_2^+$ state. 
In this connection, we notice that the $E2$ transition properties of the observed $2_2^+$ state 
in \kr{76} are different from those in \kr{74}:  
while it decays to the $2_1^+$ and $0_2^+$ states with almost equal $E2$ strengths in \kr{74},
it decays almost totally to the $0_2^+$ state in \kr{76}.   
This point was emphasized by Cl\'{e}ment et~al.~\cite{Clement2007}.  
This difference is qualitatively reproduced in our calculation 
if we assume that, in \kr{76}, the band consisting of the experimental 
$0_2^+,2_2^+,4_2^+$ and $6_2^+$ states corresponds to the band 
comprised of the calculated 
$0_2^+,2_3^+,4_3^+$ and $6_3^+$ states.

 \subsection{Discussion}
 
\begin{figure}[h]
\begin{center}
\includegraphics[width=0.5\textwidth,keepaspectratio,clip]{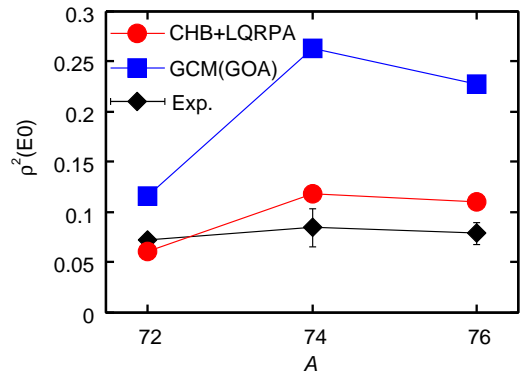}
\caption{Electric monopole transition matrix element $\rho^2(E0)$ 
calculated using Eq.~(\ref{eq:rhoE0})
in comparison with 
the experimental data \cite{Chandler1997,Bouchez2003} 
and the result of the HFB-based GCM (GOA) calculation \cite{Girod2009}.}
\label{fig:rho2E0}
\end{center}
\end{figure}

In addition to the $E2$ transition properties and the quadrupole moments of the 
low-lying states, we have also calculated $E0$ transition strengths $\rho^2 (E0)$.  
The result of calculation is compared with the experimental data in Fig.~\ref{fig:rho2E0}. 
We see that the calculated result well reproduces the experimental data 
both qualitatively and quantitatively.
The $\rho^2 (E0)$ takes a maximal value at $A=74$, which reflects the shape transition.
These $E0$ transition strengths are evaluated using Eq.~(\ref{eq:rhoE0})  
where the phenomenological Bohr relation \cite{Bohr1975, Kumar1975} 
between the $E0$ matrix elements and the quadrupole deformation is assumed.  
In this sense, this way of calculating  $\rho (E0)$ is semi-microscopic,  
in contrast to the calculation of $B(E2)$ and spectroscopic quadrupole moments 
which are carried out in a fully microscopic way using the expectation values of  
the quadrupole operators at each point on the ($\bg$) deformation space, 
as explained in Section \ref{subsec:Electric}.
For a fully microscopic calculation of $\rho(E0)$ without assuming Bohr's relation, 
we need to use a more realistic effective interaction than Baranger-Kumar's version of 
the P+Q force model \cite{Baranger1968}.  This task is left for future works.
Note that $\rho (E0)$ vanishes in Baranger-Kumar's P+Q force model,   
because this force is designed such that the radial matrix elements of 
the monopole operators take the same value for all single-particle states.

Let us add a few remarks concerning the results of calculation for $E2$ transition properties  
shown in Figs.~\ref{fig:spectra74} and \ref{fig:spectra76}. 
Although we have succeeded in reproducing a number of features observed in experiments, 
some aspects in experimental data remains unexplained. 
In particular, the large value of $B(E2;0_3^+ \rightarrow 2_1^+)$ 
in \kr{74} is not reproduced in the calculation.
In this connection, it should be emphasized that the properties of the excited $0^+$ states 
are quite sensitive to the mixing of the $\beta$~vibrations
and the large-amplitude $\gamma$~vibrations 
(large-amplitude fluctuations in the triaxial shape degree of freedom). 
In this study, we have not adjusted the interaction parameters by fitting the theoretical 
result to the experiment,
and the interaction we have employed itself is rather simple.
Therefore, it remains for future to examine whether or not these disagreements 
with experiments are improved when a better interaction is used.

As we mentioned in Introduction, 
the low-lying states of light Kr isotopes have recently been studied also 
by Girod et al. \cite{Girod2009} by means of the CHFB-based GCM (GOA) method.  
Consistent with our results, their calculation also  
demonstrates that the large-amplitude fluctuation in the triaxial shape degree of freedom 
plays an important role in the description of the shape coexistence 
and the shape transition in the light Kr isotopes. 
They have used the Gogny D1S force, which is more realistic than the P+Q force.
However, their inertial masses are derived with the cranking formula, 
which ignores the contribution from the time-odd mean field.
We would like to emphasize that there is no problem to use a more realistic interaction
within the framework of the CHFB+LQRPA method, 
although numerical calculation becomes much more demanding. 
The reason why we have used the schematic P+Q interaction in this paper is 
simply because this is the first application of the CHFB+LQRPA method to 
real nuclear collective phenomena.  
We shall employ a more realistic interaction like Skyrme interactions in future works. 
It will also become possible to use the CHFB+LQRPA method in conjunction 
with modern nuclear energy density functionals. 

\section{Conclusion}
\label{sec:conclusion}

We have studied the oblate-prolate shape coexistence/mixing in the low-lying states 
of \kr{72,74,76} using 
the 5D quadrupole Hamiltonian determined microscopically by means of 
the CHFB+LQRPA method, which is based on the 2D ASCC method.
Our results indicates 
a shape transition from oblate in the ground state of \kr{72} to prolate in \kr{74,76}, 
which is consistent with the experimental data.
We have shown that the basic features of the low-lying spectra in these nuclei 
are determined by the interplay of the large-amplitude vibrations 
in the triaxial shape degree of freedom, 
the $\beta$-vibrational excitations and the rotational motions. 
We have furthermore shown that the rotational motion plays an important role 
for the growth of localization of the vibrational wave functions 
in the ($\beta, \gamma$) deformation space.


\vspace{2em}
The authors thank T. Nakatsukasa, M. Matsuo, and K. Matsuyanagi for discussions.
K. S. and N. H. are supported by the Junior Research Associate Program and Special postdoctoral
Researcher Program of RIKEN, respectively.
The numerical calculations were carried out on Altix3700 BX2 at Yukawa 
Institute for Theoretical Physics in Kyoto University
and RIKEN Cluster of Clusters (RICC) facility.
This work is supported by
the Core-to-Core Program 
of the Japan Society for the Promotion of Science
"International Research Network for Exotic Femto Systems".

\bibliographystyle{model1a-num-names}
\bibliography{Ref4Krpaper}

\clearpage

\end{document}